\documentclass[showpacs,preprintnumbers,amsmath,amssymb,12pt]{revtex4}

\usepackage{amsmath, amssymb, mathrsfs, bbm}
\usepackage{graphicx}
\usepackage{dcolumn}
\usepackage{bm}

\usepackage{bbm}

\newcommand{\beqs}{\begin{equation*}}
\def\beq{\begin{equation}}

\newcommand{\eeqs}{\end{equation*}}
\def\eeq{\end{equation}}

\newcommand{\beqas}{\begin{eqnarray*}}
\newcommand{\beqa}{\begin{eqnarray}}

\newcommand{\eeqas}{\end{eqnarray*}}
\newcommand{\eeqa}{\end{eqnarray}}

\newcommand{\al}{\alpha}

\newcommand{\de}{\delta}

\newcommand{\ka}{\kappa}

\newcommand{\si}{\sigma}

\newcommand{\Ga}{\Gamma}

\newcommand{\Om}{\Omega}

\newcommand{\La}{\Lambda}
\newcommand{\Si}{\Sigma}

\newcommand{\blist}{\begin{itemize}}

\newcommand{\elist}{\end{itemize}}

\providecommand{\href}[2]{#2}

\DeclareFontFamily{OT1}{rsfs}{}
\DeclareFontShape{OT1}{rsfs}{m}{n}{ <-7> rsfs5 <7-10> rsfs7 <10->rsfs10}{} 
\DeclareMathAlphabet{\mycal}{OT1}{rsfs}{m}{n}

\def\cV{{\cal V}}
\def\cL{{\cal L}}

\def\cM{{\cal M}}

\def\cA{{\cal A}}

\def\TrL2{{\rm Tr}_{L^2}}

\def\atbdry{\Big|_{\partial \cM}}
\def\atbdry0{\Big|_{\partial \cM_0}}
\def\atbdry1{\Big|_{\partial \cM_1}}

\newcommand{\bra}[1]{\langle #1 |}
\newcommand{\ket}[1]{| #1 \rangle}

\newcommand{\braket}[2]{\langle #1 | #2 \rangle}

\begin{document}


\title{Vertex Operators and Scattering Amplitudes of the Bosonic Open String Theory in the Linear Dilaton Background}


\author{Chuan-Tsung Chan}
\email{ctchan@thu.edu.tw}
\author{Wei-Ming Chen}
\email{tainist@gmail.com}
\affiliation{Department of Physics, Tunghai University, Taiwan, 40704}


\date{\today}

\begin{abstract}
The operator formalism of the first quantized string theory is applied to the stringy excitations in the linear dilaton background. In particular, the normal-ordered vertex operators in the old-covariant spectrum of the bosonic open string, which correspond to the physical state solutions of the Virasoro constraints, are shown to satisfy the conformal algebra. Tree-level scattering amplitudes among different stringy states are computed using the coherent-state method and the modified inner product of the Hilbert space. Decoupling of the zero-norm states, i.e., the on-shell stringy Ward identities are shown to hold for all tree-level amplitudes under consideration.
\end{abstract}


\maketitle
\clearpage
\section{Introduction}
\par Our present understanding of the quantum theory of string often relies on two approximation schemes: in the low energy approximation ($\al^\prime E^2\rightarrow 0$), we ignore the massive stringy excitations and work with supergravity degrees of freedom \cite{Green:1987sp,Lust:1989tj,Scherk:1974jj,Yoneya:1974jg}, or we assume that the string coupling constants is small and expand all correlation functions in power series of $g_s$ (genus expansion) \cite{Polyakov:1981rd,Polyakov:1981re,Hatfield:1992rz}. The need of a non-perturbative formulation of string theory is further hampered by another practical issue: the background independence of the current formulation. Like the case in quantum field theory, we are always free to choose a particular classical background to approximate the physical situation of interest and only treat the quantum effect as small corrections. A "good" choice of background implies a fast convergence of the perturbative expansion. But if we know how to sum up all perturbative series (if this is possible) for a given background, we would expect any choice of background should lead to the same answer. While this expectation can be partially realized in the context of string field theory, where one can prove the physical content are equivalent for two continuously connected backgrounds \cite{Sen:1990hh,Sen:1990na,Sen:1992pw}, one would still hope that there are more direct approaches to resolve this issue, in analogous to the formulation of the electroweak theory in the manifestly symmetric vacuum.
\par Since string theory naturally includes gravity and gauge theories, we can follow the past wisdom and look for any possible symmetry principle underlying the string theory \cite{Gross:1988ue,Witten:1988sy}. In the bottom-up approach, one would hope to extract such a symmetry principle based on the patterns in the scattering amplitudes. Nevertheless, at least within the present framework, any such attempt,must face the problem of background dependence mentioned above. For a given conformal invariant background, we need to solve for the spectrum of stringy excitations and calculate the scattering amplitudes separately. One can gather many useful information and interesting patterns from different backgrounds and various kinematic limits \cite{Gross:1987kza,Gross:1987ar,Gross:1989ge,Chan:2003ee,Chan:2004yz,Chan:2005ne,Ho:2006zu,Chan:2006qf,Chan:2006pf,Lee:2006fp,Lee:2007cx,Lee:2007dp,Lee:2008ba,Ko:2008re,Ko:2008ft}. However, if we wish to look for a symmetry principle underlying string theory, we should be able to relate these symmetry patterns and extract a universal property from these calculations. For this reason, it would be of interest to study a continuous family of conformal invariant backgrounds and examine the formulation of stringy spectrum and dynamics. Specifically, we would like to study the spectral flow of massive stringy excitations and the deformation of any string global symmetry. Bosonic open string theory in the linear dilaton background \cite{Polchinski:1998rq,Ho:2007ar} is one of the simplest examples that suits this purpose and we will make a detailed study of its quantum dynamics in this paper. Here the gradient of the dilaton field $V^\mu\equiv\partial^\mu\Phi$ serves as the spectral parameters, ($V^\mu=0$ corresponds to the flat space-time) and we shall uncover some interesting phenomena as we turn up these moduli parameters.\par
There are other independent interests for studying the string theory in the linear dilaton background. For one thing, this is the simplest time-dependent background for stringy theory, and one hope to extract useful lesson from these studies for cosmology. There are also calculations based on functional integral approach where the curvature at world-sheet boundary leads to a modified energy-momentum conservation rule. In this paper, not only we extend the detailed calculations to the massive stringy excitations, but we also wish to provide a different perspective for illustrating this issue. For this reason, we adopt the operator formalism in the first quantized string theory. The explicit solutions of massive vertex operators and the relation to the covariant quantum spectrum hopefully will provide a detailed explanation of the concise derivation in the standard textbook \cite{Green:1987sp,Chan:2009vx}.

This paper is organized as follows: we first review the first quantized bosonic open string theory in the linear dilaton background in section \ref{IIA}. The Virasoro constraints and the covariant physical state solutions are discussed in section \ref{IIB}.
In section \ref{III}, based on the results, we write down the covariant vertex operators for each physical state, and show that the normal-ordered form of these vertex operators indeed satisfy the conformal algebra. With these explicit solutions of the vertex operators, we calculate various scattering amplitudes of bosonic open string excitations in the linear dilaton background. Finally, in section \ref{IV}, we verify the stringy Ward identities by showing the decoupling of zero-norm states in the stringy scattering amplitudes under consideration.

\section{First Quantization of Bosonic Open String Theory in the Linear Dilaton Background}
\subsection{Polyakov action for the bosonic open string theory in the linear dilaton background}
Since our starting point is very similar to that of \cite{Ho:2007ar}, we shall follow the notations in \cite{Ho:2007ar} closely. The Polyakov action for the bosonic open string theory in the linear dilaton background is given by
\beqa
\notag S&=&\frac{1}{4\pi\al^\prime}\int_\Si d^2 \si \sqrt{g}g^{ab} \partial_aX(\si)\cdot\partial_bX(\si)
+\frac{1}{4\pi}\int_\Si d^2\si \sqrt{g}R(\si)V\cdot X(\si)\\
&&+\frac{1}{2\pi}\int_{\partial\Si}ds\ka(\xi)V\cdot X(\xi),
\eeqa
here $R(\si)$ is Ricci scalar of the world-sheet $\Si$, $\ka(\xi)$ is the geodesic curvature along the boundary of the world-sheet $\partial \Si$, and $V^\mu\equiv\partial^\mu\Phi$ is the gradient vector of the dilaton field $\Phi$.
From this we can extract the energy-momentum tensor,
\beqa
\label{EM}T_{zz}=-\frac{1}{\al^\prime}:\partial X\cdot\partial X:+V\cdot\partial^2X.
\eeqa
The string coordinates in the oscillator representation are
\beqa
X^\mu(z,\bar{z})&=&x^\mu-i\al^\prime p^\mu \ln|z|^2+i\sqrt{\frac{\al^\prime}{2}}\sum_{m=-\infty, m\neq0}^{m=\infty}\frac{\al_m^\mu}{m}(z^{-m}+\bar{z}^{-m}).
\eeqa
The Virasoro generators of the conformal transformation is defined as the Fourier modes of the energy-momentum tensor $T_{zz}$,
\beqa
\label{cLm}\cL_m\equiv\oint dz z^{m+1}T_{zz}=\frac{1}{2}\sum_{n=-\infty}^{\infty}:\al_{m-n}\al_n:+i\sqrt{\frac{\al^\prime}{2}}(m+1)V\cdot\al_m.
\eeqa

Using the basic commutation relation among oscillators, $\big[\al_m,\al_n\big]=m\de_{m+n}$, one can check that the Virasoro generators satisfy the following algebra relation,
\beqa
\big[\cL_m,\cL_n\big]=\big(m-n\big)\cL_{m+n}+\frac{D+6\al^\prime V^2}{12}m\big(m^2-1\big)\de_{m+n}.
\eeqa
Notice that the central charge $c\equiv D+6\al^\prime V^2$ includes a term which is in proportion to $V^2\equiv V^\mu V_\mu$ and conformal symmetry implies that we can have different space-time dimension $D$ depending on the sign of $V^2$ (space-like $V^2>0$ $\Rightarrow$ $D<26$, time-like $V^2<0$ $\Rightarrow D>26$). Even though the effect of Liouville potential is ignored in this paper, we make an effort to maintain all the $V^2$ terms throughout our calculations. In our later calculations of the vertex operators and stringy scattering amplitudes, we find the following notations useful:
\begin{enumerate}
\item First of all, the world-sheet time $\tau$ is related to the complex variable $z$ at the open string boundary as $y=e^{i\tau}$ and the derivative with respect to $\tau$ is denoted as a dot over operators.
\item All the operators in mathcal letters refer to bosonic open string operators in the linear dilaton background. Operators with capitalized Roman letters stand for bosonic open string operators in flat space-time. For instance: (1) the shifted string coordinates $\mathcal{X}^\mu$ is defined as $\mathcal{X}^\mu\equiv X^\mu+i\al^\prime V^\mu$, (2) the Virasoro generators are separated into two terms,  $\cL_m=L_m+i\sqrt{\frac{\al^\prime}{2}}(m+1)V\cdot\al_m$, (3) mathcal letters of vertex operators corresponds to bosonic open string physical state in the linear dilaton background and capital Roman fonts are reserved for flat space-time vertex operators.
\end{enumerate} \label{IIA}
\subsection{Virasoro Constraints of physical states in the bosonic open string theory in the linear dilaton background}
The physical spectrum of the bosonic open string theory in the linear dilaton background is defined similarly to that of flat space-time \cite{Lee:1994wp,Chan:2005qd}. In the oscillator representation, we solve all possible linear combinations of creation operators acting on a Fock vacuum, subject to the Virasoro constraints:
\beqa
\cL_0\ket{\Phi(k)}=\ket{\Phi(k)},\quad \mbox{and} \quad \cL_n \ket{\Phi(k)}=0,\quad n\geqslant1.
\eeqa
These constraints in general lead to the generalized on-shell condition for the center of mass momenta, and restrict the polarization tensors to be transverse and traceless.
In the covariant spectrum, where physical states consist of linear combinations of oscillators with various polarization tensors, Virasoro constraints implies lower spin polarization tensors are given by the projections of higher spin polarization tensors.
In this paper, we shall focus on the physical states up to the first massive level, and we shall use capital letter to represent the particles. For instance, the tachyon state (T) is defined as
\beqa
\ket{T(k)}\equiv\ket{0,k}.
\eeqa
The $\cL_0$ condition
\beqa
\cL_0\ket{T(k)}=\Big(\frac{1}{2}\al_0^2+i\sqrt\frac{\al^\prime}{2}V\cdot\al_0\Big)\ket{0,k}=0,
\eeqa
together with the eigenvalue condition for $\al_0^\mu$, $\al_0^\mu\ket{k,0}=\sqrt{2\al^\prime}k^\mu\ket{k,0}$ leads to generalized on-shell condition
\beqa
\label{L0T}\al^\prime k\cdot(k+iV)=1.
\eeqa
The $\cL_1$ and $\cL_2$ conditions are trivial for the tachyon state.
\par At massless level, we have a photon state (P) with polarization vector $\zeta(k)$,
\beqa
\ket{P(\zeta, k)}\equiv\zeta\cdot\al_{-1}\ket{0,k}.
\eeqa
One can check that the $\cL_0$ condition,
\beqa
\cL_0\ket{P(\zeta, k)}=\Big(\al_{-1}\cdot\al_{1}+i\sqrt{\frac{\al^\prime}{2}}V\cdot\al_0\Big)\zeta\cdot\al_{-1}\ket{0,k}=0,
\eeqa
leads to
\beqa
\label{L0P}\al^\prime k\cdot(k+iV)=0.
\eeqa
On the other hand, the $\cL_1$ condition
\beqa
\cL_1\ket{P(\zeta, k)}=\Big(\al_{0}\cdot\al_{1}+i\sqrt{2\al^\prime}V\cdot\al_1\Big)\zeta\cdot\al_{-1}\ket{0,k}=0,
\eeqa
gives the generalized transverse condition,
\beqa
\label{L1P}\zeta\cdot(k+iV)=0.
\eeqa
The $\cL_2$ condition is trivial for the photon state.
\par At the first massive level (M), we have a tensor particle with spin-two, and it is written as
\beqa
\ket{M(\epsilon_{\mu\nu},k)}=\big(\epsilon_{\mu\nu}\al^\mu_{-1}\al^\nu_{-1}+\epsilon_{\mu}\al^\mu_{-2}\big)\ket{0,k}.
\eeqa
The generalized on-shell condition, as derived from the $\cL_0$ condition
\beqa
\cL_0\ket{M(\epsilon_{\mu\nu},k)}=\Big(\al_{-1}\cdot\al_{1}+\al_{-2}\cdot\al_{2}+i\sqrt{\frac{\al^\prime}{2}}V\cdot\al_0\Big)\Big(\epsilon_{\mu\nu}\al^\mu_{-1}\al^\nu_{-1}+\epsilon_{\mu}\al_{-2}^\mu\Big)\ket{0,k}=0,
\eeqa
is
\beqa
\label{L0M}\al^\prime k\cdot(k+iV)=1.
\eeqa
The $\cL_{1}$ conditions
\beqa
\cL_{1}\ket{M(\epsilon_{\mu\nu},k)}=
\Big(\al_{0}\cdot\al_{1}+\al_{-1}\cdot\al_{2}+i\sqrt{2\al^\prime}V\cdot\al_1\Big)\Big(\epsilon_{\mu\nu}\al^\mu_{-1}\al^\nu_{-1}+\epsilon_{\mu}\al_{-2}^\mu\Big)\ket{0,k}=0,
\eeqa
implies the polarization vector, $\epsilon_\mu$, can be written as a projection of the spin-two polarization tensor $\epsilon_{\mu\nu}$,
\beqa
\label{L1M}\sqrt{2\al^\prime}\epsilon_{\mu\nu}(k^\nu+iV^\nu)+\epsilon_\mu=0.
\eeqa
For this reason, we suppress the $\epsilon_\mu$ dependence in the notation of $M(\epsilon_{\mu\nu},k)$.
Finally, the $\cL_2$ conditions, $\cL_{2}\ket{M(\epsilon_{\mu\nu},k)}=0$,
\beqa
\Big(\al_{1}\cdot\al_{1}+\al_{0}\cdot\al_{2}+3i\sqrt{\frac{\al^\prime}{2}}V\cdot\al_2\Big)\Big(\epsilon_{\mu\nu}\al^\mu_{-1}\al^\nu_{-1}+\epsilon_{\mu}\al_{-2}^\mu\Big)\ket{0,k}=0
\eeqa
gives
\beqa
\label{L2M}\epsilon_{\mu\nu}\eta^{\mu\nu}+\sqrt{2\al^\prime}\epsilon_\mu\big(2k^\mu+3iV^\mu\big)=0.
\eeqa
Substituting $\epsilon_\mu$ from Eq.\eqref{L1M} to Eq.\eqref{L2M}, we get
\beqa
\label{Vra}2\al^\prime\epsilon_{\mu\nu}\big(k^\mu+iV^\mu\big)\big(2k^\nu+3iV^\nu\big)-\epsilon_{\mu\nu}\eta^{\mu\nu}=0.
\eeqa
Note that all of these relations contain explicit dependence on the linear dilaton gradient $V^\mu$, and one can verify that as $V^\mu$ goes to zero, we recover the previous results on physical spectrum for bosonic open string in flat space-time \cite{Chan:2005qd}. Consequently, it is natural to treat $V^\mu$ as moduli parameters and identify the solutions to the Virasoro constraints as a spectral flow. In addition to the interpretation of physical spectrum deformation (as a function of $V^\mu$), it is also crucial to emphasize that the inner product in the one string Fock space in the linear dilaton background is also deformed. Here we follow the prescription in \cite{Ho:2007ar} and define the inner product for the center of mass degrees of freedom of any stringy excitation,
\beqa
\label{in}\braket{k^\prime}{k}\equiv\big(2\pi\big)^D\de^{(D)}\big(k^{\prime\ast}-k-iV\big).
\eeqa
In our later derivation of the vertex operators, we shall see that how the requirement of the conformal invariance, in the form of conformal algebra, Eq.\eqref{ComforAlg}, will naturally reproduce the physical state conditions, Eqs.\eqref{L0T}, \eqref{L0P}, \eqref{L1P}, \eqref{L0M}, \eqref{L1M}, \eqref{L2M}.
\label{IIB}
\subsection{Zero-norm State Spectrum of the Bosonic Open String Theory in the Linear Dilaton Background}
Having derived the Virasoro constraints of physical states in the bosonic open string theory, it is natural to look for explicit solutions to these algebraic relations and to identify a set of basis for physical states at a given mass level. Fortunately, for our present purpose, we do not need these explicit solutions for calculating stringy scattering amplitudes or proving stringy Ward identities. However, for the verifications of stringy Ward identities,
 we need the explicit form of zero-norm state polarization tensors up to the first massive level, which we discuss in the following: 
\par At massless level, we have a type I singlet states
\beqa
\ket{\Phi}\equiv \cL_{-1}\ket{0,k}=(k\cdot\al_{-1})\ket{0,k},
\eeqa
with $\al^\prime k\cdot (k+iV)=0$. If we identify the momentum $k$ as polarization $\zeta$, then it is clear that the $\cL_1$ condition is the same as the $\cL_0$ condition. One can also check that the normalization of this state is zero, if we assume the modified inner product,
\beqa
\notag\braket{\Phi(k^\prime)}{\Phi(k)}&=&\bra{0,k^\prime}(k^\ast\cdot \al_1)(k\cdot \al_{-1})\ket{0,k}\\
&=&(k^{\prime\ast}\cdot k\cdot)\de(k^{\prime\ast}-k-iV)=k\cdot(k+iV)=0.
\eeqa
\par
At the first massive level, we have two types of zero-norm states (ZNS):
\begin{itemize}
  \item Type I vector ZNS\\
  \beqa
\cL_{-1}\ket{\chi}\equiv\cL_{-1}\big(\epsilon_\mu\al_{-1}^\mu\big)\ket{0,k},
\eeqa
where the "seed state" $\ket\chi\equiv\epsilon\cdot\al_{-1}\ket{0,k}$, satisfies the following conditions,
\beqa
  \cL_{0}\ket{\chi}=0 & \Rightarrow & \al^\prime k\cdot\big(k+iV\big)=-1 \quad\mbox{(on-shell condition)},\\
  \label{L1}\cL_{1}\ket{\chi}=0 & \Rightarrow & \epsilon\cdot \big(k+iV\big)=0 \qquad\mbox{(transverse condition)},
\eeqa
and $\cL_{2}\ket{\chi}=0$ holds automatically.
If we use the oscillator representation of Virasoro generator,
\beqas
\cL_{-1}\sim\al_{-1}\al_0+\al_{-2}\al_1,
\eeqas
we can read out the polarizations of the vector zero-norm states,
\beqa
\label{ZNSI}\cL_{-1}\ket \chi=\big(\epsilon_{\mu\nu}\al^\mu_{-1}\al^\nu_{-1}+\epsilon_\mu\al^\mu_{-2}\big)\ket{0,k},
\eeqa
where $\displaystyle\epsilon_{\mu\nu}=\sqrt\frac{\al^\prime}{2}\big(\epsilon_\mu k_\nu+\epsilon_\nu k_\mu\big)$.
  \item Type II singlet ZNS\\
  The type II ZNS at the first massive level can be calculated by the same formula as that in the flat space-time. We have
\beqa
\notag\ket{\varphi(k)}&\equiv&\big(2\cL_{-2}+3\cL_{-1}^2\big)\ket{0,k}\\
\label{ZNSII}&=&\Big[\epsilon_{\mu\nu}\al_{-1}^\mu\al_{-1}^\nu+\epsilon_\mu\al_{-2}^\mu\Big]\ket{0,k},\\
\notag\mbox{where}\qquad \epsilon_{\mu\nu}&=&6\al^\prime k_\mu k_\nu+\eta_{\mu\nu},\\
\notag\mbox{and}\qquad\quad \epsilon_\mu&=&\sqrt{2\al^\prime}\big(5k_\mu-iV_\mu\big).
\eeqa
One can check that the normalization of $\ket \varphi$ is
\beqas
\braket{\varphi(k^\prime)}{\varphi(k)}&=&2\Big[\epsilon_{\mu\nu}(k)\epsilon^{\ast\mu\nu}(k^\prime)
+\epsilon_\mu(k)\epsilon^{\ast\mu}(k^\prime)\Big]\de(k^{\prime\ast}-k-iV)\\
&=&4\left[\begin{split}&18\al^{\prime2}\big(k\cdot k^{\prime\ast}\big)^2+3\al^\prime k^2+3\al^\prime\big(k^{\prime\ast}\big)^2\\
&+25\al^\prime\big(k\cdot k^{\prime\ast}\big)+5i\al^\prime\big(k-k^{\prime\ast}\big)V+\frac{D^2}{2}+\al^\prime V^2\end{split}\right]\de(k^{\prime\ast}-k-iV).
\eeqas
Substitute the on-shell condition $\al^\prime k\cdot\big(k+iV\big)=-1$ and $\al^\prime k^\ast\cdot\big(k^\ast-iV\big)=-1$, one can verify that $\ket{\varphi(k)}$ is indeed a zero-norm state for all $D$ and $V^\mu$.
\end{itemize}
\par
One should be careful that, in general,
the choice of basis for the physical states depends on the kinematic set up. We refer the reader to \cite{Chan:2009kb} for the explicit solutions of the physical state spectrum of bosonic 
open string in the light-like linear dilaton background.  

\section{Normal-ordered Covariant Vertex Operators of Bosonic Open String Theory in the Linear Dilaton Background}\label{III}
\subsection{State-operator Correspondence}
Our main goal in this paper is to solve for the covariant vertex operators of bosonic open string theory and use them to calculate stringy scattering amplitudes. To construct general vertex operators, it is useful to recall the correspondence between states and operators in conformal field theory. In the oscillator representation for the one string Fock space, we have
\beqas
\al_{m}^\mu&=&\sqrt\frac{2}{\al^\prime}\oint\frac{dz}{2\pi}z^{-m}\partial^m X^\mu(z)\\
&\rightarrow& \sqrt\frac{2}{\al^\prime}\frac{i}{(m-1)!}\partial^mX^\mu(0)\\
x_0^\mu&\rightarrow&X^\mu(0)\qquad.
\eeqas
From this correspondence, we have the following dictionary for bosonic open string states and the correspondence vertex operators:
\beqas
\begin{array}{rrcl}
\mbox{tachyon (T)}:&\ket{0,k}&\sim&:e^{ik\cdot X}:\\
\mbox{photon (P)}:&\zeta\cdot\al^{-1}\ket{k,0}&\sim&:\zeta\cdot \partial X e^{ik\cdot X}:\\
\mbox{first massive state (M)}:\quad&\big(\epsilon_{\mu\nu}\al^\mu_{-1}\al^\nu_{-1}+\epsilon_{\mu}\al_{-2}^{\mu}\big)\ket{k,0}
&\sim&:\Big[\epsilon_{\mu\nu}(\partial X^\mu)(\partial X^\nu)+\epsilon_{\mu}\partial^2 X^\mu\Big]e^{ik\cdot X}:\quad.
\end{array}
\eeqas
Note that the partial derivative $\partial$ denote the differentiation with respect to complex world-sheet variable in the upper half plane $\partial\equiv\frac{\partial}{\partial z}$. In the strip diagram for open string world-sheet, we need to make a change of variable $z=e^{i\tau}$ to facilitate operator calculations (this also leads to some factors of $i$ in the expression, see Eq.\eqref{cVM}).

\subsection{Normal-ordering of the Covariant Vertex Operators}
In the following, we first derive the descending formulae \cite{Chan:2009vx} among nontrivial-ordered operators. The key equations are Eqs.\eqref{NorOrdVP} and \eqref{NorOrdVM}. First of all, we define the mode-decomposition of the string coordinates:
\beqas
X^\mu\equiv X^\mu_+ +X^\mu_0+X^\mu_-,
\eeqas
where
\beqa
\mbox{annihilation part of $X^\mu$} \qquad&\Rightarrow&\quad X^\mu_+\equiv\sqrt{2\al^\prime}\sum_{n=1}^\infty\frac{i}{n}\al_{n}e^{-in\tau}=\sqrt{2\al^\prime}\sum_{n=1}^\infty\frac{i}{n}\al_{n}y^{-n},\\
\mbox{zero mode part of $X^\mu$} \qquad&\Rightarrow&\quad X^\mu_0\equiv x^\mu+2\al^\prime p^\mu\tau,\\
\mbox{creation part of $X^\mu$}\qquad&\Rightarrow&\quad X^\mu_-\equiv-\sqrt{2\al^\prime}\sum_{n=1}^\infty\frac{i}{n}\al_{-n}e^{in\tau}=-\sqrt{2\al^\prime}\sum_{n=1}^\infty\frac{i}{n}\al_{n}y^n.
\eeqa
Some of the useful commutators are collected in Appendix \ref{ApxB}.\par
The vertex operator for tachyon (T) in flat space-time is defined as
\beqa
V_T\equiv:e^{ik\cdot X}:\equiv\exp\big(ik\cdot X_-\big)\exp\big(ik\cdot X_0\big)\exp\big(ik\cdot X_+\big).
\eeqa
The vertex operator for photon (P) in flat space-time is defined as
\beqa
\notag V_P&\equiv& :\frac{\zeta\cdot\dot{X}}{\sqrt{2\al^\prime}}V_T:\\
\notag&=&\frac{\zeta}{\sqrt{2\al^\prime}}\cdot\big(\dot{X}_-V_T+\dot{X}_0V_T+V_T\dot{X}_+\big)+
\label{NorOrdVP}\frac{\zeta_\mu}{\sqrt{2\al^\prime}}\big[\dot X_+^\mu, V_T\big]\\
&=&\frac{\zeta}{\sqrt{2\al^\prime}}\cdot V_{P_1}-\sqrt\frac{\al^\prime}{2}(\zeta\cdot k)V_T
\eeqa
Here we define
\beqa
\label{VP1}V_{P_1}^\mu\equiv\dot{X}_-^\mu V_T+\dot{X}_0^\mu V_T+V_T\dot{X}_+^\mu
\eeqa
and $V_{P_1}=V_P$ if the polarization vector satisfies flat space-time $L_1$ condition $\zeta\cdot k=0$. We shall refer to Eq.\eqref{NorOrdVP} as the first descend formula. \par
The vertex operator for spin two tensor in flat space-time is defined as
\beqas
V_M&\equiv&:\frac{\epsilon_{\mu\nu}}{2\al^\prime}\dot X^\mu\dot X^\nu V_T-\frac{i\epsilon}{\sqrt{2\al^\prime}}\cdot \ddot XV_T:\\
&=&\frac{\epsilon_{\mu\nu}}{2\al^\prime}\left(\begin{array}{rl}&\dot{X}_-^\mu\dot{X}_-^\nu V_T
+\dot{X}_-^\mu\dot{X}_0^\nu V_T+\dot{X}_0^\mu\dot{X}_-^\nu V_T\\
+&\dot{X}_0^\mu\dot{X}_0^\nu V_T+\dot{X}_0^\mu V_T\dot{X}_+^\nu
+\dot{X}_0^\nu V_T\dot{X}_+^\mu\\
+&{X}_-^\mu V_T \dot{X}_+^\nu
+\dot{X}_-^\nu V_T\dot{X}_+^\mu+ V_T \dot{X}_+^\mu\dot{X}_+^\nu
\end{array}\right)-\frac{i\epsilon}{\sqrt{2\al^\prime}}\cdot\big(\ddot{X}_-V_T+V_T\ddot{X}_+\big)\\
&&+\frac{\epsilon_{\mu\nu}}{2\al^\prime}\left\{\begin{array}{rl}&\big[\dot X_+^\mu \dot X_+^\nu, V_T\big]+\big[\dot X_+^\mu,\big(\dot X_0^\nu+\dot X_-^\nu\big)V_T\big]\\
+&\big(\dot X_0^\nu+\dot X_-^\nu\big)\big[\dot X_+^\nu,V_T\big]\end{array}\right\}-\frac{i\epsilon}{\sqrt{2\al^\prime}}\big[\ddot X_+^\nu,V_T\big].
\eeqas
The contribution from the commutator terms can be calculated using the formulae in Appendix \ref{ApxB}. The results are
\beqas
\big[\dot X_+^\mu \dot X_+^\nu, V_T\big]&=&-\al^\prime k^\mu V_TX_+^\nu-\al^\prime k^\nu V_T X_+^\mu+\al^\prime k^\mu k^\nu V_T,\\
\big[\dot X_+^\mu,\big(\dot X_0^\nu+\dot X_-^\nu\big)V_T\big]&=&-\al^\prime k^\mu \big(\dot X_0^\nu+\dot X_-^\nu\big)V_T-\frac{\al^\prime}{6}\eta^{\mu\nu}V_T\\
\big(\dot X_0^\nu+\dot X_-^\nu\big)\big[\dot X_+^\nu,V_T\big]&=&-\al^\prime k^\nu\big(\dot X_0^\mu+\dot X_-^\mu\big)V_T
\eeqas
After simplify the commutator terms, we get the second descending formula for the normal-ordered vertex operator
\beqa
\label{NorOrdVM}
V_M=V_{M_1}-\epsilon_{\mu\nu}k^\mu V_{P_1}^\nu+\Big(\frac{\al^\prime}{2}\epsilon_{\mu\nu}k^\mu k^\nu
-\frac{1}{12}\epsilon_{\mu\nu}g^{\mu\nu}+\frac{\sqrt{2\al^\prime}k\cdot \epsilon}{12}\Big)V_T.
\eeqa
Here
\beqa
\label{VM1}\hspace{-0.5cm}V_{M_1}(\epsilon_{\mu\nu},\epsilon_\mu)\equiv\frac{\epsilon_{\mu\nu}}{2\al^\prime}\left(\begin{array}{rl}&\dot{X}_-^\mu\dot{X}_-^\nu V_T
+\dot{X}_-^\mu\dot{X}_0^\nu V_T+\dot{X}_0^\mu\dot{X}_-^\nu V_T\\
+&\dot{X}_0^\mu\dot{X}_0^\nu V_T+\dot{X}_0^\mu V_T\dot{X}_+^\nu
+\dot{X}_0^\nu V_T\dot{X}_+^\mu\\
+&{X}_-^\mu V_T \dot{X}_+^\nu
+\dot{X}_-^\nu V_T\dot{X}_+^\mu+ V_T \dot{X}_+^\mu\dot{X}_+^\nu
\end{array}\right)-\frac{i\epsilon}{\sqrt{2\al^\prime}}\cdot\big(\ddot{X}_-V_T+V_T\ddot{X}_+\big).
\eeqa
and $V_{P_1}$ is defined in Eq.\eqref{VP1}. Notice that the coefficient in front of $V_T$ in Eq.\eqref{NorOrdVM} vanishes if we assume $L_1$ and $L_2$ conditions for spin-two particles in flat space-time. However, $V_M$ and $V_{M_1}$ are not the same operator in general covariant gauges. This is a new feature for all massive vertex operators and one should use $V_M$ instead $V_{M_1}$ in the operator formalism of string theory calculations. 
\subsection{Conformal Algebra of the Normal-ordered Covariant Vertex Operators}
\par In the calculations of scattering amplitudes (correlation functions) in any conformal field theory, the use of vertex operators ensures that the final results are conformal invariant. For string theory, in particular, we use integrated vertex operators to allow for all possible particle emissions (or injections)
\beqas
\cV_{string}\equiv\int d\tau \cV_{string}(\tau).
\eeqas
Here to compensate for the conformal transformation of the integration measure
$d\tau$ $\rightarrow$ $d\tau^\prime\Big(\frac{d\tau}{d\tau^\prime}\Big)$, we need to impose the condition that the unintegrated vertex operator $\cV_{string}(\tau)$ to transform like $\cV_{string}(\tau)$ $\rightarrow$ $\Big(\frac{d\tau^\prime}{d\tau}\Big)\cV_{string}(\tau)$. Consequently, we require all unintegrated vertex operators to carry conformal weight $J=1$. More precisely, if we check the action of conformal transformation induced by energy-momentum tensor on any unintegrated vertex operator, we must have the following algebraic relation,
\beqa
\label{ComforAlg}\big[\cL_m,\cV_{string}(\tau)\big]=e^{im\tau}\Big(-i\frac{d}{d\tau}+mJ\Big)\cV_{string}(\tau).
\eeqa
In the following, we show that the solutions of unintegrated vertex operators to the conformal algebra, Eq.\eqref{ComforAlg} must satisfy the Virasoro constraints. Our calculations is a straightforward generalization of the textbook \cite{Green:1987sp} method. However, there are two new ingredients which deserve careful examinations:
\begin{itemize}
  \item [(1)]Due to the linear dilaton background, we need to consider the shift of the string coordinates $X^\mu$ $\rightarrow$ $\mathcal{X}^\mu$
      .
  \item [(2)]For the massive string excitations, the procedure of normal-ordering introduces some new patterns of cancelation, which to our best knowledge, has not been studied in literature. Hence, we believe that it is instructive to present the detailed derivations of the covariant vertex operators.
\end{itemize}
The general strategy is to expand any normal-ordered vertex operators into a linear combination of normal-ordered vertex operators with lower spins, 
and we can calculate the conformal algebra recursively. One will see how this process works in the following explicit examples.

\subsubsection{Tachyon}
The vertex operator for tachyon in the linear dilaton background is
\beqa
\label{cVT}\cV_T(\tau)&\equiv& :e^{ik\cdot\mathcal{X}(\tau)}:=e^{-\al^\prime k\cdot V\tau}V_T(\tau).
\eeqa
Since
\beqas
-i\frac{d}{d\tau}\cV_T=e^{-\al^\prime k\cdot V\tau}\Big[i\al^\prime \big(k\cdot V\big) V_T-i\frac{d}{d\tau}V_T\Big],
\eeqas
and
\beqas
\big[\cL_m,V_T\big]&=&\big[L_m,V_T\big]+i\sqrt{\frac{\al^\prime}{2}}(m+1)V^\mu\big[\al_m^\mu, V_T\big]\\
&=&e^{im\tau}\bigg\{-i\frac{d}{d\tau}V_T+\Big[\al^\prime mk^2+i\al^\prime(m+1)k\cdot V\Big]V_T\bigg\}.
\eeqas
One can check that $\cV_T(\tau)$ satisfies the conformal algebra,
\beqa
\notag\big[\cL_m,\cV_T\big]&=&e^{-\al^\prime k\cdot V\tau}\big[\cL_m,V_T\big]\\
\notag&=&e^{-\al^\prime k\cdot V\tau+im\tau}\bigg\{-i\frac{d}{d\tau}V_T+\Big[\al^\prime mk^2+i\al^\prime(m+1)k\cdot V\Big]V_T\bigg\}\\
\notag&=&e^{-\al^\prime k\cdot V\tau+im\tau}\bigg\{-ie^{-\al^\prime (k\cdot V)\tau}\Big(\frac{d}{d\tau}\cV_T\Big)-i\al^\prime k\cdot V V_T+
\Big[\al^\prime mk^2+i\al^\prime(m+1)k\cdot V\Big]V_T\bigg\}\\
\label{ConfomAgVT}&=&e^{im\tau}\bigg\{-i\Big(\frac{d}{d\tau}\cV_T\Big)+
\Big[ m(\al^\prime k^2+i \al^\prime k\cdot V)\Big]\cV_T\bigg\}.
\eeqa
Now, if we require that the $\cV_T$ carry conformal dimension $J=1$, or equivalently,
\beqas
\big[\cL_m,\cV_T\big]=e^{im\tau}\Big(-i\frac{d}{d\tau}+mJ\Big)\cV_T.
\eeqas
We see that this naturally give rise to the $\cL_0$ condition for tachyon,
\beqas
\al^\prime k\cdot(k+iV)=1.
\eeqas 
\subsubsection{Photon}
The vertex operator for photon (P) in the linear dilaton background is
\beqa
\notag\cV_P(\zeta;\tau)&\equiv&:\frac{\zeta\cdot \mathcal{\dot X}}{\sqrt{2\al^\prime}}e^{ik\cdot \mathcal{X}}:\\
\notag&=&\La\frac{\zeta_\mu}{\sqrt{2\al^\prime}}:\Big(\dot X^\mu+i\al^\prime V^\mu\Big)e^{ik\cdot X}:\\
\label{cVP}&=&\La\frac{\zeta_\mu}{\sqrt{2\al^\prime}}V_{P_1}^\mu
-\sqrt\frac{\al^\prime}{2}\zeta\cdot(k-iV)\cV_T.
\eeqa
Here we define $\La\equiv e^{-\al^\prime k\cdot V\tau}$ and we have used the descending formula, Eq.\eqref{cVP}, in the last line of Eq.\eqref{cVP}.
Here we have used the results of normal-ordered photon vertex operator and defined $V_{P_1}^\mu$ and $\cV_T$ in Eqs.\eqref{VP1}, \eqref{cVT} separately.
\par
To check the conformal algebra, we separate the commutator of Virasoro generators $\cL_m$ and photon vertex operator $\cV_P$ can be separated into two terms:
\beqa
\label{cLmcVP}\big[\cL_m,\cV_P(\xi,\tau)\big]=\La\frac{\zeta_\mu}{\sqrt{2\al^\prime}}\big[\cL_m,V_{P_1}^\mu\big]
-\sqrt\frac{\al^\prime}{2}\zeta\cdot(k-iV)\big[\cL_m,\cV_T\big],
\eeqa
where
\beqas
\big[\cL_m,V_{P_1}^\mu\big]&=&e^{im\tau}\Big[-i\frac{d}{d\tau}+
  m(\al^\prime k^2+1)+i\al^\prime (m+1)k\cdot V\Big]V_{P_1}^\mu\\
 &&+ e^{im\tau} m^2(k^\mu+iV^\mu)V_T-\al^\prime e^{im\tau} m(k^\mu-iV^\mu)V_T.
\eeqas
 Now we have the first term in Eq.\eqref{cLmcVP} as
\beqas
\La\frac{\zeta_\mu}{\sqrt{2\al^\prime}}\big[\cL_m,V_{P_1}^\mu\big]&=&e^{im\tau}\Big[-i\frac{d}{d\tau}+
  m (\al^\prime k^2+i\al^\prime k\cdot V+1)\Big]\Big(\La \frac{\zeta_\mu}{\sqrt{2\al^\prime}}V_{P_1}^\mu\Big)\\
 &&+ e^{im\tau}m^2 \sqrt\frac{\al^\prime}{2}\zeta\cdot(k+iV)\cV_T-e^{im\tau}m\sqrt\frac{\al^\prime}{2}\zeta\cdot(k-iV)\cV_T,
\eeqas
The second term in Eq.\eqref{cLmcVP} can be derived from Eq.\eqref{ConfomAgVT},
\beqas
&&-\sqrt\frac{\al^\prime}{2}\zeta\cdot\big(k-iV\big)\big[\cL_m,\cV_T\big]\\
&=&e^{im\tau}\Big[-i\frac{d}{d\tau}+
  m(\al^\prime k^2+i\al^\prime k\cdot V+1)\Big]\Big[-\sqrt\frac{\al^\prime}{2}\zeta\cdot\big(k-iV\big)\cV_T\Big]+e^{im\tau}m\sqrt\frac{\al^\prime}{2}\zeta\cdot(k-iV)\cV_T.
\eeqas
Combining these two terms, we find cancelation for $m^1$ term and
\beqas
\label{ConfomAgcVP}\big[\cL_m,\cV_P\big]&=&e^{im\tau}\Big[-i\frac{d}{d\tau}+
  m (\al^\prime k^2+i\al^\prime k\cdot V+1)\Big]\cV_P+ e^{im\tau} m^2 \sqrt\frac{\al^\prime}{2}\zeta\cdot\big(k+iV\big)\cV_T.
\eeqas
From this result, it is clear that if we require $\cV_P$ to have conformal dimension $J=1$, $\big[\cL_m,\cV_T\big]=e^{im\tau}\Big(-i\frac{d}{d\tau}+mJ\Big)\cV_T.$, we naturally derive
\beqas
\mbox{\begin{tabular}{rrr}
         & $\al^\prime k\cdot(k+iV)=0$,~~~~ & ($\cL_0$ condition) \\
         &&\\
        and~~~~& $\al^\prime \zeta\cdot(k+iV)=0$,~~~~ & ($\cL_1$ condition) \\
      \end{tabular}
}
\eeqas
as on-shell and transverse conditions for the photon state. 
\subsubsection{Spin-two Tensor}
The vertex operator for spin-two tensor (M) is
\beqa
\notag\cV_M(\epsilon_{\mu\nu},\epsilon_\mu)&\equiv&:\Big(\frac{\epsilon_{\mu\nu}}{2\al^\prime}\mathcal{\dot X}^\mu\mathcal{\dot X}^\nu-\frac{i\epsilon_\mu}{\sqrt{2\al^\prime}}\mathcal{\ddot X}^\mu\Big)\cV_T:\\
\label{cVM}&=&
\La:\bigg[\frac{\epsilon_{\mu\nu}}{2\al^\prime}\Big(\dot X+i\al^\prime V\Big)^\mu\Big(\dot X+i\al^\prime V\Big)^\nu-\frac{i\epsilon_\mu}{\sqrt{2\al^\prime}}\ddot X^\mu\bigg]V_T:
\eeqa
Expand the normal-ordered form, we get
\beqa
\cV_M(\epsilon_{\mu\nu},\epsilon_\mu)=\La V_M(\epsilon_{\mu\nu},\epsilon_\mu)+i\epsilon_{\mu\nu}V^\mu \cV_P^\nu+\Om \cV_T,
\eeqa
where
\beqa
\label{Om}\Om\equiv\frac{\al^\prime\epsilon_{\mu\nu}}{2} V^\mu V^\nu+\frac{\al^\prime\epsilon_{\mu\nu}}{2} k^\mu k^\nu-\frac{\epsilon_{\mu\nu}\eta^{\mu\nu}}{12}+\frac{\sqrt{2\al^\prime}}{12}\epsilon\cdot k.
\eeqa
We shall calculate the commutator of the Virasoro generate $\cL_m$ and $\cV_M(\epsilon_{\mu\nu},\epsilon_\mu)$ in four steps:
\beqa
\label{cLmcVM}
\big[\cL_m,\cV_M(\epsilon_{\mu\nu},\epsilon_\mu)\big]=\underbrace{\big[\cL_m,\La V_M(\epsilon_{\mu\nu},\epsilon_\mu)\big]}_{\mbox{(I)}}+\underbrace{\big[\cL_m, i\epsilon_{\mu\nu}V_\nu \cV_P^\mu\big]}_{\mbox{(II)}}+\underbrace{\big[\cL_m,\Om \cV_T\big]}_{\mbox{(III)}}
\eeqa
\begin{itemize}
  \item[\underline{Step 1}:]Calculation of (I) in Eq.\eqref{cLmcVM}
  \beqa
  \Big[\cL_m,\La V_M(\epsilon_{\mu\nu},\epsilon_\mu)\Big]=\underbrace{\Big[L_m,\La V_M\Big]}_{\mbox{(I-1)}}+\underbrace{\Big[i\sqrt{\frac{\al^\prime}{2}}(m+1)V\cdot \al_m,\La V_M\Big]}_{\mbox{(I-2)}}
  \eeqa
  Here the first term, (I-1), of (I) is
  \beqas
  \big[L_m,\La V_M\big]&=&\La e^{im\tau}\Big[-i\frac{d}{d\tau}
  +m(\al^\prime k^2+2)\Big]V_M\\&&+\La e^{im\tau}\left\{\begin{array}{rl}
                                            m^3& \displaystyle  \Big(\frac{\epsilon_{\mu\nu}\eta^{\mu\nu}}{6}+\frac{\sqrt{2\al^\prime}}{3}\epsilon\cdot k\Big)V_T\\
                                           +m^2  & \displaystyle \left[\begin{array}{rl}
                                                   &\displaystyle \Big(\epsilon_{\mu\nu}k^\mu+\frac{\epsilon_\mu}{\sqrt{2\al^\prime}}\Big)V_{P_1}^\mu\\
                                                 -&\displaystyle    \Big(\al^\prime\epsilon_{\mu\nu}k^\mu k^\nu+\sqrt{\frac{\al^\prime}{2}}\epsilon\cdot k\Big)V_T
                                                 \end{array}\right]
                                             \\
                                        +m &  \displaystyle    \Big(\al^\prime \epsilon_{\mu\nu}k^\mu k^\nu+\frac{\sqrt{2\al^\prime}}{6}\epsilon\cdot k-\frac{\epsilon_{\mu\nu}\eta^{\mu\nu}}{6}\Big)V_T
                                           \end{array}\right\},
  \eeqas
  and the second term (I-2) of (I) is
  \beqas
  &&i\sqrt{\frac{\al^\prime}{2}}(m+1)V_\mu\Big[\al_m^\mu,\La V_M\Big]\\
  &=&i\sqrt{\frac{\al^\prime}{2}}(m+1)e^{im\tau}\sqrt{2\al^\prime}k\cdot V\La V_M
  +e^{im\tau}\La(m+1)\left[\begin{array}{rl}m^2&\displaystyle\Big(i\sqrt{\frac{\al^\prime}{2}}\epsilon\cdot VV_T\Big)\\
                                         +m&   \displaystyle  \left(\begin{array}{rl}
                                              &  \displaystyle  i\epsilon_{\mu\nu}V^\mu V_{P_1}^\nu\\
                                              -& \displaystyle   i\al^\prime\epsilon_{\mu\nu}k^\mu V^\nu V_T
                                               \end{array}\right)
                                           \end{array}\right]\\
                                           &=&e^{im\tau}\bigg[\Big(-i\frac{d}{d\tau}\La\Big)V_M+i\al^\prime(k\cdot V)\La V_M\bigg]
  +e^{im\tau}\La(m+1)\left[\begin{array}{rl}m^2&\displaystyle
                                              \Big(i\sqrt{\frac{\al^\prime}{2}}\epsilon\cdot VV_T\Big)\\
                                            +m&\displaystyle  \left(\begin{array}{rl}
                                                &\displaystyle  i\epsilon_{\mu\nu}V^\mu V_{P_1}^\nu\\
                                               -&\displaystyle   i\al^\prime\epsilon_{\mu\nu}k^\mu V^\nu V_T
                                               \end{array}\right)
                                           \end{array}\right]
  \eeqas
  Adding (I-1) and (I-2), we get
  \beqa
  \notag&&\big[\cL_m,\La V_M(\epsilon_{\mu\nu},\epsilon_\mu)\big]\\
  \notag&=&e^{im\tau}\Big[-i\frac{d}{d\tau}+m\big(\al^\prime k^2+i\al^\prime k\cdot V+2\big) \Big]\Big(\La V_M\Big)\\
  \label{VMI}&&+
  e^{im\tau}\La\left\{\begin{array}{rl}
                 \displaystyle m^3&\displaystyle\Big(\frac{\epsilon_{\mu\nu}\eta^{\mu\nu}}{6}+\frac{\sqrt{2\al^\prime}}{3}\epsilon\cdot k+i\sqrt{\frac{\al^\prime}{2}}\epsilon\cdot V\Big)V_T \\
                 +m^2 &\left[\begin{array}{rl}
                        &\displaystyle\Big(\epsilon_{\mu\nu}k^\nu+i\epsilon_{\mu\nu}V^\nu+\frac{\epsilon_\mu}{\sqrt{2\al^\prime}}\Big)V_{P_1}^\mu \\
                        -&\left(\begin{array}{rcl}
                            \displaystyle\al^\prime\epsilon_{\mu\nu}k^\mu k^\nu&+&\displaystyle\sqrt\frac{\al^\prime}{2}\epsilon\cdot k\\
                           \displaystyle +i\al^\prime\epsilon_{\mu\nu}k^\mu V^\nu&-&\displaystyle i\sqrt\frac{\al^\prime}{2}\epsilon\cdot V
                          \end{array}\right)V_T
                      \end{array}\right]
                 \\
                 +m &\left[\begin{array}{rl}
                     & i\epsilon_{\mu\nu}V^\nu V_{P_1}^\mu \\
                      +&\left(\begin{array}{rl}
                       \displaystyle &\displaystyle\al^\prime\epsilon_{\mu\nu}k^\mu k^\nu-i\al^\prime\epsilon_{\mu\nu}k^\mu V^\nu \\
                        \displaystyle+&\displaystyle\frac{\sqrt{2\al^\prime}}{6}\epsilon\cdot k -\frac{\epsilon_{\mu\nu}\eta^{\mu\nu}}{6}
                      \end{array}\right) V_T
                    \end{array}\right]
               \end{array}\right\}.
  \eeqa
  \item[\underline{Step 2}:]Calculation of (II) in Eq.\eqref{cLmcVM}\\
  We can apply the result for the vertex operator of photon, Eq.\eqref{cVP}, to calculation (II) by identifying the polarization vector $\frac{\zeta_\mu}{\sqrt{2\al^\prime}}$ in Eq.\eqref{cVP} as $i\epsilon_{\mu\nu}V^\nu$. Consequently,
  \beqa
  \big[\cL_m, i\epsilon_{\mu\nu}V_\nu \cV_P^\mu\big]
  \notag&=&e^{im\tau}\Big[-i\frac{d}{d\tau}+m\big(\al^\prime k^2+i\al^\prime k+1\big) \Big]\Big(i\epsilon_{\mu\nu}V^\nu\cV_P^\mu\Big)\\
  \notag&&+e^{im\tau}m^2\al^\prime\epsilon_{\mu\nu}V^\nu(k+iV)^\mu\cV_T\\
  \notag&=&e^{im\tau}\Big[-i\frac{d}{d\tau}+m\big(\al^\prime k^2+i\al^\prime k+2\big) \Big]\Big(i\epsilon_{\mu\nu}V^\nu\cV_P^\mu\Big)\\
  \label{VMII}&&+e^{im\tau}\La \left\{\begin{array}{rl}
                                                                  m^2& \al^\prime\epsilon_{\mu\nu}V^\nu(k+iV)^\mu V_T\\
                                                                  +m& \left[\begin{array}{rl}
                                                                        - & \epsilon_{\mu\nu}V^\nu V^\mu_{P_1} \\
                                                                        + & i\al^\prime \epsilon_{\mu\nu}V^\nu(k-iV)^\mu V_T
                                                                      \end{array}\right]
                                                                \end{array}  \right\}
  \eeqa
  Here in the last equality, we adjust the conformal dimension from $\al^\prime k^2+i\al^\prime k\cdot V+1$ to $\al^\prime k^2+i\al^\prime k\cdot V+2$, and we use the expanded form for $\cV_P^\mu$ of $i\epsilon_{\mu\nu}V^\nu\cV_P^\mu$ from Eq.\eqref{}.
  \item[\underline{Step 3}:]Calculation of (III) in Eq.\eqref{cLmcVM}\\
  We can apply the result for the vertex operator of the tachyon, Eq.\eqref{ConfomAgVT}, to calculation (III)
  Consequently,
  \beqa
  \notag\big[\cL_m, \Om\cV_T\big]&=&e^{im\tau}\Big[-i\frac{d}{d\tau}+m\big(\al^\prime k^2+i\al^\prime k\big) \Big]\Big(\Om\cV_T\Big)\\
  \label{VMIII}&=&e^{im\tau}\Big[-i\frac{d}{d\tau}+m\big(\al^\prime k^2+i\al^\prime k+2\big) \Big]\Big(\Om\cV_T\Big)-e^{im\tau}m(2\Om)\cV_T.
  \eeqa
  Here, in the last equality, we adjust the conformal dimension from $\al^\prime k^2+i\al^\prime k\cdot V$ to $\al^\prime k^2+i\al^\prime k\cdot V+2$, and we make up the difference with an extra term.
  \item[\underline{Step 4}:]Summing up all contribution\\
  Having calculated each contribution to the commutator $[\cL_m,\cV_M]$ in the previous three steps, we can add up all contributions,
  \beqas
  \big[\cL_m, \cV_M(\epsilon_{\mu\nu},\epsilon_\mu)\big]=\mbox{(I)}+\mbox{(II)}+\mbox{(III)}=\mbox{(A)}+\mbox{(B)}.
  \eeqas
  Here (A) is the answer we expect and it comes from the sum of all first items in (I), (II) and (III), i.e. Eqs.\eqref{VMI},\eqref{VMII} and \eqref{VMIII}.
  \beqas
  (A)=e^{im\tau}\Big[-i\frac{d}{d\tau}+m\big(\al^\prime k^2+i\al^\prime k+2\big) \Big]\cV_M.
  \eeqas
  (B) comes from the remaining items in (I), (II) and (III) of \eqref{VMI},\eqref{VMII} and \eqref{VMIII}, and we collect all these terms according to the power of $m$ and the type of lower spin vertex operators ($V_{P_1}^\mu$, $V_T$):
  \begin{itemize}
    \item[(1)]For the $m^3$ term in (B), we have only one contribution from (I) (Eq.\eqref{VMI})
    \beqas
    \Big(\underbrace{\frac{\epsilon_{\mu\nu}\eta^{\mu\nu}}{6}+\frac{\sqrt{2\al^\prime}}{3}\epsilon\cdot k}_{\mbox{from (I-1)}}+\underbrace{i\sqrt{\frac{\al^\prime}{2}}\epsilon\cdot V\Big)V_T}_{\mbox{from (I-2)}}
  \eeqas
  This is precisely the $\cL_2$ condition for the spin-two state, Eq.\eqref{L2M}.
    \item[(2)] For the $m^2$ term in (B), we have
    \beqas
    \Big[\underbrace{\epsilon_{\mu\nu}(k+iV)^\nu+\frac{\epsilon_{\mu}}{\sqrt{2\al^\prime}}}_{\mbox{from (I)}}\Big]V_{P_1}^\mu+\left[\begin{array}{l}
         \underbrace{-\al^\prime\epsilon_{\mu\nu}k^\mu(k+iV)^\nu  -\sqrt{\frac{\al^\prime}{2}}\epsilon\cdot k}_{\mbox{from (I)}} \\
         \underbrace{+i\al^\prime\epsilon_{\mu\nu}V^\mu(k+iV)^\nu}_{\mbox{from (II)}} + \underbrace{i\sqrt{\frac{\al^\prime}{2}}\epsilon\cdot V}_{\mbox{from (I)}}
       \end{array}\right]V_T.
    \eeqas
    Notice the coefficient associated with $V_{P_1}^\mu$ operator are precisely the $\cL_1$ condition for the spin-two state, Eq.\eqref{L1M} and the two sets of coefficients associated with $V_T$ are simply projection of the $\cL_1$ condition along $k^\mu$ and $iV^\mu$ respectively.
    \item[(3)]For the $m^1$ term in (B), we have
    \beqas
    \Big(\underbrace{i\epsilon_{\mu\nu}V^\nu}_{\mbox{from (I)}}\underbrace{-i\epsilon_{\mu\nu}V^\nu}_{\mbox{from (II)}}\Big)V_{P_1}^\mu+
    \left(\begin{array}{l}
    \hspace{0.3cm}\displaystyle \underbrace{\al^\prime\epsilon_{\mu\nu}k^\mu k^\nu+\frac{\sqrt{2\al^\prime}}{6}\epsilon\cdot k -\frac{\epsilon_{\mu\nu}\eta^{\mu\nu}}{6}}_{\mbox{from (I)}}\\
    \displaystyle \underbrace{-i\al^\prime\epsilon_{\mu\nu}k^\mu V^\nu}_{\mbox{from (I)}} \underbrace{+i\al^\prime\epsilon_{\mu\nu}k^\mu V^\nu}_{\mbox{from (II)}}\\
    \displaystyle \underbrace{+\al^\prime\epsilon_{\mu\nu}V^\mu V^\nu}_{\mbox{from (II)}}\underbrace{-2\Om}_{\mbox{from (III)}}                                                             \end{array}\right)V_T
    \eeqas
    It is clear that while the coefficients od $V_{P_1}^\mu$ vanishes identically, the coefficient of $V_T$ is also zero, duo to the definite of $\Om$, Eq.\eqref{Om}. Having shown that the total contribution from (B) terms gives null result, we thus verify the normal-ordered covariant vertex operator $\cV_M(\epsilon_{\mu\nu},\epsilon_\mu)$, as defined in Eq.\eqref{cVM}, satisfies the conformal algebra
    \beqa
    \big[\cL_{m},\cV_M(\epsilon_{\mu\nu},\epsilon_\mu)\big]=e^{im\tau}\Big(-i\frac{d}{d\tau}+m\Big)\cV_M,
    \eeqa
    provide the polarization tensor ($\epsilon_{\mu\nu},\epsilon_\mu$) satisfies the physical state condition Eqs.\eqref{L1M} and \eqref{L2M}.
  \end{itemize}
\end{itemize}
We emphasize that even though the vertex operators in the linear dilaton background, written in terms of the shifted string coordinates $\mathcal{X}\equiv X+i\al^\prime V\tau$, Eqs.\eqref{cVT}, \eqref{cVP}, \eqref{cVM}, are apparently of the same form as those in flat space-time. The calculations we have presented are not trivial at all. This is because the energy momentum tensor T(z), Eq.\eqref{EM}, is not of the same form as that in flat space-time. Equivalently, one can not redefine the Virasoro generators $\cL_m$ in Eq.\eqref{cLm} by a shift in $al_0$ to make them as the same form in flat space-time. 
\section{String Scattering Amplitude of the Bosonic Open String Theory in the Linear Dilaton Background}\label{IV}
In this section, we present the calculations of three-point and four-point stringy scattering amplitudes. Our method follows that of \cite{Green:1987sp}, and we collect some of useful formulae in Appendix \ref{ApxC}. Notice that we have suppressed the delta functions associated with generalized energy-momentum conservation in all amplitudes.
\subsection{Three-point Functions}

\subsubsection{Computation of the amplitude P($\zeta,k_1$)-T($k_2$)-T($k_3$)}
\beqa
\notag\cA_{PTT}&=&\bra{0,k_1}\zeta^\ast\cdot\al_1\cV_T(k_2,1)\ket{0,k_3}\\
\notag&=&\bra{0,k_1}\Big[\zeta^\ast\cdot\al_1,\cV_T(k_2,1)\Big]\ket{0,k_3}\\
&=&\sqrt{2\al^\prime}\zeta^\ast\cdot k_2
\eeqa
\subsubsection{Computation of the amplitude P($\zeta_1,k_1$)-T($k_2$)-P($\zeta_3,k_3$)}
\beqa
\notag\cA_{PTP}&=&\bra{0,k_1}\zeta_1^\ast\cdot\al_1\cV_T(k_2,1)\zeta_3\cdot\al_{-1}\ket{0,k_3}\\
\notag&=&\bra{0,k_1}\Big[\big[\zeta_1^\ast\cdot\al_1,\cV_T(k_2,1)\big],\zeta_3\cdot\al_{-1}\Big]\ket{0,k_3}+\bra{0,k_1}\cV_T(k_2,1)\zeta_1^\ast\cdot\al_1\zeta_3\cdot\al_{-1}\ket{0,k_3}\\
&=&-2\al^\prime\zeta^*_1\cdot k_2 \zeta_3\cdot k_2+\zeta_1^*\cdot\zeta_3.
\eeqa
\subsubsection{Computation of the amplitude P($\zeta_1,k_1$)-P($\zeta_2,k_2$)-T($k_3$)}
\beqa
\notag\cA_{PPT}&=&\bra{0,k_1}\zeta_1^\ast\cdot\al_1\cV_P(k_2,1)\ket{0,k_3}\\
\notag&=&\bra{0,k_1}\zeta_1^\ast\cdot\al_1\Big[\frac{\zeta_2}{\sqrt{2\al^\prime}}\cdot\big(\dot X_0-2\al^\prime k_2\big)\cV_T(k_2,1)+\zeta_2\cdot\al_{-1}\cV_T(k_2,1)\Big]\ket{0,k_3}\\
\notag&=&\sqrt{2\al^\prime}\zeta_2\cdot k_3\bra{0,k_1}\Big[\zeta_1^\ast\cdot\al_1,\cV_T(k_2,1)\Big]\ket{0,k_3}+\bra{0,k_1}\Big[\zeta_1^\ast\cdot \al_1,\zeta_2\cdot\al_{-1}\Big]\cV_T(k_2,1)\ket{0,k_3}\\
&=&2\al^\prime\zeta^*_1\cdot k_2 \zeta_2\cdot k_3+\zeta_1^*\cdot\zeta_2.
\eeqa
\subsubsection{Computation of the amplitude P($\zeta_1,k_1$)-P($\zeta_2,k_2$)-P($\zeta_3,k_3$)}
\beqa
\notag\cA_{PPP}&=&
\bra{0,k_1}\zeta_1^\ast\cdot\al_1\cV_P(k_2,1)\zeta_3\cdot\al_{-1}\ket{0,k_3}\\
\notag&=&\bra{0,k_1}\zeta_1^\ast\cdot\al_1\left[\begin{array}{rl}&\frac{\zeta_2}{\sqrt{2\al^\prime}}\cdot\big(\dot X_0-2\al^\prime k_2\big)\cV_T(k_2,1)\\+&\zeta_2\cdot\al_{-1}\cV_T(k_2,1)+\cV_T(k_2,1)\zeta_2\cdot\al_{1}\end{array}\right]\zeta_3\cdot\al_{-1}\ket{0,k_3}\\
\notag&=&\sqrt{2\al^\prime}\zeta_2\cdot k_3\bra{0,k_1}\Big[\big[\zeta_1^\ast\cdot\al_1,\cV_T(k_2,1)\big],\zeta_3\cdot\al_{-1}\Big]\ket{0,k_3}\\
\notag&&+\sqrt{2\al^\prime}\zeta_2\cdot k_3\bra{0,k_1}\cV_T(k_2,1)\Big[\zeta_1^\ast\cdot\al_1,\zeta_3\cdot\al_{-1}\Big]\ket{0,k_3}\\
\notag&&+\bra{0,k_1}\Big[\zeta_1^\ast\cdot \al_1,\zeta_2\cdot\al_{-1}\Big]\Big[\cV_T(k_2,1),\zeta_3\cdot\al_{-1}\Big]\ket{0,k_3}\\
\notag&&+\bra{0,k_1}\Big[\zeta_3\cdot\al_{1},\cV_T(k_2,1)\Big]\Big[\zeta_2\cdot \al_1,\zeta_3\cdot\al_{-1}\Big]\ket{0,k_3}\\
\notag&=&-(2\al^\prime)^{\frac{3}{2}}\zeta_1^\ast\cdot k_2\zeta_2\cdot k_3 \zeta_3\cdot k_2\\
&&+
\sqrt{2\al^\prime}\zeta^*_1\cdot\zeta_3 \zeta_2\cdot k_3-\sqrt{2\al^\prime}\zeta^*_1\cdot\zeta_2 \zeta_3\cdot k_2+
\sqrt{2\al^\prime}\zeta_2\cdot\zeta_3 \zeta_1^*\cdot k_2.
\eeqa
\subsubsection{Computation of the amplitude P($\zeta,k_1$)-M($\epsilon_{\mu\nu},k_2$)-T($k_3$)}
\beqas
\cA_{PMT}&=&\bra{0,k_1}\zeta^\ast\cdot\al_1\cV_M(k_2,1)\ket{0,k_3}\\
&=&\underbrace{\bra{0,k_1}\zeta^\ast\cdot\al_1\La_2V_{M_1}(k_2,1)\ket{0,k_3}}_{\cA_{PMT}^{(1)}}\\
&&\underbrace{-\bra{0,k_1}\zeta^\ast\cdot\al_1\La_2\epsilon_{\mu\nu}\big(k_2^\nu-iV^\nu\big)V_{P_1}^\mu(k_2,1)\ket{0,k_3}}_{\cA^{(2)}_{PMT}}\\
&&\underbrace{-\bra{0,k_1}\zeta^\ast\cdot\al_1\big(i\al^\prime\epsilon_{\mu\nu} k_2^\mu V^\nu+\al^\prime\epsilon_{\mu\nu} V^\mu V^\nu-\Om\big)\cV_T(k_2,1)\ket{0,k_3}}_{\cA_{PMT}^{(3)}}\\
\eeqas
\beqas
\cA_{PMT}^{(1)}&=&\bra{0,k_1}\zeta^\ast\cdot\al_1\Big[\epsilon_{\mu\nu}(2\al^\mu_{-1}\al_0^\nu+\al_0^\mu\al_0^\nu)+\epsilon\cdot\al_{-1}\Big]\cV_T(k_2,1)\ket{0,k_3}\\
&=&2\sqrt{2\al^\prime}\epsilon_{\mu\nu}\zeta^{*\mu}(k^\nu_2+k_3^\nu)+(2\al^\prime)^{\frac{3}{2}}\zeta^\ast\cdot k_2\epsilon_{\mu\nu}(k^\mu_2+k_3^\mu)(k^\nu_2+k_3^\nu)+\epsilon\cdot\zeta^*\\
\cA_{PMT}^{(2)}&=&-\bra{0,k_1}\zeta^\ast\cdot\al_1\sqrt{2\al^\prime}\epsilon_{\mu\nu}\big(k_2^\nu-iV^\nu\big)(\al_{-1}^\mu+\al_0^\mu)\cV_T(k_2,1)\ket{0,k_3}\\
&=&-\sqrt{2\al^\prime}\epsilon_{\mu\nu}\zeta^{*\mu}\big(k_2^\nu-iV^\nu\big)-(2\al^\prime)^{\frac{3}{2}}\zeta^\ast\cdot k_2\epsilon_{\mu\nu}\big(k_2^\nu-iV^\nu\big)(k_2^\mu+k_3^\mu)\\
\cA_{PMT}^{(3)}&=&-\sqrt{2\al^\prime}\zeta^\ast\cdot k_2\big(i\al^\prime\epsilon_{\mu\nu} k_2^\mu V^\nu+\al^\prime\epsilon_{\mu\nu} V^\mu V^\nu-\Om\big)
\eeqas
\beqa
\notag\cA_{PMT}&=&\cA_{PMT}^{(1)}+\cA_{PMT}^{(2)}+\cA_{PMT}^{(3)}\\
\notag&=&\sqrt{2\al^\prime}\epsilon_{\mu\nu}\zeta^{*\mu}(k^{\nu}_2+2k^\nu_3+iV^\nu)+\epsilon\cdot\zeta^*\\
\notag&&+\sqrt{2\al^\prime}\zeta^\ast\cdot k_2\Big[2\al^\prime\epsilon_{\mu\nu}(k_2^\mu+k_3^\mu+iV^\mu)k_3^\nu\underbrace{+\al^\prime\epsilon_{\mu\nu}(ik_2^\mu V^\nu-V^\mu V^\nu)+\Om}_{=0\mbox{~by~}\cL_1\mbox{~and~}\cL_2 \mbox{~conditions}}\Big]\\
&=&\sqrt{2\al^\prime}\epsilon_{\mu\nu}\zeta^{*\mu}(k^{*\nu}_1+k^\nu_3)+(2\al^\prime)^{\frac{3}{2}}\zeta^*\cdot k_2\epsilon_{\mu\nu}k_1^{*\mu} k_3^\nu+\epsilon\cdot\zeta^*.
\eeqa
\subsubsection{Computation of the amplitude P($\zeta_1,k_1$)-M($\epsilon_{\mu\nu},k_2$)-P($\zeta_3,k_3$)}
\beqas
\cA_{PMP}&=&\bra{0,k_1}\zeta_1^\ast\cdot\al_1\cV_M(k_2,1)\zeta_3\cdot \al_{-1}\ket{0,k_3}\\
&=&\underbrace{\bra{0,k_1}\zeta_1^\ast\cdot\al_1\La_2V_{M_1}(k_2,1)\zeta_3\cdot \al_{-1}\ket{0,k_3}}_{\cA_{PMP}^{(1)}}\\
&&\underbrace{-\bra{0,k_1}\zeta_1^\ast\cdot\al_1\La_2\epsilon_{\mu\nu}\big(k_2^\nu-iV^\nu\big)V_{P_1}^\mu(k_2,1)\zeta_3\cdot \al_{-1}\ket{0,k_3}}_{\cA^{(2)}_{PMP}}\\
&&\underbrace{-\bra{0,k_1}\zeta_1^\ast\cdot\al_1\big(i\al^\prime\epsilon_{\mu\nu} k_2^\mu V^\nu+\al^\prime\epsilon_{\mu\nu} V^\mu V^\nu-\Om\big)\cV_T(k_2,1)\zeta_3\cdot \al_{-1}\ket{0,k_3}}_{\cA_{PMP}^{(3)}}\\
\eeqas
\beqas
\cA_{PMP}^{(1)}&=&\bra{0,k_1}\zeta_1^\ast\cdot\al_1
\left[\begin{array}{rl}&(2\epsilon_{\mu\nu}\al_0^\nu+\epsilon_\mu)\al^\mu_{-1}\cV_T(k_2,1)+\epsilon_{\mu\nu}\al_0^\mu\al_0^\nu\cV_T(k_2,1)\\
+&(2\epsilon_{\mu\nu}\al_0^\nu-\epsilon_\mu)\cV_T(k_2,1)\al_{1}^\mu+
2\al_{-1}^\mu\cV_T(k_2,1)\al^\nu_{1}\end{array}\right]\zeta_3\cdot \al_{-1}\ket{0,k_3}\\
&=&-\sqrt{2\al^\prime}\zeta_3\cdot k_2\Big[2\sqrt{2\al^\prime}\epsilon_{\mu\nu}(k_2^\mu+k_3^\mu)\zeta_1^{\ast\nu}+\epsilon\cdot\zeta^\ast_1\Big]\\
&&+(2\al^\prime\zeta^*_1\cdot k_2 \zeta_3\cdot k_2-\zeta_1^*\cdot\zeta_3)\Big[-2\al^\prime\epsilon_{\mu\nu}(k_2^\mu+k_3^\mu)(k_2^\nu+k_3^\nu)\Big]\\
&&+\sqrt{2\al^\prime}\zeta_1^\ast\cdot k_2\Big[2\sqrt{2\al^\prime}\epsilon_{\mu\nu}(k_2^\mu+k_3^\mu)\zeta_3^\nu-\epsilon\cdot\zeta_3\Big]+2\epsilon_{\mu\nu}\zeta_1^{\ast\mu}\zeta_3^\nu\\
\cA_{PMP}^{(2)}&=&-\bra{0,k_1}\zeta_1^\ast\cdot\al_1\epsilon_{\mu\nu}\big(k_2^\nu-iV^\nu\big)\Big[(\al_0^\mu+\al_{-1}^\mu)\cV_T(k_2,1)+\cV_T(k_2,1)\al_{1}^\mu\Big]\zeta_3\cdot \al_{-1}\ket{0,k_3}\\
&=&\sqrt{2\al^\prime}\zeta_3\cdot k_2\Big[\sqrt{2\al^\prime}\epsilon_{\mu\nu}\big(k_2^\nu-iV^\nu\big)\zeta_1^\ast\Big]+(2\al^\prime\zeta^*_1\cdot k_2 \zeta_3\cdot k_2\\
&&-\zeta_1^*\cdot\zeta_3)\Big[2\al^\prime\epsilon_{\mu\nu}\big(k_2^\nu-iV^\nu\big)(k_2^\mu+k_3^\mu)\Big]\\
&&-\sqrt{2\al^\prime}\zeta_1^\ast\cdot k_2\Big[\sqrt{2\al^\prime}\epsilon_{\mu\nu}\big(k_2^\nu-iV^\nu\big)\zeta_3^\mu\Big]\\
\cA_{PMP}^{(3)}&=&(2\al^\prime\zeta^*_1\cdot k_2 \zeta_3\cdot k_2-\zeta_1^*\cdot\zeta_3)\big(i\al^\prime\epsilon_{\mu\nu} k_2^\mu V^\nu+\al^\prime\epsilon_{\mu\nu} V^\mu V^\nu-\Om\big)
\eeqas
\beqa
 \notag\cA_{PMP}&=&A_{PMP}^{(1)}+A_{PMP}^{(2)}+A_{PMP}^{(3)}\\
 \notag&=&\Big[-2\al^\prime\epsilon_{\mu\nu}(k_2^\mu+k_3^\mu+iV^\mu) k_3^\nu\underbrace{-\al^\prime\epsilon_{\mu\nu}(ik_2^\mu V^\nu-V^\mu V^\nu)-\Om}_{=0\mbox{~by~}\cL_1\mbox{~and~}\cL_2 \mbox{~conditions}}\Big](2\al^\prime\zeta_1^*\cdot k_2 \zeta_3 \cdot k_2-\zeta_1^*\cdot\zeta_3)\\
\notag &&+(\sqrt{2\al^\prime}\zeta_1^*\cdot k_2)\Big[\sqrt{2\al^\prime}\epsilon_{\mu\nu}\zeta_3^\mu (k_2^\nu+2k_3^\nu+iV^\nu)-\epsilon\cdot\zeta_3\Big]\\
 \notag&&-(\sqrt{2\al^\prime}\zeta_3\cdot k_2)\Big[\sqrt{2\al^\prime}\epsilon_{\mu\nu}\zeta_1^{*\mu} (k_2^\nu+2k_3^\nu+iV^\nu)+\epsilon\cdot\zeta_1^*\Big]+2\epsilon_{\mu\nu}\zeta_1^{*\mu}\zeta_3^\nu\\
\notag&=&-2\al^\prime\epsilon_{\mu\nu}k_1^{*\mu} k_3^\nu(2\al^\prime\zeta_1^*\cdot k_2 \zeta_3 \cdot k_2-\zeta_1^*\cdot\zeta_3)\\
\notag &&+(\sqrt{2\al^\prime}\zeta_1^*\cdot k_2)\Big[\sqrt{2\al^\prime}\epsilon_{\mu\nu}\zeta_3^\mu (k_1^{*\nu}+k_3^\nu)-\epsilon\cdot\zeta_3\Big]\\
 &&-(\sqrt{2\al^\prime}\zeta_3\cdot k_2)\Big[\sqrt{2\al^\prime}\epsilon_{\mu\nu}\zeta_1^{*\mu} (k_1^{*\nu}+k_3^\nu)+\epsilon\cdot\zeta_1^*\Big]+2\epsilon_{\mu\nu}\zeta_1^{*\mu}\zeta_3^\nu.
 \eeqa
\subsubsection{Computation of the amplitude P($\zeta,k_1$)-M($\epsilon^{(2)}_{\mu\nu},k_2$)-M($\epsilon^{(3)}_{\mu\nu},k_3$)}
\beqas
&&\cA_{PMM}\\
&=&\bra{0,k_1}\zeta^\ast\cdot\al_1\cV_M(k_2,1)\big(\epsilon^{(3)}_{\mu\nu}\al^\mu_{-1}\al^{\nu}_{-1}+\epsilon^{(3)}\cdot\al_{-2}\big)\ket{0,k_3}\\
&=&\underbrace{\bra{0,k_1}\zeta^\ast\cdot\al_1\Big[\La_2V_{M_1}(k_2,1)\Big]\big(\epsilon^{(3)}_{\mu\nu}\al^\mu_{-1}\al^{\nu}_{-1}+\epsilon^{(3)}\cdot\al_{-2}\big)\ket{0,k_3}}_{\cA_{PMM}^{(1)}}
\\&&\underbrace{-\bra{0,k_1}\zeta^\ast\cdot\al_1\Big[\La_2\epsilon^{(2)}_{\mu\nu}\big(k_2^\nu-iV^\nu\big)V_{P_1}^\mu(k_2,1)\Big]\big(\epsilon^{(3)}_{\mu\nu}\al^\mu_{-1}\al^{\nu}_{-1}+\epsilon^{(3)}\cdot\al_{-2}\big)\ket{0,k_3}}_{\cA^{(2)}_{PMM}}\\
&&\underbrace{-\bra{0,k_1}\zeta^\ast\cdot\al_1\Big[\big(i\al^\prime\epsilon^{(2)}_{\mu\nu} k_2^\mu V^\nu+\al^\prime\epsilon^{(2)}_{\mu\nu} V^\mu V^\nu-\Om\big)\cV_T(k_2,1)\Big]\big(\epsilon^{(3)}_{\mu\nu}\al^\mu_{-1}\al^{\nu}_{-1}+\epsilon^{(3)}\cdot\al_{-2}\big)\ket{0,k_3}}_{\cA_{PMM}^{(3)}}\\
\eeqas
To simplify the calculation, we first define
\beqa
\notag\mathbb{M}&\equiv&\bra{0,k_1}\zeta^\ast\cdot\al_1\cV_T(k_2,1)\big(\epsilon^{(3)}_{\mu\nu}\al^\mu_{-1}\al^{\nu}_{-1}+\epsilon^{(3)}\cdot\al_{-2}\big)\ket{0,k_3}\\
\notag&=&\bra{0,k_1}\Big[\big[\zeta^\ast\cdot\al_1,\cV_T(k_2,1)\big],\big(\epsilon^{(3)}_{\mu\nu}\al^\mu_{-1}\al^{\nu}_{-1}+\epsilon^{(3)}\cdot\al_{-2}\big)\Big]\ket{0,k_3}\\
\notag&&+\bra{0,k_1}\Big[\cV_T(k_2,1),\big[\zeta^\ast\cdot\al_1,\epsilon^{(3)}_{\mu\nu}\al^\mu_{-1}\al^{\nu}_{-1}\big]\Big]\ket{0,k_3}\\
\notag&=&\sqrt{2\al^\prime}\zeta^{\ast}_1\cdot k_2\big(2\al^\prime \epsilon_{\mu\nu}^{(3)}k_2\mu k_2^\nu-\sqrt{2\al^\prime}\epsilon^{(3)}\cdot k_2\big)-2\sqrt{2\al^\prime} \epsilon_{\mu\nu}^{(3)}\zeta^{\ast\mu}_1k_2^\nu\\
&=&\sqrt{2\al^\prime}\zeta^{\ast}_1\cdot k_2\mathbb{M}_1-2\sqrt{2\al^\prime} \epsilon_{\mu\nu}^{(3)}\zeta^{\ast\mu}_1k_2^\nu.
\eeqa
where
\beqa
\mathbb{M}_1\equiv 2\al^\prime \epsilon_{\mu\nu}^{(3)}k_2^\mu k_2^\nu-\sqrt{2\al^\prime}\epsilon^{(3)}\cdot k_2
\eeqa
\beqas
\cA_{PMM}^{(1)}&=&\underbrace{\bra{0,k_1}\zeta^\ast\cdot\al_1\Big[\epsilon_{\mu\nu}^{(2)}(\al_0^\mu\al_0^\nu+2\al_0^\mu\al_{-1}^\nu)\cV_T(k_2,1)\Big]\big(\epsilon^{(3)}_{\mu\nu}\al^\mu_{-1}\al^{\nu}_{-1}+\epsilon^{(3)}\cdot\al_{-2}\big)\ket{0,k_3}}_{\cA_{PMM}^{(1-1)}}\\
&&\underbrace{+\bra{0,k_1}\zeta^\ast\cdot\al_1\Big[2\epsilon_{\mu\nu}^{(2)}(\al_0^\mu+\al_{-1}^\mu)\cV_T(k_2,1)(\al_1^\nu+\al^\nu_2)\Big]\big(\epsilon^{(3)}_{\mu\nu}\al^\mu_{-1}\al^{\nu}_{-1}+\epsilon^{(3)}\cdot\al_{-2}\big)\ket{0,k_3}}_{\cA_{PMM}^{(1-2)}}\\
&&\underbrace{+\bra{0,k_1}\zeta^\ast\cdot\al_1\Big[\cV_T(k_2,1)\al_1^\mu\al_1^\nu\Big]\big(\epsilon^{(3)}_{\mu\nu}\al^\mu_{-1}\al^{\nu}_{-1}\big)\ket{0,k_3}
\Big]}_{\cA_{PMM}^{(1-3)}}\\
&&\underbrace{+\bra{0,k_1}\zeta^\ast\cdot\al_1\Big[\epsilon^{(2)}\cdot \al_{-1}\cV_T(k_2,1)\Big]\big(\epsilon^{(3)}_{\mu\nu}\al^\mu_{-1}\al^{\nu}_{-1}+\epsilon^{(3)}\cdot\al_{-2}\big)\ket{0,k_3}}_{\cA_{PMM}^{(1-4)}}\\
&&\underbrace{-\bra{0,k_1}\zeta^\ast\cdot\al_1\Big[\cV_T(k_2,1)\epsilon^{(2)}\cdot (\al_{1}+\al_2)\Big]\big(\epsilon^{(3)}_{\mu\nu}\al^\mu_{-1}\al^{\nu}_{-1}+\epsilon^{(3)}\cdot\al_{-2}\big)\ket{0,k_3}}_{\cA_{PMM}^{(1-5)}}
\eeqas
\beqas
\cA_{PMM}^{(1-1)}&=&2\al^\prime \epsilon^{(2)}_{\mu\nu}(k_2^\mu+k_3^\mu)(k_2^\nu+k_3^\nu)\mathbb{M}+2\sqrt{2\al^\prime}\epsilon^{(2)}_{\mu\nu}\zeta^{\ast\mu}(k_2^\nu+k_3^\nu)\mathbb{M}_1\\
\cA_{PMM}^{(1-2)}&=&2\sqrt{2\al^\prime}\epsilon^{(2)}_{\mu\nu}(k_2^\nu+k_3^\nu)\Big[-\sqrt{2\al^\prime}\zeta^\ast\cdot k_2\big(2\sqrt{2\al^\prime}\epsilon^{(3)\mu}_\si k_2^\si-2\epsilon^{(3)\mu}\big)+2\epsilon^{(3)\mu}_\si\zeta^{\ast\si}\Big]\\
&&-4\sqrt{2\al^\prime}\epsilon^{(2)}_{\mu\nu}\zeta^{\ast\mu}\epsilon^{(3)\nu}_\si k_2^\si+4\epsilon^{(2)}_{\mu\nu}\epsilon^{(3)\mu}\zeta^{\ast\nu}\\
\cA_{PMM}^{(1-3)}&=&2\sqrt{2\al^\prime}\zeta^{\ast}\cdot k_2\epsilon^{(2)}_{\mu\nu}\epsilon^{(3)\mu\nu}\\
\cA_{PMM}^{(1-4)}&=&\epsilon^{(2)}\cdot\zeta^{\ast}\mathbb{M}_1\\
\cA_{PMM}^{(1-5)}&=&(\sqrt{2\al^\prime}\zeta^\ast\cdot k_2)\big(2\sqrt{2\al^\prime}\epsilon^{(2)\mu}\epsilon^{(3)}_{\mu\nu}k_2^\nu-4\epsilon^{(2)}\cdot\epsilon^{(3)}\big)
-2\epsilon^{(2)\mu}\epsilon^{(3)}_{\mu\nu}\zeta^{\ast\nu}
\eeqas
\beqas
\cA^{(2)}_{PMM}&=&
\underbrace{-\bra{0,k_1}\zeta^\ast\cdot\al_1\Big[
\sqrt{2\al^\prime}\epsilon^{(2)}_{\mu\nu}\big(k_2^\nu-iV^\nu\big)(\al_0^\mu+\al^\mu_1)\cV_T(k_2,1)\Big]\big(\epsilon^{(3)}_{\mu\nu}\al^\mu_{-1}\al^{\nu}_{-1}+\epsilon^{(3)}\cdot\al_{-2}\big)\ket{0,k_3}}_{A_{PMM}^{(2-1)}}\\
&&
\underbrace{-\bra{0,k_1}\zeta^\ast\cdot\al_1\Big[
\sqrt{2\al^\prime}\epsilon^{(2)}_{\mu\nu}\big(k_2^\nu-iV^\nu\big)\cV_T(k_2,1)(\al_1^\mu+\al^\mu_2)\Big]\big(\epsilon^{(3)}_{\mu\nu}\al^\mu_{-1}\al^{\nu}_{-1}+\epsilon^{(3)}\cdot\al_{-2}\big)\ket{0,k_3}}_{A_{PMM}^{(2-2)}}\\
\cA_{PMM}^{(2-1)}&=&-2\al^\prime\epsilon^{(2)}_{\mu\nu}\big(k_2^\nu-iV^\nu\big)(k_2^\mu+k_3^\mu)\mathbb{M}-\sqrt{2\al^\prime}\epsilon^{(2)}_{\mu\nu}\big(k^\nu-iV^\nu\big)\zeta^{\ast\mu}\mathbb{M}_1\\
\cA_{PMM}^{(2-2)}&=&\sqrt{2\al^\prime}\epsilon^{(2)}_{\mu\nu}\big(k_2^\nu-iV^\nu\big)\Big[\sqrt{2\al^\prime}\zeta^\ast\cdot k_2\big(2\sqrt{2\al^\prime}\epsilon^{(3)\mu}_\si k_2^\si-2\epsilon^{(3)\mu}\big)-\epsilon^{(3)\mu}_\si\zeta^{\ast\si}\Big]\\
\cA_{PMM}^{(3)}&=&\big(-i\al^\prime\epsilon^{(2)}_{\mu\nu} k_2^\mu V^\nu-\al^\prime\epsilon^{(2)}_{\mu\nu} V^\mu V^\nu+\Om\big)\mathbb{M}
\eeqas
\beqa
\notag\cA_{PMP}&=&\cA_{PMM}^{(1)}+\cA_{PMM}^{(2)}+\cA_{PMM}^{(3)}\\
\notag&=&\left[\begin{array}{rl}&2\al^\prime \epsilon^{(2)}_{\mu\nu}k_3^\mu(k_2^\nu+k_3^\nu)+2\al^\prime i\epsilon^{(2)}_{\mu\nu}V^\mu k_3^\nu\\+&\underbrace{i\al^\prime\epsilon^{(2)}_{\mu\nu}V^\mu k_2^\nu-\al^\prime\epsilon^{(2)}_{\mu\nu}V^\mu V^\nu+\Om}_{=0\mbox{~by~}\cL_1\mbox{~and~}\cL_2 \mbox{~conditions}}\end{array}\right]\mathbb{M}\\
\notag&&+\Big[2\sqrt{2\al^\prime}\epsilon^{(2)}_{\mu\nu}\zeta^{\ast\mu}k_3^\nu
+\underbrace{\epsilon^{(2)}\cdot\zeta^\ast+\sqrt{2\al^\prime}
\epsilon^{(2)}_{\mu\nu}(k_2^\mu+iV^\mu)\zeta^{\ast\nu}}_{=0\mbox{~by~}\cL_1\mbox{~condition}}\Big]\mathbb{M}_1\\
\notag&&-4\sqrt{2\al^\prime}\epsilon^{(2)}_{\mu\nu}\zeta^{\ast\mu}\epsilon^{(3)\nu}_\si k_2^\si+4\epsilon^{(2)}_{\mu\nu}\epsilon^{(3)\mu}\zeta^{\ast\nu}-2\epsilon^{(2)\mu}\epsilon^{(3)}_{\mu\nu}\zeta^{\ast\nu}\\
\notag&&+(\sqrt{2\al^\prime}\zeta^\ast\cdot k_2)\big(2\sqrt{2\al^\prime}\epsilon^{(2)\mu}\epsilon^{(3)}_{\mu\nu}k_2^\nu-4\epsilon^{(2)}\cdot\epsilon^{(3)}+2\epsilon^{(2)}_{\mu\nu}\epsilon^{(3)\mu\nu}\big)
\\
\notag&&+\left[\begin{array}{rl}&2\sqrt{2\al^\prime}\epsilon^{(2)}_{\mu\nu}k_3^\nu\\
+&\underbrace{\sqrt{2\al^\prime}\epsilon^{(2)}_{\mu\nu}\big(k_2^\nu+iV^\nu\big)}_{=-\epsilon_\mu^{(2)}\mbox{~by~}\cL_1\mbox{~condition}}\end{array}\right]\Big[-\sqrt{2\al^\prime}\zeta^\ast\cdot k_2\big(2\sqrt{2\al^\prime}\epsilon^{(3)\mu}_\si k_2^\si-2\epsilon^{(3)\mu}\big)+2\epsilon^{(3)\mu}_\si\zeta^{\ast\si}\Big]\\
\notag&=&\sqrt{2\al^\prime}\zeta^*\cdot k_2 \big(2\al^\prime \epsilon^{(3)}_{\rho\si}k^\rho_2k^\si_2-\sqrt{2\al^\prime}\epsilon^{(3)}\cdot k_2\big)
\big(2\al^\prime \epsilon^{(2)}_{\mu\nu}k^\mu_3 k^\nu_3-\sqrt{2\al^\prime}\epsilon^{(2)}\cdot k_3\big)\\
\notag&&-2\sqrt{2\al^\prime}\epsilon^{(3)}_{\rho\si}\zeta^{*\rho}k_2^\si\big(2\al^\prime \epsilon^{(2)}_{\mu\nu}k^\mu_3 k^\nu_3-\sqrt{2\al^\prime}\epsilon^{(2)}\cdot k_3\big)\\
\notag&&+2\sqrt{2\al^\prime}\epsilon^{(2)}_{\mu\nu}\zeta^{*\mu} k^\nu_3\big(2\al^\prime \epsilon^{(3)}_{\rho\si}k^\rho_2k^\si_2-\sqrt{2\al^\prime}\epsilon^{(3)}\cdot k_2\big)\\
\notag&&+4\sqrt{2\al^\prime}\epsilon^{(2)}_{\mu\nu}\epsilon^{(3)\nu}_{\si}k_3^\mu\zeta^{*\si}-4\sqrt{2\al^\prime}\epsilon^{(2)}_{\mu\nu}\epsilon^{(3)\nu}_{\si}k_2^\si\zeta^{*\mu}\\
\notag&&+\sqrt{2\al^\prime}\zeta^{*}\cdot k_2\left(\begin{array}{rl}-&8\al^\prime \epsilon^{(2)}_{\mu\nu}\epsilon^{(3)\nu\si}k_3^\mu k_{2\si}+2\epsilon^{(2)}_{\mu\nu}\epsilon^{(3)\mu\nu}\\+&4\sqrt{2\al^\prime}
\epsilon^{(3)}_{\rho\si}\epsilon^{(2)\rho}k^\si_2+4\sqrt{2\al^\prime}
\epsilon^{(2)}_{\rho\si}\epsilon^{(3)\rho}k^\si_3-6\epsilon^{(2)}\cdot\epsilon^{(3)}\end{array}\right)\\
&&+4\epsilon^{(2)}_{\mu\nu}\zeta^{*\mu}\epsilon^{(3)\nu}-4\epsilon^{(3)}_{\mu\nu}\zeta^{*\mu}\epsilon^{(2)\nu}.
\eeqa

\subsection{Four-point Functions}
In order to express the four-point stringy scattering amplitudes in terms of kinematic invariants, we define the Mandelstam variables in the linear dilaton background,
\beqa
\notag s&=&-\big(k_1+k_2\big)\cdot\big(k_1+k_2+iV\big),\\
\label{MadV}t&=&-\big(k_2+k_3\big)\cdot\big(k_2+k_3+iV\big),\\
\notag &\quad&s+t+u=\sum_{i=1}^4 m_i^2.
\eeqa

\subsubsection{Computation of the amplitude T($k_4$)-T($k_3$)-T($k_2$)-P($\zeta,k_1$)}
In this process, the amplitude is given by
\beqas
\cA_{TTTP}=\int^1_0\frac{dy}{y}\bra{0,k_4}\cV_T(k_3,1)\cV_T(k_2,y)\zeta\cdot\al_{-1}\ket{0,k_1}
\eeqas
From Eq.\eqref{AC2} in Appendix \ref{ApxC}, we obtain
\beqa
&&\notag\cA_{TTTP}\\
&=&\notag\int^1_0\frac{dy}{y}\mathbb{P}\\
\notag&=&-\int^1_0\frac{dy}{y}\sqrt{2\al^\prime}\Big[\zeta\cdot k_2 y^{-\al^\prime s-1}(1-y)^{-\al^\prime t-2}+\zeta\cdot k_3 y^{-\al^\prime s}(1-y)^{-\al^\prime t-2}\Big]\\
\label{ATTTP}&=&\frac{\Ga(-\al^\prime t-1)\Ga(-\al^\prime s-1)}{\Ga(-\al^\prime t-\al^\prime s-1)}\Big[(\al^\prime t+\al^\prime s+2)(\sqrt{2\al^\prime}\zeta\cdot k_2)+(\al^\prime s+1)(\sqrt{2\al^\prime}\zeta\cdot k_3)\Big].
\eeqa
\subsubsection{Computation of the amplitude T($k_4$)-T($k_3$)-P($\zeta_2,k_2$)-P($\zeta_1,k_1$)}
In this process, the amplitude is given by
\beqas
\cA_{TTPP}=\int^1_0\frac{dy}{y}\bra{0,k_4}\cV_T(k_3,1)\cV_P(k_2,y)\zeta\cdot\al_{-1}\ket{0,k_1}
\eeqas
where
\beqas
&&\bra{0,k_4}\cV_T(k_3,1)\cV_P(k_2,y)\zeta\cdot\al_{-1}\ket{0,k_1}\\
&=&\bra{0,k_4}\cV_T(k_3,1)\left[\begin{array}{rl}&\displaystyle\frac{\zeta_2}{\sqrt{2\al^\prime}}\cdot\dot X_-(y)\cV_T(k_2,y)\\
+&\displaystyle\Big(\frac{\zeta_2}{\sqrt{2\al^\prime}}\cdot\dot X_0(y)-\sqrt{2\al^\prime}\zeta_2\cdot k_2\Big)\cV_T(k_2,y)\\
+&\cV_T(k_2,y)\frac{\zeta_2}{\sqrt{2\al^\prime}}\cdot \al_1y^{-1}\end{array}\right]\zeta_1\cdot\al_{-1}\ket{0,k_1}\\
  &=&-\sqrt{2\al^\prime} \zeta_2\cdot k_3y(1-y)^{-1}\mathbb{P}+\sqrt{2\al^\prime} \zeta_2\cdot k_1\mathbb{P}+\zeta_1\cdot\zeta_2 y^{-1}A_4.\\
&=&
\begin{array}{rllrll}
(\zeta_1\cdot\zeta_2-2\al^\prime \zeta_1\cdot k_2 \zeta_2\cdot k_1)&
 y^{-\al^\prime s-1}&(1-y)^{-\al^\prime t-1} & +(2\al^\prime\zeta_1\cdot k_2 \zeta_2\cdot k_3) & y^{-\al^\prime s}&(1-y)^{-\al^\prime t-2} \\
+(2\al^\prime\zeta_1\cdot k_3 \zeta_2\cdot k_1)& y^{-\al^\prime s}&(1-y)^{-\al^\prime t-1} &  +(2\al^\prime\zeta_1\cdot k_3 \zeta_2\cdot k_3)&y^{-\al^\prime s+1}&(1-y)^{-\al^\prime t-2}.
\end{array}
\eeqas
\beqa
\notag&&\label{ATTPP}\cA_{TTPP}\\
&=&\frac{\Ga(-\al^\prime t-1)\Ga(-\al^\prime s-1)}{\Ga(-\al^\prime t-\al^\prime s)}
\left[\begin{array}{rl}
  (\al^\prime t+1)(\al^\prime t+\al^\prime s+1)(\zeta_1\cdot\zeta_2-&2\al^\prime \zeta_1\cdot k_2 \zeta_2\cdot k_1)\\
   -(\al^\prime t+1)(\al^\prime s+1)&(2\al^\prime\zeta_1\cdot k_3 \zeta_2\cdot k_1)\\+(\al^\prime t+\al^\prime s+1)(\al^\prime s+1)&(2\al^\prime\zeta_1\cdot k_2 \zeta_2\cdot k_3)\\
   +(\al^\prime s+1)(\al^\prime s)&(2\al^\prime\zeta_1\cdot k_3 \zeta_2\cdot k_3)
\end{array}\right].
\eeqa
\subsubsection{Computation of the amplitude T($k_4$)-P($\zeta_3,k_3$)-P($\zeta_2,k_2$)-P($\zeta_1,k_1$)}
  In this process, the amplitude is given by
\beqas
\cA_{TPPP}=\int^1_0\frac{dy}{y}\bra{0,k_4}\cV_P(k_3,1)\cV_P(k_2,y)\zeta\cdot\al_{-1}\ket{0,k_1},
\eeqas
  where
  \beqas
  &&\bra{0,k_4}\cV_P(k_3,1)\cV_P(k_2,y)\zeta\cdot\al_{-1}\ket{0,k_1}\\
  &=&\underbrace{\bra{0,k_4}\La_3\frac{\zeta_3}{\sqrt{2\al^\prime}}\cdot V_{P_1}(k_3,1)\La_2\frac{\zeta_2}{\sqrt{2\al^\prime}}\cdot V_{P_1}(k_2,y)\zeta_1\cdot\al_{-1}\ket{0,k_1}}_{I_{TPPP}^{(1)}}\\
  &&\underbrace{-\sqrt{2\al^\prime}\zeta_2\cdot k_2\bra{0,k_4}\La_3\frac{\zeta_3}{\sqrt{2\al^\prime}}\cdot V_{P_1}(k_3,1)\cV_T(k_2,y)\zeta_1\cdot\al_{-1}\ket{0,k_1}}_{I_{TPPP}^{(2)}}\\
  &&\underbrace{-\sqrt{2\al^\prime}\zeta_3\cdot k_3\bra{0,k_4}\cV_T(k_3,1)\La_2\frac{\zeta_2}{\sqrt{2\al^\prime}}\cdot V_{P_1}(k_2,y)\zeta_1\cdot\al_{-1}\ket{0,k_1}}_{I_{TPPP}^{(3)}}\\
  &&\underbrace{+2\al^\prime\zeta_2\cdot k_2\zeta_3\cdot k_3\bra{0,k_4}\cV_T(k_3,1)\cV_T(k_2,y)\zeta_1\cdot\al_{-1}\ket{0,k_1}}_{I_{TPPP}^{(4)}}.
  \eeqas
  \beqas
  I_{TPPP}^{(1)}&=&\bra{0,k_4}\left\{\begin{array}{rll}&\displaystyle\frac{\zeta_3}{\sqrt{2\al^\prime}}\cdot\Big[\dot X_0(1)\cV_T(k_3,1)+\cV_T(k_3,1)\dot X_+(1)\Big]&\\
  &\displaystyle\frac{\zeta_2}{\sqrt{2\al^\prime}}\cdot
  \left[\begin{array}{rl}&\dot X_-(y)\cV_T(k_2,y)+\dot X_0(y)\cV_T(k_2,y)\\
  +&\cV_T(k_2,y)(\sqrt{2\al^\prime}\al_{1}y^{-1})\end{array}\right]&\end{array}\right\}\zeta_1\cdot\al_{-1}\ket{0,k_1}\\
  &=&I_{TPPP}^{(1-1)}+I_{TPPP}^{(1-2)}+I_{TPPP}^{(1-3)}+I_{TPPP}^{(1-4)}+I_{TPPP}^{(1-5)}
  \eeqas
  \beqas
  I_{TPPP}^{(1-1)}&\equiv&\bra{0,k_4}\frac{\zeta_3}{\sqrt{2\al^\prime}}\cdot\dot X_0(1)\cV_T(k_3,1)\frac{\zeta_2}{\sqrt{2\al^\prime}}\cdot\dot X_0(y)\cV_T(k_2,y)\zeta_1\cdot\al_{-1}\ket{0,k_1}\\
  &=&2\al^\prime \zeta_2\cdot (k_1+k_2)\zeta_3\cdot(k_1+k_2+k_3)\mathbb{P}.\\
  I_{TPPP}^{(1-2)}&\equiv&\bra{0,k_4}\frac{\zeta_3}{\sqrt{2\al^\prime}}\cdot\dot X_0(1)\cV_T(k_3,1)\frac{\zeta_2}{\sqrt{2\al^\prime}}\cdot\dot X_-(y)\cV_T(k_2,y)\zeta_1\cdot\al_{-1}\ket{0,k_1}\\
  &=&-2\al^\prime \zeta_2\cdot k_3\zeta_3\cdot(k_1+k_2+k_3)y(1-y)^{-1}\mathbb{P}.\\
  I_{TPPP}^{(1-3)}&\equiv&\bra{0,k_4}\cV_T(k_3,1)\frac{\zeta_3}{\sqrt{2\al^\prime}}\cdot\dot X_+(1)\frac{\zeta_2}{\sqrt{2\al^\prime}}\cdot\dot X_0(y)\cV_T(k_2,y)\zeta_1\cdot\al_{-1}\ket{0,k_1}\\
  &=&2\al^\prime \zeta_2\cdot (k_1+k_2)\zeta_3\cdot k_2y(1-y)^{-1}\mathbb{P}+\sqrt{2\al^\prime}\zeta_2\cdot(k_1+k_2)\zeta_1\cdot\zeta_3A_4.\\
  I_{TPPP}^{(1-4)}&\equiv&\bra{0,k_4}\cV_T(k_3,1)\frac{\zeta_3}{\sqrt{2\al^\prime}}\cdot\dot X_+(1)\frac{\zeta_2}{\sqrt{2\al^\prime}}\cdot\dot X_-(y)\cV_T(k_2,y)\zeta_1\cdot\al_{-1}\ket{0,k_1}\\
  &=&\zeta_2\cdot\zeta_3y(1-y)^{-2}\mathbb{P}-2\al^\prime\zeta_2 \cdot k_3\zeta_3\cdot k_2 y^2(1-y)^{-2}\mathbb{P}-\sqrt{2\al^\prime}\zeta_1\cdot\zeta_3\zeta_2\cdot k_3y(1-y)^{-1}A_4\\
  I_{TPPP}^{(1-5)}&\equiv&\bra{0,k_4}\frac{\zeta_3}{\sqrt{2\al^\prime}}\cdot\dot X_0(1)\cV_T(k_3,1)\cV_T(k_2,y)\zeta_2\cdot\al_1 y^{-1}\zeta_1\cdot\al_{-1}\ket{0,k_1}\\
  &=&\sqrt{2\al^\prime} \zeta_1\cdot \zeta_2\zeta_3\cdot(k_1+k_2+k_3)(1-y)^{-1}A_4.
  \eeqas
  \beqas
  I_{TPPP}^{(2)}&=&-\zeta_2\cdot k_2\bra{0,k_4}\Big[\zeta_3\cdot\dot X_0(1)\cV_T(k_3,1)+\cV_T(k_3,1)\zeta_3\cdot \dot X_+(1)\Big]\cV_T(k_2,y)\zeta_1\cdot\al_{-1}\ket{0,k_1}\\
  &=&-2\al^\prime \zeta_2\cdot k_2\zeta_3\cdot(k_1+k_2+k_3)\mathbb{P}-2\al^\prime\zeta_2\cdot k_2 \zeta_3\cdot k_2y(1-y)^{-1}\mathbb{P}-\sqrt{2\al^\prime}\zeta_1\cdot\zeta_3\zeta_2\cdot k_2A_4.\\
  I_{TPPP}^{(3)}&=&-\zeta_3\cdot k_3\bra{0,k_4}\cV_T(k_3,1)\left[\begin{array}{rl}&\zeta_2\cdot\dot X_-(y)\cV_T(k_2,y)+\zeta_2\cdot\dot X_0(y)\cV_T(k_2,y)\\
  +&\cV_T(k_2,y)\zeta_2\cdot \al_1y^{-1}\end{array}\right]\zeta_1\cdot\al_{-1}\ket{0,k_1}\\
  &=&-2\al^\prime \zeta_2\cdot(k_1+k_2)\zeta_3\cdot k_3\mathbb{P}+2\al^\prime \zeta_2\cdot k_3\zeta_3\cdot k_3y(1-y)^{-1}\mathbb{P}-\sqrt{2\al^\prime}\zeta_1\cdot\zeta_2 \zeta_3\cdot k_3y^{-1}A_4.\\
  I_{TPPP}^{(4)}&=&2\al^\prime\zeta_2\cdot k_2\zeta_3\cdot k_3\mathbb{P}.
  \eeqas
  Finally, we have
  \beqas
  &&I_{TPPP}^{(1)}+I_{TPPP}^{(2)}+I_{TPPP}^{(3)}+I_{TPPP}^{(4)}\\
  &=&(\zeta_1\cdot\zeta_2-2\al^\prime\zeta_1\cdot k_2\zeta_2\cdot k_1)(\sqrt{2\al^\prime}\zeta_3\cdot k_1) y^{-\al^\prime s-1}(1-y)^{-\al^\prime t}\\
  &&+(\zeta_1\cdot\zeta_3-2\al^\prime\zeta_1\cdot k_3\zeta_3\cdot k_1)(\sqrt{2\al^\prime}\zeta_2\cdot k_1)   y^{-\al^\prime s}(1-y)^{-\al^\prime t}\\
  &&+(\zeta_1\cdot\zeta_2-2\al^\prime\zeta_1\cdot k_2\zeta_2\cdot k_1)(\sqrt{2\al^\prime}\zeta_3\cdot k_2)  y^{-\al^\prime s-1}(1-y)^{-\al^\prime t-1}\\
  &&+(2\al^\prime)^{\frac{3}{2}}(\zeta_1\cdot k_2\zeta_2\cdot k_3\zeta_3\cdot k_1-\zeta_1\cdot k_3\zeta_2\cdot k_2\zeta_3\cdot k_3)y^{-\al^\prime s}(1-y)^{-\al^\prime t-1}\\
  &&-(\zeta_1\cdot\zeta_3-2\al^\prime\zeta_1\cdot k_3\zeta_3\cdot k_1)(\sqrt{2\al^\prime}\zeta_2\cdot k_3) y^{-\al^\prime s+1}(1-y)^{-\al^\prime t-1}\\
    &&-(\zeta_2\cdot\zeta_3-2\al^\prime\zeta_2\cdot k_3\zeta_3\cdot k_2)(\sqrt{2\al^\prime}\zeta_1\cdot k_2)  y^{-\al^\prime s}(1-y)^{-\al^\prime t-2}  \\
   &&-(\zeta_2\cdot\zeta_3-2\al^\prime\zeta_2\cdot k_3\zeta_3\cdot k_2)(\sqrt{2\al^\prime}\zeta_1\cdot k_3)y^{-\al^\prime s+1}(1-y)^{-\al^\prime t-2}
  \eeqas
  \beqa
&&\label{ATPPP}\hspace{-1.5cm}\cA_{TPPP}
=\frac{\Ga(-\al^\prime t-1)\Ga(-\al^\prime s-1)}{\Ga(-\al^\prime t-\al^\prime s+1)}\times\\
\notag&&\hspace{-1.5cm}\times\left\{\begin{array}{r}
                 (\al^\prime t+\al^\prime s)\left\{
                 \begin{array}{rll}
                   (\al^\prime t+1)(\al^\prime t)&\Big[&(2\al^\prime)^{\frac{3}{2}}\zeta_1\cdot k_2\zeta_2\cdot k_1\zeta_3\cdot k_1-\sqrt{2\al^\prime}\zeta_3\cdot k_1\zeta_1\cdot\zeta_2\Big]\\
                   +(\al^\prime t+1)(\al^\prime s+1)&\Big[-& (2\al^\prime)^{\frac{3}{2}}\zeta_1\cdot k_2\zeta_2\cdot k_3\zeta_3\cdot k_1
+(2\al^\prime)^{\frac{3}{2}}\zeta_1\cdot k_3\zeta_2\cdot k_1\zeta_3\cdot k_2\Big]\\
                   +(\al^\prime s+1)(\al^\prime s)&\Big[-& (2\al^\prime)^{\frac{3}{2}}\zeta_1\cdot k_3\zeta_2\cdot k_3\zeta_3\cdot k_2+\sqrt{2\al^\prime}\zeta_1\cdot k_3\zeta_2\cdot\zeta_3\big)\Big]
                 \end{array}\right\}
                 \\
                 +(\al^\prime t+1)(\al^\prime s+1)\left\{\begin{array}{rll}
                 (\al^\prime t)&\Big[&\hspace{-2mm}(2\al^\prime)^{\frac{3}{2}}\zeta_1\cdot k_3\zeta_2\cdot k_1\zeta_3\cdot k_1-\sqrt{2\al^\prime}\zeta_2\cdot k_1\zeta_1\cdot\zeta_3\Big]\\
                 +(\al^\prime s)&\Big[-&\hspace{-2mm}(2\al^\prime)^{\frac{3}{2}}\zeta_1\cdot k_3\zeta_2\cdot k_3\zeta_3\cdot k_1+\sqrt{2\al^\prime}\zeta_2\cdot k_3\zeta_1\cdot\zeta_3\Big]
                 \end{array}\right\}
                  \\
                  +(\al^\prime t+\al^\prime s+1)(\al^\prime t+\al^\prime s)\left\{
                  \begin{array}{rll}
                  (\al^\prime t+1)&\Big[&(2\al^\prime)^{\frac{3}{2}}\zeta_1\cdot k_2\zeta_2\cdot k_1\zeta_3\cdot k_2-\sqrt{2\al^\prime}\zeta_3\cdot k_2\zeta_1\cdot\zeta_2\Big]\\
                  +(\al^\prime s+1)&\Big[-&(2\al^\prime)^{\frac{3}{2}}\zeta_1\cdot k_2\zeta_2\cdot k_3\zeta_3\cdot k_2+\sqrt{2\al^\prime}\zeta_1\cdot k_2\zeta_2\cdot\zeta_3\Big]
                  \end{array}
                  \right\}
               \end{array}\right\}.
\eeqa
\subsubsection{Computation of the amplitude T($k_4$)-T($k_3$)-M($\epsilon_{\mu\nu},k_2$)-P($\zeta,k_1$)}
In this process, the amplitude is given by
\beqas
\cA_{TTMP}=\int^1_0\frac{dy}{y}\bra{0,k_4}\cV_T(k_3,1)\cV_M(k_2,y)\zeta\cdot\al_{-1}\ket{0,k_1},
\eeqas
where
\beqas
&&\bra{0,k_4}\cV_T(k_3,1)\cV_M(k_2,y)\zeta\cdot\al_{-1}\ket{0,k_1}\\
&=&\underbrace{\bra{0,k_4}\cV_T(k_3,1)\La_2V_{M_1}(k_2,y)\zeta\cdot\al_{-1}\ket{0,k_1}}_{I_{TTMP}^{(1)}}\\
&&\underbrace{-\bra{0,k_4}\cV_T(k_3,1)\La_2\epsilon_{\mu\nu}\big(k_2^\nu-iV^\nu\big)V_{P_1}^\mu(k_2,y)\zeta\cdot\al_{-1}\ket{0,k_1}}_{I^{(2)}_{TTMP}}\\
&&\underbrace{-\bra{0,k_4}\cV_T(k_3,1)\big(i\al^\prime\epsilon_{\mu\nu} k^\mu_2 V^\nu+\al^\prime\epsilon_{\mu\nu} V^\mu V^\nu-\Om\big)\cV_T(k_2,y)\zeta\cdot\al_{-1}\ket{0,k_1}}_{I_{TTMP}^{(3)}}.
\eeqas
\beqas
&&I_{TTMP}^{(1)}\\
&=&\bra{0,k_4}\cV_T(k_3,1)\left[\begin{array}{rl}
                            \displaystyle\frac{\epsilon_{\mu\nu}}{2\al^\prime} &  \left(\begin{array}{rl}
                                                                       \dot X_-^\mu(y)\dot X_-^\nu(y)\cV_T(k_2,y) &+ \dot 2X_-^\mu(y)\dot X_0^\nu(y)\cV_T(k_2,y)\\
                                                                       +X_0^\mu(y)\dot X_0^\nu(y)\cV_T(k_2,y) &
                                                                       +2\dot X_-^\mu(y)\cV_T(k_2,y)\dot X_+^\nu(y)\\ + 2\dot X_0^\mu(y)\cV_T(k_2,y)\dot X_+^\nu(y)&
                                                                     \end{array}\right)
                            \\
                            - & \displaystyle\frac{i\epsilon}{\sqrt{2\al^\prime}}\cdot\ddot X_-(y)\cV_T(k_2,y)-\cV_T(k_2,y)\epsilon\cdot\al_1y^{-1}
                          \end{array}\right]\zeta\cdot\al_{-1}\ket{0,k_1}\\
&=&2\al^\prime\epsilon_{\mu\nu}k^\mu_3k^\nu_3y^2(1-y)^{-2}\mathbb{P}-4\al^\prime \epsilon_{\mu\nu}(k^\mu_1+k^\mu_2)k^\nu_3y(1-y)^{-1}\mathbb{P}+2\al^\prime\epsilon_{\mu\nu}(k^\mu_1+k^\mu_2)(k^\nu_1+k^\nu_2)\mathbb{P}\\
&&-2\sqrt{2\al^\prime}\epsilon_{\mu\nu}\zeta^\mu k^\nu_3(1-y)^{-1}A_4+2\sqrt{2\al^\prime}\epsilon_{\mu\nu}\zeta^\mu (k^\nu_1+k^\nu_2)y^{-1}A_4\\
&&-\sqrt{2\al^\prime}\epsilon\cdot k_3y(1-y)^{-2}\mathbb{P}-\epsilon\cdot\zeta y^{-1}A_4
\eeqas
\beqas
I_{TTMP}^{(2)}&=&-\bra{0,k_4}\cV_T(k_3,1)\epsilon_{\mu\nu}\big(k_2^\nu-iV^\nu\big)\left[\begin{array}{rl}&\dot X_-(y)^\mu\cV_T(k_2,y)+\dot X_0^\mu(y)\cV_T(k_2,y)\\
+&\sqrt{2\al^\prime}\cV_T(k_2,y)\al_1^\mu y^{-1}\end{array}\right]\zeta\cdot\al_{-1}\ket{0,k_1}\\
&=&2\al^\prime\epsilon_{\mu\nu}k_3^\mu\big(k_2^\nu-iV^\nu\big) y(1-y)^{-1}\mathbb{P}-2\al^\prime\epsilon_{\mu\nu}(k_1^\mu+k_2^\mu)\big(k_1^\nu-iV^\nu\big)\mathbb{P}\\
&&-\sqrt{2\al^\prime}\epsilon_{\mu\nu}\zeta^\mu\big(k_2^\nu-iV^\nu\big)y^{-1}A_4\\
I_{TTMP}^{(3)}&=&\big(i\al^\prime\epsilon_{\mu\nu} k^\mu V^\nu+\al^\prime\epsilon_{\mu\nu} V^\mu V^\nu-\Om\big)\mathbb{P}.
\eeqas
\beqas
&&I_{TTMP}^{(1)}+I_{TTMP}^{(2)}+I_{TTMP}^{(3)}\\
&=&
\Big[-(2\al^\prime \epsilon_{\mu\nu} k_1^\mu k_1^\nu-\sqrt{2\al^\prime}\epsilon\cdot k_1)(\sqrt{\al^\prime}\zeta\cdot k_2)+2\sqrt{2\al^\prime}\epsilon_{\mu\nu}k_1^\mu\zeta^\nu-2\epsilon\cdot\zeta\Big]y^{-\al^\prime s-1}(1-y)^{-\al^\prime t}\\
&&+\Big[-(2\al^\prime \epsilon_{\mu\nu} k_1^\mu k_1^\nu-\sqrt{2\al^\prime}\epsilon\cdot k_1)(\sqrt{\al^\prime}\zeta\cdot k_3)\Big]y^{-\al^\prime s}(1-y)^{-\al^\prime t}\\
&&+\Big[2(2\al^\prime)^{\frac{3}{2}}\zeta\cdot k_3\epsilon_{\mu\nu}k_1^\mu k_3^\nu\Big]y^{-\al^\prime s+1}(1-y)^{-\al^\prime t-1}\\
&&+\Big[2(2\al^\prime)^{\frac{3}{2}}\zeta\cdot k_2\epsilon_{\mu\nu}k_1^\mu k_3^\nu-2\sqrt{2\al^\prime}\epsilon_{\mu\nu}\zeta^\mu k_3^\nu\Big]y^{-\al^\prime s}(1-y)^{-\al^\prime t-1}\\
&&+\Big[(2\al^\prime\epsilon_{\mu\nu}k^\mu_3k^\nu_3-\sqrt{2\al^\prime}\epsilon\cdot k_3)(\sqrt{2\al^\prime}\zeta\cdot k_2)\Big]y^{-\al^\prime s+1}(1-y)^{-\al^\prime t-2}\\
&&+\Big[(2\al^\prime\epsilon_{\mu\nu}k^\mu_3k^\nu_3-\sqrt{2\al^\prime}\epsilon\cdot k_3)(\sqrt{2\al^\prime}\zeta\cdot k_3)\Big]y^{-\al^\prime s+2}(1-y)^{-\al^\prime t-2}
\eeqas
To obtain result above, we have imposed $\cL_1$ condition of the spin-two particle to eliminate terms which consist of the contractions between spin-two tensor $\epsilon_{\mu\nu}$ and the momentum $k_2^\mu$.
  \beqa
&&\label{ATTMP}\cA_{TTMP}
=\frac{\Ga(-\al^\prime t-1)\Ga(-\al^\prime s-1)}{\Ga(-\al^\prime t-\al^\prime s+1)}\times\\
\notag&&\times\left\{\begin{array}{r}
(\al^\prime t+\al^\prime s)\left\{\begin{array}{rl}(\al^\prime t+1)(\al^\prime t)
&\left[\begin{array}{l}~~\sqrt{2\al^\prime}\zeta\cdot k_2\big(2\al^\prime\epsilon_{\mu\nu}k^\mu_1k^\nu_1-\sqrt{2\al^\prime}\epsilon\cdot k_1\big)\\
-2\sqrt{2\al^\prime}\epsilon_{\mu\nu}k_1^\mu\zeta^\nu+2\epsilon\cdot\zeta\end{array}\right]\\
+(\al^\prime t+1)(\al^\prime s+1)&\Big[-2(2\al^\prime)^{\frac{3}{2}}\zeta\cdot k_2\epsilon_{\mu\nu}k_1^\mu k_3^\nu+2\sqrt{2\al^\prime}\epsilon_{\mu\nu}\zeta^\mu k_3^\nu\Big]\\
+(\al^\prime s+1)(\al^\prime s)&\Big[~~\sqrt{2\al^\prime}\zeta\cdot k_2\big(2\al^\prime\epsilon_{\mu\nu}k^\mu_3k^\nu_3-\sqrt{2\al^\prime}\epsilon\cdot k_3\big)\Big]\end{array}\right\}\\
+(\al^\prime s+1)\left\{\begin{array}{rll}(\al^\prime t+1)(\al^\prime t)&\Big[&\sqrt{2\al^\prime}\zeta\cdot k_3\big(2\al^\prime\epsilon_{\mu\nu}k^\mu_1k^\nu_1-\sqrt{2\al^\prime}\epsilon\cdot k_1\big)\Big]\\
+(\al^\prime t+1)(\al^\prime s)&\Big[-&2(2\al^\prime)^{\frac{3}{2}}\zeta\cdot k_3\epsilon_{\mu\nu}k_1^\mu k_3^\nu\Big]\\
+(\al^\prime s)(\al^\prime s-1)&\Big[&\sqrt{2\al^\prime}\zeta\cdot k_3\big(2\al^\prime\epsilon_{\mu\nu}k^\mu_3k^\nu_3-\sqrt{2\al^\prime}\epsilon\cdot k_3\big)\Big]\end{array}\right\}
\end{array}\right\}.
\eeqa
\subsubsection{Computation of the amplitude T($k_4$)-P($\zeta_3,k_3$)-M($\epsilon_{\mu\nu},k_2$)-P($\zeta_1,k_1$)}
In this process, the amplitude is given by
\beqas
\cA_{TPMP}=\int^1_0\frac{dy}{y}\bra{0,k_4}\cV_P(k_3,1)\cV_M(k_2,y)\zeta_1\cdot\al_{-1}\ket{0,k_1},
\eeqas
where
\beqas
&&\bra{0,k_4}\cV_P(k_3,1)\cV_M(k_2,y)\zeta\cdot\al_{-1}\ket{0,k_1}\\
&=&\underbrace{\bra{0,k_4}\zeta_3\cdot\Big[\frac{\dot X_0(1)}{\sqrt{2\al^\prime}}-\sqrt{2\al^\prime}k_3\Big]\cV_T(k_3,1)\cV_M(k_2,y)\zeta_1\cdot\al_{-1}\ket{0,k_1}}_{I_{TPMP}^{(1)}}\\
&&\underbrace{+\bra{0,k_4}\cV_T(k_3,1)\frac{\zeta_3}{\sqrt{2\al^\prime}}\cdot\dot X_+(1)\cV_M(k_2,y)\zeta\cdot\al_{-1}\ket{0,k_1}}_{I_{TPMP}^{(2)}}
\eeqas
\beqas
I_{TPMP}^{(1)}&=&I_{TPMP}^{(1-1)}+I_{TPMP}^{(1-2)}+I_{TPMP}^{(1-3)}+I_{TPMP}^{(1-4)}+I_{TPMP}^{(1-5)}+I_{TPMP}^{(1-6)}+I_{TPMP}^{(1-7)}\\
I_{TPMP}^{(2)}&=&I_{TPMP}^{(2-1)}+I_{TPMP}^{(2-2)}+I_{TPMP}^{(2-3)}+I_{TPMP}^{(2-4)}+I_{TPMP}^{(2-5)}+I_{TPMP}^{(2-6)}+I_{TPMP}^{(2-7)}
\eeqas
\beqas
&&I_{TPMP}^{(1-1)}\\
&=&\bra{0,k_4}\zeta_3\cdot\Big[\frac{\dot X_0(1)}{\sqrt{2\al^\prime}}-\sqrt{2\al^\prime}k_3\Big]\cV_T(k_3,1)
\left[\begin{array}{rl}&\displaystyle\frac{\epsilon_{\mu\nu}}{2\al^\prime}\dot X^\mu_0(y) \dot X^\nu_0(y)\\-&\epsilon_{\mu\nu}\big(k_2^\mu-iV^\mu\big)\dot X_0^\nu(y)\\
-&\al^\prime\epsilon_{\mu\nu}V^\mu k_2^\nu\\-&\al^\prime\epsilon_{\mu\nu}V^\mu V^\nu+\Om\end{array}\right]\cV_T(k_2,y)\zeta_1\cdot\al_{-1}\ket{0,k_1}\\
&=&\sqrt{2\al^\prime}\zeta_3\cdot(k_1+k_2)\left[\begin{array}{rl}&2\al^\prime\epsilon_{\mu\nu}(k_1^\mu+k_2^\mu)(k_1^\nu+k_2^\nu) -2\al^\prime\epsilon_{\mu\nu}\big(k_2^\mu-iV^\mu\big)(k_1^\nu+k_2^\nu)\\
-&\al^\prime\epsilon_{\mu\nu}V^\mu k_2^\nu-\al^\prime\epsilon_{\mu\nu}V^\mu V^\nu+\Om\end{array}\right]\mathbb{P}\\
&=&\sqrt{2\al^\prime}\zeta_3\cdot(k_1+k_2)(2\al^\prime\epsilon_{\mu\nu}k_1^\mu k_1^\nu-\sqrt{2\al^\prime}\epsilon\cdot k_1)\mathbb{P}.
\eeqas
In the last equality, we use the $\cL_1$ condition for spin-two particles.
\beqas
&&I_{TPMP}^{(1-2)}\\
&=&\bra{0,k_4}\zeta_3\cdot\Big[\frac{\dot X_0(1)}{\sqrt{2\al^\prime}}-\sqrt{2\al^\prime}k_3\Big]\cV_T(k_3,1)\epsilon_{\mu\nu}\Big[\frac{2\dot X_0^\mu(y)}{2\al^\prime}-k_2^\mu+iV^\mu\Big]\dot X^\nu_-(y)\cV_T(k_2,y)\zeta_1\cdot\al_{-1}\ket{0,k_1}\\
&=&\sqrt{2\al^\prime}\zeta_3\cdot(k_1+k_2)\Big(2\epsilon_{\mu\nu}k_3^\nu-\frac{\epsilon_{\mu}}{\sqrt{2\al^\prime}}\Big)\bra{0,k_4}\cV_T(k_3,1)\dot X^\nu_-(y)\cV_T(k_2,y)\zeta_1\cdot\al_{-1}\ket{0,k_1}\\
&=&\sqrt{2\al^\prime}\zeta_3\cdot(k_1+k_2)\big(-4\al^\prime \epsilon_{\mu\nu}k_1^\mu k_3^\nu +\sqrt{2\al^\prime}\epsilon\cdot k_3\big)y(1-y)^{-1}\mathbb{P}.
\eeqas
Again, in the second equality, we have used $\cL_1$ condition for spin-two particle. Similarly, we have
\beqas
I_{TPMP}^{(1-3)}&=&\bra{0,k_4}\zeta_3\cdot\Big[\frac{\dot X_0(1)}{\sqrt{2\al^\prime}}-\sqrt{2\al^\prime}k_3\Big]\cV_T(k_3,1)\epsilon_{\mu\nu}\left[\begin{array}{rl}&\displaystyle\frac{2\dot X_0^\mu(y)}{2\al^\prime}\\-&k_2^\mu+iV^\mu\end{array}\right]\cV_T(k_2,y)\dot X^\nu_+(y)\zeta_1\cdot\al_{-1}\ket{0,k_1}\\
&=&\sqrt{2\al^\prime}\zeta_3\cdot(k_1+k_2)\big(2\sqrt{2\al^\prime} \epsilon_{\mu\nu}\zeta_1^\mu k_1^\nu -\epsilon\cdot \zeta_1\big)y^{-1}A_4\\
I_{TPMP}^{(1-4)}&=&\bra{0,k_4}\zeta_3\cdot\Big[\frac{\dot X_0(1)}{\sqrt{2\al^\prime}}-\sqrt{2\al^\prime}k_3\Big]\cV_T(k_3,1)\frac{\epsilon_{\mu\nu}}{2\al^\prime}\dot X_-^\mu(y)\dot X_-^\nu(y)\cV_T(k_2,y)\zeta_1\cdot\al_{-1}\ket{0,k_1}\\
&=&\sqrt{2\al^\prime}\zeta_3\cdot(k_1+k_2)(2\al^\prime \epsilon_{\mu\nu}k_3^\mu k_3^\nu)y^2(1-y)^{-2}\mathbb{P}\\
I_{TPMP}^{(1-5)}&=&\bra{0,k_4}\zeta_3\cdot\Big[\frac{\dot X_0(1)}{\sqrt{2\al^\prime}}-\sqrt{2\al^\prime}k_3\Big]\cV_T(k_3,1)\frac{\epsilon_{\mu\nu}}{2\al^\prime}2\dot X_-^\mu(y)\cV_T(k_2,y)\dot X_+^\nu(y)\zeta_1\cdot\al_{-1}\ket{0,k_1}\\
&=&\sqrt{2\al^\prime}\zeta_3\cdot(k_1+k_2)\big(-2\sqrt{2\al^\prime} \epsilon_{\mu\nu}\zeta_1^\mu k_3^\nu \big)(1-y)^{-1}A_4\\
I_{TPMP}^{(1-6)}&=&\bra{0,k_4}\zeta_3\cdot\Big[\frac{\dot X_0(1)}{\sqrt{2\al^\prime}}-\sqrt{2\al^\prime}k_3\Big]\cV_T(k_3,1)\Big[-\frac{i\epsilon}{\sqrt{2\al^\prime}}\cdot \ddot X_-^\mu(y)\cV_T(k_2,y)\Big]\zeta_1\cdot\al_{-1}\ket{0,k_1}\\
&=&\sqrt{2\al^\prime}\zeta_3\cdot(k_1+k_2)\big(-\sqrt{2\al^\prime}\epsilon\cdot k_3\big)y(1-y)^{-2}\mathbb{P}\\
I_{TPMP}^{(1-7)}&=&\bra{0,k_4}\zeta_3\cdot\Big[\frac{\dot X_0(1)}{\sqrt{2\al^\prime}}-\sqrt{2\al^\prime}k_3\Big]\cV_T(k_3,1)\Big[-\cV_T(k_2,y)\frac{i\epsilon}{\sqrt{2\al^\prime}}\cdot \ddot X_+^\mu(y)\Big]\zeta_1\cdot\al_{-1}\ket{0,k_1}\\
&=&\sqrt{2\al^\prime}\zeta_3\cdot(k_1+k_2)\big(-\epsilon\cdot\zeta_1\big)y^{-1}A_4.
\eeqas
\beqas
I_{TPMP}^{(2-1)}&=&\bra{0,k_4}\cV_T(k_3,1)\frac{\zeta_3}{\sqrt{2\al^\prime}}\cdot \dot X_+(1)\left[\begin{array}{rl}&\displaystyle\frac{\epsilon_{\mu\nu}}{2\al^\prime}\dot X^\mu_0(y) \dot X^\nu_0(y)\\-&\epsilon_{\mu\nu}\big(k_2^\mu-iV^\mu\big)\dot X_0^\nu(y)\\
-&\al^\prime\epsilon_{\mu\nu}V^\mu k_2^\nu-\al^\prime\epsilon_{\mu\nu}V^\mu V^\nu\\+&\Om\end{array}\right]\cV_T(k_2,y)\zeta_1\cdot\al_{-1}\ket{0,k_1}\\
&=&(2\al^\prime\epsilon_{\mu\nu}k^\mu_1k^\nu_1-\sqrt{2\al^\prime}\epsilon\cdot k_1)\bra{0,k_4}\cV_T(k_3,1)\frac{\zeta_3}{\sqrt{2\al^\prime}}\cdot \dot X_+(1)\cV_T(k_2,y)\zeta_1\cdot\al_{-1}\ket{0,k_1}\\
&=&(2\al^\prime\epsilon_{\mu\nu}k^\mu_1k^\nu_1-\sqrt{2\al^\prime}\epsilon\cdot k_1)\big(\sqrt{2\al^\prime}\zeta_3\cdot k_2 y(1-y)^{-1}\mathbb{P}+\zeta_1\cdot\zeta_3A_4\big)\\
I_{TPMP}^{(2-2)}&=&\bra{0,k_4}\cV_T(k_3,1)\frac{\zeta_3}{\sqrt{2\al^\prime}}\cdot \dot X_+(1)\epsilon_{\mu\nu}\Big[\frac{2\dot X_0^\mu(y)}{2\al^\prime}-k_2^\mu+iV^\mu\Big]\dot X^\nu_-(y)\cV_T(k_2,y)\zeta_1\cdot\al_{-1}\ket{0,k_1}\\
&=&\Big(2\epsilon_{\mu\nu}k_3^\nu-\frac{\epsilon_{\mu}}{\sqrt{2\al^\prime}}\Big)\bra{0,k_4}\cV_T(k_3,1)\frac{\zeta_3}{\sqrt{2\al^\prime}}\cdot \dot X_+(1)\dot X^\nu_-(y)\cV_T(k_2,y)\zeta_1\cdot\al_{-1}\ket{0,k_1}\\
&=&(2\sqrt{2\al^\prime}\epsilon_{\mu\nu}\zeta_3^\mu k_1^\nu-\epsilon\cdot\zeta_3)y(1-y)^{-2}\mathbb{P}\\
&&+(-4\al^\prime\epsilon_{\mu\nu}k^\mu_1k^\nu_1+\sqrt{2\al^\prime}\epsilon\cdot k_3)\big[\sqrt{2\al^\prime}\zeta_3\cdot k_2 y^2(1-y)^{-2}\mathbb{P}+\zeta_1\cdot\zeta_3y(1-y)^{-1}A_4\big]
\eeqas
\beqas
I_{TPMP}^{(2-3)}&=&\bra{0,k_4}\cV_T(k_3,1)\frac{\zeta_3}{\sqrt{2\al^\prime}}\cdot \dot X_+(1)\epsilon_{\mu\nu}\Big[\frac{2\dot X_0^\mu(y)}{2\al^\prime}-k_2^\mu+iV^\mu\Big]\cV_T(k_2,y)\dot X^\nu_+(y)\zeta_1\cdot\al_{-1}\ket{0,k_1}\\
&=&(2\sqrt{2\al^\prime}\epsilon_{\mu\nu}\zeta_1^\mu k_1^\nu-\epsilon\cdot\zeta_1)(\sqrt{2\al^\prime}\zeta_3\cdot k_2)(1-y)^{-1}A_4\\
I_{TPMP}^{(2-4)}&=&\bra{0,k_4}\cV_T(k_3,1)\frac{\zeta_3}{\sqrt{2\al^\prime}}\cdot \dot X_+(1)\frac{\epsilon_{\mu\nu}}{2\al^\prime}\dot X_-^\mu(y)\dot X_-^\nu(y)\cV_T(k_2,y)\zeta_1\cdot\al_{-1}\ket{0,k_1}\\
&=&\frac{2\epsilon_{\mu\nu}\zeta_3^\mu}{2\al^\prime}y(1-y)^{-2}\bra{0,k_4}\cV_T(k_3,1)\dot X_-^\nu(y)\cV_T(k_2,y)\zeta_1\cdot\al_{-1}\ket{0,k_1}\\
&&+\bra{0,k_4}\cV_T(k_3,1)\frac{\epsilon_{\mu\nu}}{2\al^\prime}\dot X_-^\mu(y)\dot X_-^\nu(y)\frac{\zeta_3}{\sqrt{2\al^\prime}}\cdot \dot X_+(1)\cV_T(k_2,y)\zeta_1\cdot\al_{-1}\ket{0,k_1}\\
&=&-2\sqrt{2\al^\prime}\epsilon_{\mu\nu}\zeta_3^\mu k_3^\nu y^2(1-y)^{-3}\mathbb{P}\\
&&+(2\al^\prime\epsilon_{\mu\nu}k_3^\mu k_3^\nu)\big[\sqrt{2\al^\prime}\zeta_3\cdot k_2 y^3(1-y)^{-3}\mathbb{P}+\zeta_1\cdot\zeta_3 y^2(1-y)^{-2}\big]\\
I_{TPMP}^{(2-5)}&=&\bra{0,k_4}\cV_T(k_3,1)\frac{\zeta_3}{\sqrt{2\al^\prime}}\cdot \dot X_+(1)\frac{\epsilon_{\mu\nu}}{2\al^\prime}2\dot X_-^\mu(y)\cV_T(k_2,y)\dot X_+^\nu(y)\zeta_1\cdot\al_{-1}\ket{0,k_1}\\
&=&\big[2\epsilon_{\mu\nu}\zeta_1^\mu \zeta_3^\nu(1-y)^{-2}-(2\sqrt{2\al^\prime}\epsilon_{\mu\nu}\zeta_1^\mu k_3^\nu)(\sqrt{2\al^\prime}\zeta_3\cdot k_2)y(1-y)^{-2}\big]A_4\\
I_{TPMP}^{(2-6)}&=&\bra{0,k_4}\cV_T(k_3,1)\frac{\zeta_3}{\sqrt{2\al^\prime}}\cdot \dot X_+(1)\Big[-\frac{i\epsilon}{\sqrt{2\al^\prime}}\cdot \ddot X_-^\mu(y)\cV_T(k_2,y)\Big]\zeta_1\cdot\al_{-1}\ket{0,k_1}\\
&=&\epsilon\cdot\zeta_3(1+y)y(1-y)^{-3}\mathbb{P}-(\sqrt{2\al^\prime}\zeta_3\cdot k_2)(\sqrt{2\al^\prime}\epsilon\cdot k_3) y^2(1-y)^{-3}\mathbb{P}\\&&-\sqrt{2\al^\prime}\epsilon\cdot k_3\zeta_1\cdot \zeta_3 y(1-y)^{-2}A_4\\
I_{TPMP}^{(2-7)}&=&\bra{0,k_4}\cV_T(k_3,1)\frac{\zeta_3}{\sqrt{2\al^\prime}}\cdot \dot X_+(1)\Big[-\cV_T(k_2,y)\frac{i\epsilon}{\sqrt{2\al^\prime}}\cdot \ddot X_+^\mu(y)\Big]\zeta_1\cdot\al_{-1}\ket{0,k_1}\\
&=&-\sqrt{2\al^\prime}\zeta_3\cdot k_2\epsilon\cdot\zeta_1(1-y)^{-1}A_4.
\eeqas
Sum up the total contribution, we have
\beqas
&&I_{TPMP}^{(1)}+I_{TPMP}^{(2)}\\
&=&\big[\sqrt{2\al^\prime}\zeta_3\cdot k_1+\sqrt{2\al^\prime}\zeta_3\cdot k_2(1-y)^{-1}\big]
\left[\begin{array}{rl}
& \big(2\al^\prime\epsilon_{\mu\nu}k^\mu_1k^\nu_1-\sqrt{2\al^\prime}\epsilon\cdot k_1\big) \\
+ & \big(2\al^\prime\epsilon_{\mu\nu}k^\mu_3k^\nu_3-\sqrt{2\al^\prime}\epsilon\cdot k_3\big)y^2(1-y)^{-2} \\
- &   \big(4\al^\prime\epsilon_{\mu\nu}k^\mu_1k^\nu_3\big)y(1-y)^{-1}
 \end{array}\right]\mathbb{P}\\
 &&+\Big[\big(2\sqrt{2\al^\prime}\epsilon_{\mu\nu}k^\mu_1\zeta^\nu_3\big)y(1-y)^{-2}
 +\big(-2\sqrt{2\al^\prime}\epsilon_{\mu\nu}k^\mu_3\zeta^\nu_3+2\epsilon\cdot\zeta_3\big)y^2(1-y)^{-3}\Big]\mathbb{P}\\
 &&+\big[\sqrt{2\al^\prime}\zeta_3\cdot k_1+\sqrt{2\al^\prime}\zeta_3\cdot k_2(1-y)^{-1}\big]
 \left[\begin{array}{rl}&\big(-2\sqrt{2\al^\prime}\epsilon_{\mu\nu}k^\mu_3\zeta^\nu_1\big)y(1-y)^{-2}\\
 +&\big(2\sqrt{2\al^\prime}\epsilon_{\mu\nu}k^\mu_1\zeta^\nu_1-2\epsilon\cdot\zeta_1\big)y^{-1}\end{array}\right]A_4\\
 &&+\zeta_1\cdot\zeta_3\left[\begin{array}{rl}
& \big(2\al^\prime\epsilon_{\mu\nu}k^\mu_1k^\nu_1-\sqrt{2\al^\prime}\epsilon\cdot k_1\big) \\
+ & \big(2\al^\prime\epsilon_{\mu\nu}k^\mu_3k^\nu_3-\sqrt{2\al^\prime}\epsilon\cdot k_3\big)y^2(1-y)^{-2} \\
- &   \big(4\al^\prime\epsilon_{\mu\nu}k^\mu_1k^\nu_3\big)y(1-y)^{-1}
 \end{array}\right]A_4+2\epsilon_{\mu\nu}\zeta_1^\mu\zeta_3^\nu(1-y)^{-2}A_4\\
 &=&\big(\zeta_1\cdot\zeta_3-2\al^\prime\zeta_1\cdot k_3 \zeta_3\cdot k_1\big)\big(2\al^\prime\epsilon_{\mu\nu}k^\mu_1k^\nu_1-\sqrt{2\al^\prime}\epsilon\cdot k_1\big)y^{-\al^\prime s}(1-y)^{-\al^\prime t-1}\\
 &&+\big(\zeta_1\cdot\zeta_3-2\al^\prime\zeta_1\cdot k_3 \zeta_3\cdot k_1\big)\big(-4\al^\prime \epsilon_{\mu\nu}k_1^\mu k_3^\nu\big)y^{-\al^\prime s+1}(1-y)^{-\al^\prime t}\\
 &&+\big(\zeta_1\cdot\zeta_3-2\al^\prime\zeta_1\cdot k_3 \zeta_3\cdot k_1\big)\big(2\al^\prime\epsilon_{\mu\nu}k^\mu_3k^\nu_3-\sqrt{2\al^\prime}\epsilon\cdot k_3\big)y^{-\al^\prime s-2}(1-y)^{-\al^\prime t+3}\\
 &&+\left[\begin{array}{r}
-2\al^\prime\zeta_1\cdot k_2\zeta_3\cdot k_1\big(2\al^\prime\epsilon_{\mu\nu}k^\mu_1k^\nu_1-\sqrt{2\al^\prime}\epsilon\cdot k_1\big)\\
+2\sqrt{2\al^\prime} \zeta_3\cdot k_1\big(\sqrt{2\al^\prime}\epsilon_{\mu\nu}k^\mu_1\zeta_1^\nu-\epsilon\cdot\zeta_1\big)\end{array}\right]y^{-\al^\prime s-1}(1-y)^{-\al^\prime t+1}\\
&&+\left[\begin{array}{r}-2\al^\prime\zeta_1\cdot k_3\zeta_3\cdot k_2\big(2\al^\prime\epsilon_{\mu\nu}k^\mu_1k^\nu_1-\sqrt{2\al^\prime}\epsilon\cdot k_1\big)\\
+2(2\al^\prime)^2\zeta_1\cdot k_2 \zeta_3\cdot k_1\epsilon_{\mu\nu}k^\mu_1k^\nu_3-4\al^\prime\zeta_3\cdot k_1\epsilon_{\mu\nu}\zeta^\mu_1k^\nu_3\end{array}\right]y^{-\al^\prime s}(1-y)^{-\al^\prime t}\\
&&+\left[\begin{array}{r}-2\al^\prime\zeta_1\cdot k_2\zeta_3\cdot k_1\big(2\al^\prime\epsilon_{\mu\nu}k^\mu_3k^\nu_3-\sqrt{2\al^\prime}\epsilon\cdot k_3\big)
\\+2(2\al^\prime)^2\zeta_1\cdot k_3 \zeta_3\cdot k_2\epsilon_{\mu\nu}k^\mu_1k^\nu_3-4\al^\prime\zeta_1\cdot k_3\epsilon_{\mu\nu}k^\mu_1\zeta^\nu_3\end{array}\right]y^{-\al^\prime s+1}(1-y)^{-\al^\prime t-1}\\
&&+\left[\begin{array}{r}-2\al^\prime\zeta_1\cdot k_3\zeta_3\cdot k_2\big(2\al^\prime\epsilon_{\mu\nu}k^\mu_3k^\nu_3-\sqrt{2\al^\prime}\epsilon\cdot k_3\big)
\\+2\sqrt{2\al^\prime} \zeta_1\cdot k_3\big(\sqrt{2\al^\prime}\epsilon_{\mu\nu}k^\mu_3\zeta_3^\nu-\epsilon\cdot\zeta_3\big)\end{array}\right]y^{-\al^\prime s+2}(1-y)^{-\al^\prime t-2}\\
&&+\left[\begin{array}{r}-2\al^\prime\zeta_1\cdot k_2\zeta_3\cdot k_2\big(2\al^\prime\epsilon_{\mu\nu}k^\mu_1k^\nu_1-\sqrt{2\al^\prime}\epsilon\cdot k_1\big)
\\+2\sqrt{2\al^\prime} \zeta_3\cdot k_2\big(\sqrt{2\al^\prime}\epsilon_{\mu\nu}k^\mu_1\zeta_1^\nu-\epsilon\cdot\zeta_1\big)\end{array}\right]y^{-\al^\prime s-1}(1-y)^{-\al^\prime t}\\
&&+\left[\begin{array}{r}2(2\al^\prime)^2\zeta_1\cdot k_2\zeta_3\cdot k_2 \epsilon_{\mu\nu}k^\mu_1k^\nu_3-4\al^\prime \zeta_1\cdot k_2 \epsilon_{\mu\nu}k^\mu_1\zeta^\nu_3\\
+2\epsilon_{\mu\nu}\zeta_1^\mu\zeta_3^\nu-4\al^\prime \zeta_3\cdot k_2 \epsilon_{\mu\nu}\zeta^\mu_1 k^\nu_3\end{array}\right]y^{-\al^\prime s-2}(1-y)^{-\al^\prime t+1}\\
&&+\left[\begin{array}{r}-2\al^\prime\zeta_1\cdot k_2\zeta_3\cdot k_2\big(2\al^\prime\epsilon_{\mu\nu}k^\mu_3k^\nu_3-\sqrt{2\al^\prime}\epsilon\cdot k_3\big)
\\+2\sqrt{2\al^\prime} \zeta_1\cdot k_2\big(\sqrt{2\al^\prime}\epsilon_{\mu\nu}k^\mu_3\zeta_3^\nu-\epsilon\cdot\zeta_3\big)\end{array}\right]y^{-\al^\prime s+1}(1-y)^{-\al^\prime t-2}.
\eeqas
\beqa
&&\label{ATPMP}\hspace{-2.3cm}\cA_{TPMP}=\frac{\Ga(-\al^\prime t-1)\Ga(-\al^\prime s-1)}{\Ga(-\al^\prime t-\al^\prime s+2)}\times\\
&&\notag\hspace{-2.3cm}\times\left\{\begin{array}{r}
(\al^\prime t+1)(\al^\prime s+1)
\left\{\begin{array}{rl}(\al^\prime t)(\al^\prime t-1)&\Big[\big(\zeta_1\cdot\zeta_3-2\al^\prime\zeta_1\cdot k_3 \zeta_3\cdot k_1\big)\big(2\al^\prime\epsilon_{\mu\nu}k^\mu_1k^\nu_1-\sqrt{2\al^\prime}\epsilon\cdot k_1\big)\Big]\\
+(\al^\prime t)(\al^\prime s)&\Big[\big(\zeta_1\cdot\zeta_3-2\al^\prime\zeta_1\cdot k_3 \zeta_3\cdot k_1\big)\big(-4\al^\prime \epsilon_{\mu\nu}k_1^\mu k_3^\nu\big)\Big]\\
+(\al^\prime s)(\al^\prime s-1)&\Big[\big(\zeta_1\cdot\zeta_3-2\al^\prime\zeta_1\cdot k_3 \zeta_3\cdot k_1\big)\big(2\al^\prime\epsilon_{\mu\nu}k^\mu_3k^\nu_3-\sqrt{2\al^\prime}\epsilon\cdot k_3\big)\Big]
\end{array}\right\}\\
+(\al^\prime t+\al^\prime s-1)\left\{\begin{array}{rl}
(\al^\prime t+1)(\al^\prime t)(\al^\prime t-1)&\left[\begin{array}{r}
-2\al^\prime\zeta_1\cdot k_2\zeta_3\cdot k_1\big(2\al^\prime\epsilon_{\mu\nu}k^\mu_1k^\nu_1-\sqrt{2\al^\prime}\epsilon\cdot k_1\big)\\
+2\sqrt{2\al^\prime} \zeta_3\cdot k_1\big(\sqrt{2\al^\prime}\epsilon_{\mu\nu}k^\mu_1\zeta_1^\nu-\epsilon\cdot\zeta_1\big)\end{array}\right]\\
+(\al^\prime t+1)(\al^\prime t)(\al^\prime s+1)&\left[\begin{array}{r}-2\al^\prime\zeta_1\cdot k_3\zeta_3\cdot k_2\big(2\al^\prime\epsilon_{\mu\nu}k^\mu_1k^\nu_1-\sqrt{2\al^\prime}\epsilon\cdot k_1\big)\\
+2(2\al^\prime)^2\zeta_1\cdot k_2 \zeta_3\cdot k_1\epsilon_{\mu\nu}k^\mu_1k^\nu_3-4\al^\prime\zeta_3\cdot k_1\epsilon_{\mu\nu}\zeta^\mu_1k^\nu_3\end{array}\right]\\
+(\al^\prime t+1)(\al^\prime s+1)(\al^\prime s)&\left[\begin{array}{r}-2\al^\prime\zeta_1\cdot k_2\zeta_3\cdot k_1\big(2\al^\prime\epsilon_{\mu\nu}k^\mu_3k^\nu_3-\sqrt{2\al^\prime}\epsilon\cdot k_3\big)
\\+2(2\al^\prime)^2\zeta_1\cdot k_3 \zeta_3\cdot k_2\epsilon_{\mu\nu}k^\mu_1k^\nu_3-4\al^\prime\zeta_1\cdot k_3\epsilon_{\mu\nu}k^\mu_1\zeta^\nu_3\end{array}\right]\\
+(\al^\prime s+1)(\al^\prime s)(\al^\prime s-1)&\left[\begin{array}{r}-2\al^\prime\zeta_1\cdot k_3\zeta_3\cdot k_2\big(2\al^\prime\epsilon_{\mu\nu}k^\mu_3k^\nu_3-\sqrt{2\al^\prime}\epsilon\cdot k_3\big)
\\+2\sqrt{2\al^\prime} \zeta_1\cdot k_3\big(\sqrt{2\al^\prime}\epsilon_{\mu\nu}k^\mu_3\zeta_3^\nu-\epsilon\cdot\zeta_3\big)\end{array}\right]
\end{array}\right\}\\
+(\al^\prime t+\al^\prime s)(\al^\prime t+\al^\prime s-1)\left\{\begin{array}{rl}
(\al^\prime t+1)(\al^\prime t)&\left[\begin{array}{r}-2\al^\prime\zeta_1\cdot k_2\zeta_3\cdot k_2\big(2\al^\prime\epsilon_{\mu\nu}k^\mu_1k^\nu_1-\sqrt{2\al^\prime}\epsilon\cdot k_1\big)
\\+2\sqrt{2\al^\prime} \zeta_3\cdot k_2\big(\sqrt{2\al^\prime}\epsilon_{\mu\nu}k^\mu_1\zeta_1^\nu-\epsilon\cdot\zeta_1\big)\end{array}\right]\\
+(\al^\prime t+1)(\al^\prime s+1)&\left[\begin{array}{r}2(2\al^\prime)^2\zeta_1\cdot k_2\zeta_3\cdot k_2 \epsilon_{\mu\nu}k^\mu_1k^\nu_3-4\al^\prime \zeta_1\cdot k_2 \epsilon_{\mu\nu}k^\mu_1\zeta^\nu_3\\
+2\epsilon_{\mu\nu}\zeta_1^\mu\zeta_3^\nu-4\al^\prime \zeta_3\cdot k_2 \epsilon_{\mu\nu}\zeta^\mu_1 k^\nu_3\end{array}\right]\\
+(\al^\prime s+1)(\al^\prime s)&\left[\begin{array}{r}-2\al^\prime\zeta_1\cdot k_2\zeta_3\cdot k_2\big(2\al^\prime\epsilon_{\mu\nu}k^\mu_3k^\nu_3-\sqrt{2\al^\prime}\epsilon\cdot k_3\big)
\\+2\sqrt{2\al^\prime} \zeta_1\cdot k_2\big(\sqrt{2\al^\prime}\epsilon_{\mu\nu}k^\mu_3\zeta_3^\nu-\epsilon\cdot\zeta_3\big)\end{array}\right]
\end{array}\right\}\end{array}
\right\}.
\eeqa

\section{Stringy Ward Identities of the Bosonic Open String Theory in the Linear Dilaton Background}
This section is devoted to the verification of stringy Ward identities \cite{Lee:1994wp}, 
or equivalently, we check the decoupling of zero norm states in all processes under consideration. While the explicit calculations could be lengthy, the essential idea is the same. We first substitute all possible polarization tensors, Eqs.\eqref{ZNSI}, \eqref{ZNSII}, for zero-norm states into the stringy scattering amplitudes and replace all momentum contractions in terms of Mandelstam variables. All the amplitudes become sums of polarization projections, weighted by different kinematic invariants. One can then check that all terms with the same independent polarization projection cancel exactly. Hence we obtain a proof for stringy Ward identities.     
\subsection{Three-point Functions}
To prove the Ward identities of three point function of bosonic open string in linear dilaton background, it is useful first to know the contractions of momentum between different particles,
\beqa
\label{KContar3}\left\{\begin{array}{rl}
  \al^\prime k_1^\ast\cdot k_2+\al^\prime k_1^\ast\cdot k_3= & -m_1^2 \\
  \al^\prime k_1^\ast\cdot k_2-\al^\prime k_2\cdot k_3= & -m_2^2 \\
  \al^\prime k_1^\ast\cdot k_3-\al^\prime k_2\cdot k_3= & -m_3^2
\end{array}\right.
\quad\Rightarrow\quad\left\{\begin{array}{rl}
               \al^\prime k_1^\ast\cdot k_2= & \displaystyle\frac{-m_1^2-m_2^2+m_3^2}{2} \\
               \al^\prime k_1^\ast\cdot k_3= & \displaystyle\frac{-m_1^2+m_2^2-m_3^2}{2} \\
               \al^\prime k_2\cdot k_3= & \displaystyle\frac{-m_1^2+m_2^2+m_3^2}{2}
             \end{array}\right..
\eeqa

\subsubsection{Ward identity for P($\zeta,k_1$)-T($k_2$)-T($k_3$)}

From Eq.\eqref{KContar3}, we have
\beqas
k_1^\ast\cdot k_2=0.
\eeqas
\begin{itemize}
  \item Case 1: $\zeta^\ast\rightarrow\sqrt{2\al^\prime}k_1^\ast$
\beqa
\cA_{PTT}=\quad2\al^\prime k_1^\ast\cdot k_2=0.
\eeqa
\end{itemize}

\subsubsection{Ward identities for P($\zeta_1,k_1$)-T($k_2$)-P($\zeta_3,k_3$)}
From Eq.\eqref{KContar3}, we have
\beqas
 2\al^\prime k_1^\ast\cdot k_2=-2\al^\prime k_2\cdot k_3=1.
\eeqas
From polarization conditions and momentum conservation, we have
\beqas
\zeta_3\cdot(k_1^\ast-k_2-k_3-iV)=0\quad &\Rightarrow&\quad\zeta_3 \cdot k_1^\ast=\zeta_3 \cdot k_2\quad \\ \mbox{and}\quad\zeta_1^\ast\cdot(k_1^\ast-k_2-k_3-iV)=0\quad &\Rightarrow&\quad \zeta^\ast_1\cdot k_2=-\zeta^\ast_1\cdot k_3.
\eeqas
\begin{itemize}
  \item Case 1: $\zeta_1^\ast\rightarrow\sqrt{2\al^\prime}k_1^\ast$
  \beqa
\cA_{PTP}=\sqrt{2\al^\prime}\zeta_3\cdot k_1^\ast-(2\al^\prime)^{\frac{3}{2}}\zeta_3\cdot k_2 k_1^\ast\cdot k_2=0.
\eeqa
  \item Case 2: $\zeta_3\rightarrow\sqrt{2\al^\prime}k_3$
  \beqa
\cA_{PTP}=\sqrt{2\al^\prime}\zeta_1^\ast\cdot k_3 -(2\al^\prime)^{\frac{3}{2}}\zeta_1^\ast\cdot k_2 k_2\cdot k_3=0.
\eeqa
\end{itemize}

\subsubsection{Ward identities for P($\zeta_1,k_1$)-P($\zeta_2,k_2$)-T($k_3$)}
From Eq.\eqref{KContar3}, we have
\beqas
2\al^\prime k_1^\ast\cdot k_2=2\al^\prime k_2\cdot k_3=-1.
\eeqas
From polarization condition and momentum conservation, we have
\beqas
\zeta_2\cdot(k_1^\ast-k_2-k_3-iV)=0\quad \Rightarrow\quad \zeta_2 \cdot k_1^\ast=\zeta_2 \cdot k_3.
\eeqas

\begin{itemize}
  \item Case 1: $\zeta_1^\ast\rightarrow\sqrt{2\al^\prime}k_1^\ast$
  \beqa
\cA_{PPT}=\quad\sqrt{2\al^\prime}\zeta_2\cdot k_1^\ast+(2\al^\prime)^{\frac{3}{2}}\zeta_2\cdot k_3 k_1^\ast\cdot k_2=0.
\eeqa
  \item Case 2: $\zeta_2\rightarrow\sqrt{2\al^\prime}k_2$
  \beqa
\cA_{PPT}=\sqrt{2\al^\prime}\zeta_1^\ast\cdot k_2 +(2\al^\prime)^{\frac{3}{2}}\zeta_1^\ast\cdot k_2 k_2\cdot k_3=0.
\eeqa
\end{itemize}

\subsubsection{Ward identities for P($\zeta_1,k_1$)-P($\zeta_2,k_2$)-P($\zeta_3,k_3$)}
From Eq.\eqref{KContar3}, we have
\beqas
k_1^\ast\cdot k_2=k_1^\ast\cdot k_3=k_2\cdot k_3=0.
\eeqas
From polarization conditions and momentum conservation, we have
\beqas
\zeta_2\cdot(k_1^\ast-k_2-k_3-iV)=0\quad &\rightarrow&\quad
\zeta_2\cdot k_1^\ast = \zeta_2\cdot k_3 \\
\zeta_1^\ast\cdot(k_1^\ast-k_2-k_3-iV)=0\quad &\rightarrow&\quad\quad\mbox{ and}\quad \zeta_1^\ast\cdot(k_2+k_3)=0.
\eeqas

\begin{itemize}
  \item Case 1: $\zeta_1^\ast\rightarrow\sqrt{2\al^\prime}k_1^\ast$
  \beqa
  \notag\cA_{PPP}&=&-(2\al^\prime)^{2} \underbrace{k_1^\ast\cdot k_2}_{=0}\zeta_2\cdot k_3 \zeta_3\cdot k_2-2\al^\prime\zeta_2\cdot k_1^\ast\zeta_3\cdot k_2\\
  &&+2\al^\prime\zeta_2\cdot \zeta_3 \underbrace{k_1^\ast\cdot k_2}_{=0}+2\al^\prime\zeta_2\cdot k_3 \zeta_3 \cdot k_1^\ast= 0.
\eeqa
  \item Case 2: $\zeta_2\rightarrow\sqrt{2\al^\prime}k_2$
  \beqa
 \notag\cA_{PPP}&=&-(2\al^\prime)^{2}\zeta_1^\ast\cdot k_2 \underbrace{k_2\cdot k_3}_{=0} \zeta_3\cdot k_2-2\al^\prime\zeta_1^\ast\cdot k_2\zeta_3\cdot k_2\\
  &&+2\al^\prime\zeta_1^\ast\cdot k_2\zeta_3\cdot k_2+2\al^\prime \zeta_1^\ast \cdot \zeta_3 \underbrace{k_2\cdot k_3}_{=0} = 0.
  \eeqa
  \item Case 3: $\zeta_3\rightarrow\sqrt{2\al^\prime}k_3$
  \beqa
  \notag\cA_{PPP}&=&-(2\al^\prime)^{2}\zeta_1^\ast\cdot k_2\zeta_2\cdot k_3 \underbrace{k_2\cdot k_3}_{=0}-2\al^\prime\zeta_1^\ast\cdot \zeta_2 \underbrace{k_2\cdot k_3}_{=0}\\
  &&+2\al^\prime \zeta_1^\ast\cdot k_2\zeta_2\cdot k_3 +2\al^\prime \zeta_1^\ast \cdot k
  _3 \zeta_2\cdot k_3 = 0.
  \eeqa
\end{itemize}
\subsubsection{Ward identities for P($\zeta,k_1$)-M($\epsilon_{\mu\nu},k_2$)-T($k_3$)}
From Eq.\eqref{KContar3}, we have
\beqas
\al^\prime k_1^\ast\cdot k_2=-\al^\prime k_1^\ast\cdot k_3=-1 \quad\mbox{and}\quad \al^\prime k_2\cdot k_3=0.
\eeqas

\begin{itemize}
  \item Case 1: $\zeta^\ast\rightarrow\sqrt{2\al^\prime}k_1$
  \beqa
  \notag\cA_{PMT}&=& 2\al^\prime \epsilon_{\mu\nu}k^{\ast\mu}_1(k^{\ast\nu}_1+k_3^\nu)+\sqrt{2\al^\prime}\epsilon\cdot k_1^\ast +(2\al^\prime)^2 k_1^\ast\cdot k_2 \epsilon_{\mu\nu}k^{\ast\mu}_1k_3^\nu\\
  \notag&=&2\al^\prime\epsilon_{\mu\nu}k^{\ast\mu}_1(k^{\ast\nu}_1-k_3^\nu)+\sqrt{2\al^\prime}\epsilon\cdot k_1^\ast\\
  &=&\underbrace{\Big[\sqrt{2\al^\prime}\epsilon_{\mu\nu}(k_2+iV)^\nu+\epsilon_\mu\Big]}_{=0 \quad\mbox{by}\quad \cL_1 \quad\mbox{condition}}\sqrt{2\al^\prime} k_1^{\ast\mu}=0
  \eeqa
  \item Case 2: Type I vector zero-norm state \\
  In this case, we replace the polarization tensors as follows: $\displaystyle\epsilon_{\mu\nu} \rightarrow\sqrt{\frac{\al^\prime}{2}}(e_\mu k_{2\nu}+e_\nu k_{2\mu})$ and $\epsilon_\mu\rightarrow e_\mu$.
  \beqa
      \notag\cA_{PMT}&=& \al^\prime\Big[\zeta^\ast\cdot k_2 e\cdot(k_1+k_3)+\zeta^\ast\cdot e k_2\cdot(k_1+k_3)\Big]+\zeta^\ast\cdot e \\
      \notag&&+2\al^{\prime 2}\zeta^\ast\cdot k_2 \Big(e\cdot k_1^\ast k_2\cdot k_3+ e\cdot k_3 k_1^\ast\cdot k_2\Big)\\
      &=&\al^\prime \zeta^\ast\cdot k_2 e\cdot(k_1-k_3)=\al^\prime \zeta^\ast\cdot k_2 e\cdot(k_2+iV)=0.
\eeqa
  \item Case 3: Type II singlet zero-norm state\\
  In this case, we replace the polarization tensors as follows: $\displaystyle\epsilon_{\mu\nu} \rightarrow3\al^\prime k_{2\mu}k_{2\nu}+\frac{1}{2}\eta_{\mu\nu}$ and $\displaystyle\epsilon_\mu\rightarrow \sqrt{\frac{\al^\prime}{2}}\Big(5k_{2\mu}-iV_\mu\Big)$
  \beqa
      \notag\cA_{PMT}&=&3\sqrt{2}\al^{\prime\frac{3}{2}}\zeta^\ast\dot k_2 k_2\cdot(k_1^\ast+k_3)+\sqrt{\frac{\al^\prime}{2}}\zeta^\ast\cdot (k_1^\ast+k_3)\\
      \notag&&+5\sqrt{\frac{\al^\prime}{2}}\zeta^\ast\cdot k_2-i\sqrt{\frac{\al^\prime}{2}}\zeta^\ast\cdot V\\
      \notag&&+6\sqrt{2}\al^{\prime\frac{5}{2}}\zeta^\ast\cdot k_2 k_1^\ast\cdot k_2 k_2\cdot k_3+\sqrt 2 \al^{\prime\frac{3}{2}}\zeta^\ast\cdot k_2 k_1^\ast\cdot k_3\\
      &=&\sqrt{\frac{\al^\prime}{2}}\zeta^\ast\cdot(k_1^\ast+k_2+k_3-iV)=\sqrt{2\al^\prime}\zeta^\ast\cdot(k_1^\ast-iV)=0
      \eeqa
\end{itemize}

\subsubsection{Ward identities for P($\zeta_1,k_1$)-M($\epsilon_{\mu\nu},k_2$)-P($\zeta_3,k_3$)}
From Eq.\eqref{KContar3}, we have
\beqas
2\al^\prime k_1^\ast\cdot k_2=-2\al^\prime k_2\cdot k_3=-2\al^\prime k_1^\ast\cdot k_3=-1 .
\eeqas
From polarization conditions and momentum conservation, we have
\beqas
\zeta_1^\ast\cdot(k_1^\ast-k_2-k_3-iV)=0\quad &\rightarrow&\quad
\zeta_1^\ast\cdot k_2 = -\zeta_1^\ast\cdot k_3 \quad\\
\mbox{ and}\quad\zeta_3\cdot(k_1^\ast-k_2-k_3-iV)=0\quad &\rightarrow&\quad \zeta_3 \cdot k_1^\ast=\zeta_3\cdot k_2.
\eeqas

\begin{itemize}
  \item Case 1: $\zeta_1^\ast\rightarrow\sqrt{2\al^\prime}k_1^\ast$
  \beqa
 \notag\cA_{PMP}&=& 2\sqrt{2\al^\prime}\epsilon_{\mu\nu}\zeta_3^\nu k_1^{\ast\mu}-(2\al^\prime)^{\frac{3}{2}}\epsilon_{\mu\nu}k_1^{\ast\mu} (k_1^{\ast\nu}+k_3^\nu) \zeta_3\cdot k_2
 +(2\al^\prime)^{\frac{3}{2}}\epsilon_{\mu\nu}\zeta_3^\mu (k_1^{\ast\nu}+k_3^\nu)k_1^\ast \cdot k_2\\
 \notag&&+(2\al^\prime)^{\frac{3}{2}}\epsilon_{\mu\nu}k_1^{\ast\mu} k_3^\nu(\zeta_3\cdot k_1^\ast-2\al^\prime k_1^\ast\cdot k_2 \zeta_3 \cdot k_2)-2\al^\prime\epsilon\cdot k_1^\ast \zeta_3\cdot k_2-2\al^\prime\epsilon\cdot\zeta_3 k_1^\ast\cdot k_2 \\
 \notag&=&-\Big[\sqrt{2\al^\prime} \epsilon_{\mu\nu}(k_1^{\ast}-k_3)^\nu+\epsilon_\mu\Big]\zeta_3^\mu
 -\Big[\sqrt{2\al^\prime}\epsilon_{\mu\nu} (k_1^{\ast}-k_3)^\nu+\epsilon_\mu\Big] (2\al^\prime k_1^{\ast\mu}\zeta_3\cdot k_2)\\
 \notag&=&-\underbrace{\Big[\sqrt{2\al^\prime} \epsilon_{\mu\nu}(k_2+iV)^\nu+\epsilon_\mu\Big]}_{=0 \quad\mbox{by}\quad \cL_1 \quad\mbox{condition}}\zeta_3^\mu
 -\underbrace{\Big[\sqrt{2\al^\prime}\epsilon_{\mu\nu} (k_2+iV)^\nu+\epsilon_\mu\Big]}_{=0 \quad\mbox{by}\quad \cL_1 \quad\mbox{condition}} (2\al^\prime k_1^{\ast\mu}\zeta_3\cdot k_2)\\
 &=&0.
  \eeqa
  \item Case 2: $\zeta_3\rightarrow\sqrt{2\al^\prime}k_3$
  \beqa
  \notag\cA_{PMP}&=&2\sqrt{2\al^\prime}\epsilon_{\mu\nu}\zeta_1^{*\mu}k_3^\nu-(2\al^\prime)^{\frac{3}{2}}\epsilon_{\mu\nu}\zeta_1^{*\mu} (k_1^{*\nu}+k_3^\nu) k_2 \cdot k_3
 +(2\al^\prime)^{\frac{3}{2}}\epsilon_{\mu\nu}k_3^\mu (k_1^{*\nu}+k_3^\nu) \zeta_1^*\cdot k_2\\
 \notag&&+(2\al^\prime)^{\frac{3}{2}}\epsilon_{\mu\nu}k_1^{*\mu} k_3^\nu(\zeta_1^*\cdot k_3-2\al^\prime\zeta_1^*\cdot k_2 k_2 \cdot k_3)-2\al^\prime\epsilon\cdot\zeta_1^* k_2\cdot k_3-2\al^\prime\epsilon\cdot k_3 \zeta_1^*\cdot k_2\\
 \notag&=&-\Big[\sqrt{2\al^\prime} \epsilon_{\mu\nu}(k_1^{\ast}-k_3)^\nu+\epsilon_\mu\Big]\zeta_1^{\ast\mu}
 -\Big[\sqrt{2\al^\prime}\epsilon_{\mu\nu} (k_1^{\ast}-k_3)^\nu+\epsilon_\mu\Big] (2\al^\prime k_3^{\mu}\zeta_1^\ast\cdot k_2)\\
 \notag&=&-\underbrace{\Big[\sqrt{2\al^\prime} \epsilon_{\mu\nu}(k_2+iV)^\nu+\epsilon_\mu\Big]}_{=0 \quad\mbox{by}\quad \cL_1 \quad\mbox{condition}}\zeta_1^{\mu\ast}
 -\underbrace{\Big[\sqrt{2\al^\prime}\epsilon_{\mu\nu} (k_2+iV)^\nu+\epsilon_\mu\Big]}_{=0 \quad\mbox{by}\quad \cL_1 \quad\mbox{condition}} (2\al^\prime k_3^{\mu}\zeta_1^\ast\cdot k_2)\\
 &=&0.
  \eeqa
  \item Case 3: Type I vector zero-norm state\\
  In this case, we replace the polarization tensors as follows: $\displaystyle\epsilon_{\mu\nu} \rightarrow\sqrt{\frac{\al^\prime}{2}}(e_\mu k_{2\nu}+e_\nu k_{2\mu})$ and $\epsilon_\mu\rightarrow e_\mu$
  \beqa
      \notag\cA_{PMP}&=&\sqrt{2\al^\prime}(\zeta_1^\ast\cdot e \zeta_3\cdot k_2+\zeta_1^\ast\cdot k_2 \zeta_3 \cdot e)\\
      \notag&&-\sqrt 2(\al^\prime)^{\frac{3}{2}}\Big[\zeta_1^\ast\cdot e (k_1^{\ast}+k_3)\cdot k_2+\zeta_1^\ast\cdot k_2 (k_1^{\ast}+k_3)\cdot e\Big]\zeta_3\cdot k_2\\
      \notag&&+\sqrt 2(\al^\prime)^{\frac{3}{2}}\zeta_1^\ast\cdot k_2\Big[\zeta_3\cdot e (k_1^{\ast}+k_3)\cdot k_2+\zeta_3\cdot k_2 (k_1^{\ast}+k_3)\cdot e\Big]\\
      \notag&&+\sqrt 2(\al^\prime)^{\frac{3}{2}}(e\cdot k_1^\ast k_2\cdot k_3+k_1^\ast\cdot k_2 e\cdot k_3)(\zeta_1^*\cdot\zeta_3-2\al^\prime\zeta_1^*\cdot k_2 \zeta_3 \cdot k_2)\\
      \notag&&-\sqrt{2\al^\prime}\zeta_1^*\cdot e \zeta_3\cdot k_2-\sqrt{2\al^\prime}\zeta_1^*\cdot k_2\zeta_3 \cdot e\\
      \notag&=&\sqrt{\frac{\al^\prime}{2}}e\cdot (k_1^\ast-k_3)(\zeta_1^*\cdot\zeta_3-2\al^\prime\zeta_1^*\cdot k_2 \zeta_3 \cdot k_2)\\
      &=&\sqrt{\frac{\al^\prime}{2}}e\cdot (k_2+iV)(\zeta_1^*\cdot\zeta_3-2\al^\prime\zeta_1^*\cdot k_2 \zeta_3 \cdot k_2)=0.
      \eeqa
  \item Case4: Type II singlet zero-norm state\\
  In this case, we replace the polarization tensors as follows: $\displaystyle\epsilon_{\mu\nu} \rightarrow3\al^\prime k_{2\mu}k_{2\nu}+\frac{1}{2}\eta_{\mu\nu}$ and $\displaystyle\epsilon_\mu\rightarrow \sqrt{\frac{\al^\prime}{2}}\Big(5k_{2\mu}-iV_\mu\Big)$
  \beqa
        \notag  \cA_{PMP}&=&6\al^\prime\zeta_1^\ast\cdot k_2 \zeta_3 \cdot k_2 +\zeta_1^\ast\cdot \zeta_3\\
        \notag  &&-6\al^{\prime2}\zeta_1^\ast \cdot k_2\zeta_3\cdot k_2 (k_1^\ast+k_3)\cdot k_2-\al^\prime \zeta_1^\ast\cdot (k_1^\ast+k_3)\zeta_3\cdot k_2\\
         \notag &&+6\al^{\prime2}\zeta_1^\ast \cdot k_2\zeta_3\cdot k_2 (k_1^\ast+k_3)\cdot k_2+\al^\prime \zeta_1^\ast\cdot k_2\zeta_3\cdot (k_1^\ast+k_3)\\
         \notag &&(6\al^{\prime 2}k_1^\ast \cdot k_2 k_2\cdot k_3+\al^\prime k_1^\ast\cdot k_3)(\zeta_1^*\cdot\zeta_3-2\al^\prime\zeta_1^*\cdot k_2 \zeta_3 \cdot k_2)\\
         \notag &&-5\al^\prime \zeta_1^\ast \cdot k_{2} \zeta_3\cdot k_2+i\al^\prime \zeta_1^\ast\cdot V \zeta_3\cdot k_2-5\al^\prime \zeta_1^\ast \cdot k_{2} \zeta_3\cdot k_2+i\al^\prime \zeta_1^\ast\cdot k_2\zeta_3\cdot V\\
         \notag &=&-\al^\prime \zeta_1^\ast\cdot(k_1^\ast+k_2+k_3-iV)\zeta_3\cdot k_2+\al^\prime \zeta_1^\ast\cdot k_2\zeta_3\cdot (k_1^\ast-k_2+k_3+iV)\\
          &=&-2\al^\prime \zeta_1^\ast\cdot(k_1^\ast-iV)\zeta_3\cdot k_2+2\al^\prime \zeta_1^\ast\cdot k_2\zeta_3\cdot (k_3+iV)=0.
          \eeqa 
\end{itemize}
\subsubsection{Ward identities for P($\zeta,k_1$)-M($\epsilon^{(2)}_{\mu\nu},k_2$)-M($\epsilon^{(3)}_{\mu\nu},k_3$)}
From Eq.\eqref{KContar3}, we have
\beqas
\al^\prime k_1^\ast\cdot k_2=0 \quad\mbox{and}\quad \al^\prime k_2\cdot k_3=1.
\eeqas
\begin{itemize}
  \item Case 1: $\zeta_1^\ast\rightarrow\sqrt{2\al^\prime}k_1^\ast$
  \beqa
\notag\cA_{PMM}\\
\notag&=&2\al^\prime k_1^*\cdot k_2 \big(2\al^\prime\epsilon^{(3)}_{\mu\nu}k^\mu_2k^\nu_2-\sqrt{2\al^\prime}\epsilon^{(3)}\cdot k_2\big)
\big(2\al^\prime \epsilon^{(2)}_{\rho\si}k^\rho_3 k^\si_3-\sqrt{2\al^\prime}\epsilon^{(2)}\cdot k_3\big)\\
\notag&&+4\al^\prime\epsilon^{(2)}_{\mu\nu}k_1^{*\mu} k^\nu_3\big(2\al^\prime \epsilon^{(3)}_{\rho\si}k^\rho_2k^\si_2-\sqrt{2\al^\prime}\epsilon^{(3)}\cdot k_2\big)\\
\notag&&-4\al^\prime\epsilon^{(3)}_{\mu\nu}k_1^{*\mu}k_2^\nu\big(2\al^\prime \epsilon^{(2)}_{\rho\si}k^\rho_3 k^\si_3-\sqrt{2\al^\prime}\epsilon^{(2)}\cdot k_3\big)\\
\notag&&+2\al^\prime k_1^{*}\cdot k_2\left(\begin{array}{rl}&4\sqrt{2\al^\prime}
\epsilon^{(3)}_{\mu\nu}\epsilon^{(2)\mu}k^\nu_2+4\sqrt{2\al^\prime}
\epsilon^{(2)}_{\mu\nu}\epsilon^{(3)\mu}k^\nu_3\\-&8\al^\prime \epsilon^{(2)}_{\mu\nu}\epsilon^{(3)\nu\si}k_{2\si}k_3^\mu +2\epsilon^{(2)}_{\mu\nu}\epsilon^{(3)\mu\nu}-6\epsilon^{(2)}\cdot\epsilon^{(3)}\end{array}\right)\\
\notag&&+4\sqrt{2\al^\prime}\epsilon^{(2)}_{\mu\nu}\epsilon^{(3)\nu}k_1^{*\mu}-4\sqrt{2\al^\prime}\epsilon^{(3)}_{\mu\nu}\epsilon^{(2)\nu}k_1^{*\mu}\\
\notag&&+8\al^\prime\epsilon^{(2)}_{\mu\nu}\epsilon^{(3)\nu}_{\si}k_1^{*\si}k_3^\mu -8\al^\prime\epsilon^{(2)}_{\mu\nu}\epsilon^{(3)\nu}_{\si}k_1^{*\mu}k_2^\si \\
\notag&=&
-4\al^\prime\epsilon^{(3)}_{\mu\nu}k_1^{*\mu}k_2^\nu\big(2\al^\prime \epsilon^{(2)}_{\rho\si}k^\rho_3 k^\si_3-\sqrt{2\al^\prime}\epsilon^{(2)}\cdot k_3\big)+4\al^\prime\epsilon^{(2)}_{\mu\nu}k_1^{*\mu} k^\nu_3\big(2\al^\prime \epsilon^{(3)}_{\rho\si}k^\rho_2k^\si_2-\sqrt{2\al^\prime}\epsilon^{(3)}\cdot k_2\big)\\
\notag&&+4\sqrt{2\al^\prime}\epsilon^{(2)}_{\mu\nu}\hspace{-1.2cm}\underbrace{\epsilon^{(3)\nu}}_{=-\sqrt{2\al^\prime}\epsilon^{(3)\nu}_\si(k_3+iV)^\si}\hspace{-1.2cm}k_1^{*\mu}
-4\sqrt{2\al^\prime}\epsilon^{(3)}_{\mu\nu}\hspace{-1.2cm}\underbrace{\epsilon^{(2)\nu}}_{=-\sqrt{2\al^\prime}\epsilon_\si^{(2)\nu}(k_2+iV)^\si}\hspace{-1.2cm}k_1^{*\mu}+8\al^\prime\epsilon^{(2)}_{\mu\nu}\epsilon^{(3)\nu}_{\si}k_1^{*\si}k_3^\mu -8\al^\prime\epsilon^{(2)}_{\mu\nu}\epsilon^{(3)\nu}_{\si}k_1^{*\mu}k_2^\si \\
\notag&=&
-4\al^\prime\epsilon^{(3)}_{\mu}k_1^{*\mu\nu}k_2^\nu\underbrace{\big[2\al^\prime \epsilon^{(2)}_{\rho\si}(k_1^\ast-k_2-iV)^\si -\sqrt{2\al^\prime}\epsilon^{(2)}_\rho\big]}_{=2\al^\prime\epsilon^{(2)}_{\rho\si}k_1^\si
 \quad \mbox{by}\quad \cL_1\quad\mbox{condition}}k_3^\rho\\
\notag&&+4\al^\prime\epsilon^{(2)}_{\mu\nu}k_1^{*\mu} k^\nu_3\underbrace{\big[2\al^\prime \epsilon^{(3)}_{\rho\si}(k_1^\ast-k_3-iV)^\si-\sqrt{2\al^\prime}\epsilon^{(3)}_\rho \big]}_{=2\al^\prime\epsilon^{(3)}_{\rho\si}k_1^\si  \quad \mbox{by}\quad \cL_1\quad\mbox{condition}} k_2^\rho\\
\notag&&-8\al^\prime\epsilon^{(2)}_{\mu\nu}\epsilon^{(3)\nu}_{\si}k_1^{*\mu}k_1^\si  +8\al^\prime\epsilon^{(2)}_{\mu\nu}\epsilon^{(3)\nu}_{\si}k_1^{*\mu}k_1^\si\\
&=&0.
\eeqa
  \item Case 2: Type I vector zero norm state for $M^{(2)}$\\
  In this case, we replace the polarization tensors as follows: $\displaystyle\epsilon^{(2)}_{\mu\nu} \rightarrow\sqrt{\frac{\al^\prime}{2}}
  (e_\mu^{(2)} k_{2\nu}+e_\nu^{(2)} k_{2\mu})$ and $\epsilon_\mu^{(2)}\rightarrow e_{\mu}^{(2)}$
  \beqa
\notag&&\cA_{PMM}\\
\notag&=&\Big[\sqrt{2\al^\prime}\zeta^*\cdot k_2 \big(2\al^\prime \epsilon^{(3)}_{\mu\nu}k^\mu_2k^\nu_2-\sqrt{2\al^\prime}\epsilon^{(3)}\cdot k_2\big)
-2\sqrt{2\al^\prime}\epsilon^{(3)}_{\mu\nu}\zeta^{*\mu}k_2^\nu\Big]\left[\begin{array}{rl}(2\al^\prime)^{\frac{3}{2}}& e^{(2)}\cdot k_3 k_2\cdot k_3\\-\sqrt{2\al^\prime}&e^{(2)} \cdot k_3\end{array}\right]\\
\notag&&+\sqrt{2\al^\prime}\zeta^{*}\cdot k_2
\left[\begin{array}{rl}
   & 2\sqrt{2\al^\prime}\epsilon_{\mu\nu}^{(3)}e^{(2)\mu}k_2^\nu + 4\sqrt{2\al^\prime}
\epsilon^{(3)}_{\mu\nu}e^{(2)\mu}k^\nu_2\\+&4\al^\prime(
e^{(2)}\cdot\epsilon^{(3)}k_2\cdot k_3 +e^{(2)}\cdot k_3\epsilon^{(3)}\cdot k_2)\\
-&2(2\al^\prime)^{\frac{3}{2}} (\epsilon^{(3)}_{\mu\nu}e^{(2)\mu} k^{\nu}_2 k_2\cdot k_3+e^{(2)}\cdot k_3\epsilon^{(3)}_{\mu\nu} k_2^{\mu}k_2^\nu)-6e^{(2)}\cdot \epsilon^{(3)}
\end{array}\right]\\
\notag&&+2\al^\prime\big(\zeta^{*}\cdot e^{(2)} k_2\cdot k_3+\zeta^{*}\cdot k_2 e^{(2)}\cdot k_3\big)\big(2\al^\prime \epsilon^{(3)}_{\mu\nu}k^\mu_2k^\nu_2-\sqrt{2\al^\prime}\epsilon^{(3)}\cdot k_2\big)\\
\notag&&+2\sqrt{2\al^\prime}(\zeta^{*}\cdot e^{(2)}\epsilon^{(3)}\cdot k_2+\zeta^{*}\cdot k_2 e^{(2)}\cdot \epsilon^{(3)} )-4\epsilon^{(3)}_{\mu\nu}\zeta^{*\mu}e^{(2)\nu}\\
\notag&&+4\al^\prime(\epsilon^{(3)}_{\mu\nu}e^{(2)\mu}\zeta^{*\nu}k_2\cdot k_3+e^{(2)}\cdot k_3\epsilon^{(3)}_{\mu\nu}\zeta^{*\nu}k^{\mu}_2 )-4\al^\prime(\epsilon^{(3)}_{\mu\nu}e^{(2)\mu}k_2^\nu \zeta^{*}\cdot k_2+e^{(2)}\cdot \zeta^\ast\epsilon^{(3)}_{\mu\nu}k_2^\mu k_2^\nu)\\
\notag&=&\Big[\sqrt{2\al^\prime}\zeta^*\cdot k_2 \big(2\al^\prime \epsilon^{(3)}_{\mu\nu}k^\mu_2k^\nu_2-\sqrt{2\al^\prime}\epsilon^{(3)}\cdot k_2\big)
-2\sqrt{2\al^\prime}\epsilon^{(3)}_{\mu\nu}\zeta^{*\mu}k_2^\nu\Big](\sqrt{2\al^\prime}e^{(2)} \cdot k_3)\\
\notag&&+\sqrt{2\al^\prime}\zeta^{*}\cdot k_2
\left[\begin{array}{rl}
   & 6\sqrt{2\al^\prime}\epsilon_{\mu\nu}^{(3)}e^{(2)\mu}k_2^\nu+4
e^{(2)}\cdot\epsilon^{(3)} +4\al^\prime e^{(2)}\cdot k_3\epsilon^{(3)}\cdot k_2\\
-& 4\sqrt{2\al^\prime}\epsilon^{(3)}_{\mu\nu}e^{(2)\mu} k^{\nu}_2 -2(2\al^\prime)^{\frac{3}{2}}e^{(2)}\cdot k_3\epsilon^{(3)}_{\mu\nu} k_2^{\mu}k_2^\nu-6e^{(2)}\cdot \epsilon^{(3)}
\end{array}\right]\\
\notag&&+\big(2\zeta^{*}\cdot e^{(2)}+2\al^\prime\zeta^{*}\cdot k_2 e^{(2)}\cdot k_3\big)\big(2\al^\prime \epsilon^{(3)}_{\mu\nu}k^\mu_2k^\nu_2-\sqrt{2\al^\prime}\epsilon^{(3)}\cdot k_2\big)\\
\notag&&+2\sqrt{2\al^\prime}(\zeta^{*}\cdot e^{(2)}\epsilon^{(3)}\cdot k_2+\zeta^{*}\cdot k_2 e^{(2)}\cdot \epsilon^{(3)} )-4\epsilon^{(3)}_{\mu\nu}\zeta^{*\mu}e^{(2)\nu}\\
&&+4\epsilon^{(3)}_{\mu\nu}e^{(2)\mu}\zeta^{*\nu}+4\al^\prime e^{(2)}\cdot k_3\epsilon^{(3)}_{\mu\nu}\zeta^{*\nu}k^{\mu}_2 -4\al^\prime(\epsilon^{(3)}_{\mu\nu}e^{(2)\mu}k_2^\nu \zeta^{*}\cdot k_2+e^{(2)}\cdot \zeta^\ast\epsilon^{(3)}_{\mu\nu}k_2^\mu k_2^\nu)=0.
\eeqa
  \item Case 3: Type I vector zero norm state for $M^{(3)}$\\
  In this case, we replace the polarization tensors as follows: $\displaystyle\epsilon^{(3)}_{\mu\nu} \rightarrow\sqrt{\frac{\al^\prime}{2}}(e_\mu k_{3\nu}+e_\nu k_{3\mu})$ and $\epsilon_\mu\rightarrow e^{(3)}_\mu$
  \beqa
     \notag&& \cA_{PMM}\\
     \notag&=&\left\{\begin{array}{rl}
      &\sqrt{2\al^\prime}\zeta^\ast\cdot k_2 \Big[(2\al^\prime)^{\frac{3}{2}}e^{(3)}\cdot k_2 k_2\cdot k_3-\sqrt{2\al^\prime}e^{(3)}\cdot k_2\Big]\\
      -&2\al^\prime \zeta^\ast \cdot k_3 e^{(3)}\cdot k_2-2\al^\prime e^{(3)}\cdot \zeta^\ast k_2\cdot k_3\end{array}\right\}\big(2\al^\prime \epsilon^{(2)}_{\mu\nu}k^\mu_3 k^\nu_3-\sqrt{2\al^\prime}\epsilon^{(2)}\cdot k_3\big)\\
     \notag &&+\sqrt{2\al^\prime}\zeta^{*}\cdot k_2
\left[\begin{array}{rl}
   & 2\sqrt{2\al^\prime}\epsilon_{\mu\nu}^{(2)}e^{(3)\mu}k_3^\nu +4\al^\prime(
\epsilon^{(2)}\cdot e^{(3)}k_2\cdot k_3 +\epsilon^{(2)}\cdot k_3 e^{(3)}\cdot k_2) \\+&4\sqrt{2\al^\prime}
\epsilon^{(2)}_{\mu\nu}e^{(3)\mu}k^\nu_3
-2(2\al^\prime)^{\frac{3}{2}} (\epsilon^{(2)}_{\mu\nu}e^{(3)\mu} k^{\nu}_3 k_2\cdot k_3+\epsilon^{(2)}_{\mu\nu} k_3^{\mu}k_3^\nu e^{(3)}\cdot k_2)\\-&6\epsilon^{(2) }\cdot e^{(3)}
\end{array}\right]\\
\notag&&+2\sqrt{2\al^\prime}\epsilon^{(2)}_{\mu\nu}\zeta^{*\mu} k^\nu_3\Big[(2\al^\prime)^{\frac{3}{2}}e^{(3)}\cdot k_2 k_2\cdot k_3-\sqrt{2\al^\prime}e^{(3)}\cdot k_2\Big]\\
\notag&&+4\epsilon^{(2)}_{\mu\nu}\zeta^{*\mu}e^{(3)\nu}-2\sqrt{2\al^\prime}(\epsilon^{(2)}\cdot k_3\zeta^{*}\cdot e^{(3)}+\zeta^{*}\cdot k_3 \epsilon^{(2)}\cdot e^{(3)}  )\\
\notag&&+4\al^\prime(\epsilon^{(2)}_{\mu\nu}e^{(3)\mu}k_3^\nu\zeta^*\cdot k_3+\epsilon^{(2)}_{\mu\nu}k^{\mu}_3k^\nu_3 \zeta^{*}\cdot e^{(3)})-4\al^\prime(\epsilon^{(2)}_{\mu\nu}e^{(3)\mu}\zeta^{\ast\nu}k_2\cdot k_3+\epsilon^{(2)}_{\mu\nu}\zeta^{\mu}k_3^\nu e^{(3)}\cdot k_2^\nu)\\
      \notag&=&\Big[2\al^\prime\zeta^\ast\cdot (k_2-k_3) e^{(3)}\cdot k_2-2 e^{(3)}\cdot \zeta^\ast\Big] \big(2\al^\prime \epsilon^{(2)}_{\mu\nu}k^\mu_3 k^\nu_3-\sqrt{2\al^\prime}\epsilon^{(2)}\cdot k_3\big)\\
     \notag &&+\sqrt{2\al^\prime}\zeta^{*}\cdot k_2
\left[\begin{array}{rl}
   & 6\sqrt{2\al^\prime}\epsilon_{\mu\nu}^{(2)}e^{(3)\mu}k_3^\nu +4
\epsilon^{(2)}\cdot e^{(3)} +4\al^\prime\epsilon^{(2)}\cdot k_3 e^{(3)}\cdot k_2 \\
-& 4\sqrt{2\al^\prime}\epsilon^{(2)}_{\mu\nu}e^{(3)\mu} k^{\nu}_3-2(2\al^\prime)^{\frac{3}{2}}\epsilon^{(2)}_{\mu\nu} k_3^{\mu}k_3^\nu e^{(3)}\cdot k_2-6\epsilon^{(2) }\cdot e^{(3)}
\end{array}\right]\\
\notag&&+4\al^\prime\epsilon^{(2)}_{\mu\nu}\zeta^{*\mu} k^\nu_3e^{(3)}\cdot k_2 +4\epsilon^{(2)}_{\mu\nu}\zeta^{*\mu}e^{(3)\nu}-2\sqrt{2\al^\prime}(\epsilon^{(2)}\cdot k_3\zeta^{*}\cdot e^{(3)}+\zeta^{*}\cdot k_3 \epsilon^{(2)}\cdot e^{(3)}  )\\
\notag&&+4\al^\prime(\epsilon^{(2)}_{\mu\nu}e^{(3)\mu}k_3^\nu\zeta^*\cdot k_3+\epsilon^{(2)}_{\mu\nu}k^{\mu}_3k^\nu_3 \zeta^{*}\cdot e^{(3)})-4\epsilon^{(2)}_{\mu\nu}e^{(3)\mu}\zeta^{\ast\nu}-4\al^\prime\epsilon^{(2)}_{\mu\nu}\zeta^{\mu}k_3^\nu e^{(3)}\cdot k_2^\nu\\
\notag&=&\sqrt{2\al^\prime}\zeta^{*}\cdot (k_2+k_3)\left[\begin{array}{rl}&2\sqrt{2\al^\prime}\epsilon^{(2)}_{\mu\nu}e^{(3)\mu}k_3^\nu-2 \epsilon^{(2)}\cdot e^{(3)}\\
-&\sqrt{2\al^\prime} e^{(3)}\cdot k_2 \big(2\al^\prime \epsilon^{(2)}_{\mu\nu}k^\mu_3 k^\nu_3-\sqrt{2\al^\prime}\epsilon^{(2)}\cdot k_3\big)\end{array}\right]\\
\notag&=&\sqrt{2\al^\prime}\zeta^{*}\cdot (k_1^\ast-iV)\left[\begin{array}{rl}&2\sqrt{2\al^\prime}\epsilon^{(2)}_{\mu\nu}e^{(3)\mu}k_3^\nu-2 \epsilon^{(2)}\cdot e^{(3)}\\
-&\sqrt{2\al^\prime} e^{(3)}\cdot k_2 \big(2\al^\prime \epsilon^{(2)}_{\mu\nu}k^\mu_3 k^\nu_3-\sqrt{2\al^\prime}\epsilon^{(2)}\cdot k_3\big)\end{array}\right]\\
&=&0.
      \eeqa
  \item Case 4: Type II singlet zero-norm state for $M^{(2)}$\\
  In this case, we replace the polarization tensors as follows: 
  $\displaystyle\epsilon_{\mu\nu}^{(2)} \rightarrow3\al^\prime k_{2\mu}k_{2\nu}+\frac{1}{2}\eta_{\mu\nu}$ and $\displaystyle\epsilon^{(2)}_\mu\rightarrow \sqrt{\frac{\al^\prime}{2}}\Big(5k_{2\mu}-iV_\mu\Big)$
  First note that
          \beqas
          &&(2\al^\prime \epsilon^{(2)}_{\mu\nu}k^\mu_3 k^\nu_3-\sqrt{2\al^\prime}\epsilon^{(2)}\cdot k_3\big)\\
          &\Rightarrow\quad&6\al^{\prime2}(k_2\cdot k_3)^2+\al^\prime k_3^2-5\al^\prime k_2\cdot k_3+i\al^\prime k_3\cdot V=0.
          \eeqas
          
          \beqa
\notag&&\cA_{PMM}\\
\notag&=&\sqrt{2\al^\prime}\zeta^{*}\cdot k_2
\left[\begin{array}{rl}
&6\al^\prime\epsilon^{(3)}_{\mu\nu}k_2^\mu k_2^\nu+\epsilon^{(3)}_{\mu\nu}\eta^{\mu\nu}+20 \al^\prime\epsilon^{(3)}_{\mu\nu}k_2^\mu k_2^\nu-4\al^\prime i\epsilon^{(3)}_{\mu\nu}k_2^\mu V^\nu\\
+&12\al^\prime\sqrt{2\al^\prime} \epsilon^{(3)}\cdot k_2 k_2\cdot k_3+2\sqrt{2\al^\prime} \epsilon^{(3)}\cdot k_3\\
-&24\al^{\prime2} \epsilon^{(3)}_{\mu\nu}k_2^\mu k_2^\nu k_2\cdot k_3-4\al^\prime \epsilon^{(3)}_{\mu\nu} k_2^\mu k_3^\nu
-15\sqrt{2\al^\prime} \epsilon^{(3)}\cdot k_2+3\sqrt{2\al^\prime}i \epsilon^{(3)}\cdot V
\end{array}\right]\\
\notag&&+\big(6\al^\prime\sqrt{2\al^\prime}\zeta^{*}\cdot k_2 k_2\cdot k_3+\sqrt{2\al^\prime}\zeta^{*}\cdot k_3\big)\big(2\al^\prime \epsilon^{(3)}_{\mu\nu}k^\mu_2k^\nu_2-\sqrt{2\al^\prime}\epsilon^{(3)}\cdot k_2\big)\\
\notag&&+12\al^\prime\zeta^*\cdot k_2\epsilon^{(3)}\cdot k_2+2\epsilon^{(3)}\cdot \zeta^*-10\sqrt{2\al^\prime}\epsilon^{(3)}_{\mu\nu}\zeta^{*\mu}k^{\nu}_2+2\sqrt{2\al^\prime}
i\epsilon^{(3)}_{\mu\nu}\zeta^{*\mu}V^{\nu}\\
\notag&&+12\al^\prime\sqrt{2\al^\prime}\epsilon^{(3)}_{\mu\nu}\zeta^{*\mu}k_2^\nu k_2\cdot k_3 +2\sqrt{2\al^\prime}\epsilon^{(3)}_{\mu\nu}\zeta^{*\mu}k_3^\nu -12\al^\prime\sqrt{2\al^\prime}\epsilon^{(3)}_{\mu\nu}k_2^\mu k_2^\nu\zeta^{*}\cdot k_2-2\sqrt{2\al^\prime}\epsilon^{(3)}_{\mu\nu}\zeta^{*\mu}k_2^\nu\\
\notag&=&\sqrt{2\al^\prime}\zeta^{*}\cdot k_2
\left[\begin{array}{rl}
&26\al^\prime\epsilon^{(3)}_{\mu\nu}k_2^\mu k_2^\nu+\epsilon^{(3)}_{\mu\nu}\eta^{\mu\nu}-4\al^\prime i \epsilon^{(3)}_{\mu\nu}k_2^\mu V^\nu\\
+&12\sqrt{2\al^\prime} \epsilon^{(3)}\cdot k_2 +2\sqrt{2\al^\prime} \epsilon^{(3)}\cdot k_3
-24\al^{\prime} \epsilon^{(3)}_{\mu\nu}k_2^\mu k_2^\nu \\-&4\al^\prime \epsilon^{(3)}_{\mu\nu} k_2^\mu k_3^\nu
-15\sqrt{2\al^\prime} \epsilon^{(3)}\cdot k_2+3\sqrt{2\al^\prime}i \epsilon^{(3)}\cdot V
\end{array}\right]\\
\notag&&+\big(6\sqrt{2\al^\prime}\zeta^{*}\cdot k_2 +\sqrt{2\al^\prime}\zeta^{*}\cdot k_3\big)\big(2\al^\prime \epsilon^{(3)}_{\mu\nu}k^\mu_2k^\nu_2-\sqrt{2\al^\prime}\epsilon^{(3)}\cdot k_2\big)\\
\notag&&+12\al^\prime\zeta^*\cdot k_2\epsilon^{(3)}\cdot k_2+2\epsilon^{(3)}\cdot \zeta^*-10\sqrt{2\al^\prime}\epsilon^{(3)}_{\mu\nu}\zeta^{*\mu}k^{\nu}_2+2\sqrt{2\al^\prime}
i\epsilon^{(3)}_{\mu\nu}\zeta^{*\mu}V^{\nu}\\
\notag&&+12\sqrt{2\al^\prime}\epsilon^{(3)}_{\mu\nu}\zeta^{*\mu}k_2^\nu +2\sqrt{2\al^\prime}\epsilon^{(3)}_{\mu\nu}\zeta^{*\mu}k_3^\nu -12\al^\prime\sqrt{2\al^\prime}\epsilon^{(3)}_{\mu\nu}k_2^\mu k_2^\nu\zeta^{*}\cdot k_2-2\sqrt{2\al^\prime}\epsilon^{(3)}_{\mu\nu}\zeta^{*\mu}k_2^\nu\\
\notag&=&\sqrt{2\al^\prime}\zeta^{*}\cdot k_2
\left[\begin{array}{rl}
&2\al^\prime\epsilon^{(3)}_{\mu\nu}k_2^\mu k_2^\nu\underbrace{-4\al^\prime \epsilon^{(3)}_{\mu\nu} k_2^\mu (k_3^\nu+iV)}_{=-2\sqrt{2\al^\prime}\epsilon^{(3)}\cdot k_2}-3\sqrt{2\al^\prime} \epsilon^{(3)}\cdot k_2\\ +&2\sqrt{2\al^\prime} \epsilon^{(3)}\cdot k_3
+\epsilon^{(3)}_{\mu\nu}\eta^{\mu\nu}+3\sqrt{2\al^\prime}i \epsilon^{(3)}\cdot V
\end{array}\right]\\
\notag&&+\sqrt{2\al^\prime}\zeta^{*}\cdot k_3\big(2\al^\prime \epsilon^{(3)}_{\mu\nu}k^\mu_2k^\nu_2-\sqrt{2\al^\prime}\epsilon^{(3)}\cdot k_2\big)+2\zeta^{*\mu}\Big[\sqrt{2\al^\prime}
\epsilon^{(3)}_{\mu\nu}(k_3+iV)^\nu+\epsilon^{(3)}_\mu\Big] \\
\notag&=&\sqrt{2\al^\prime}\underbrace{\zeta^{*}\cdot (k_2+k_3)}_{=\zeta\cdot (k_1^\ast-iV)=0}\big(2\al^\prime \epsilon^{(3)}_{\mu\nu}k^\mu_2k^\nu_2-\sqrt{2\al^\prime}\epsilon^{(3)}\cdot k_2\big)+2\zeta^{*\mu}\underbrace{\Big[\sqrt{2\al^\prime}
\epsilon^{(3)}_{\mu\nu}(k_3+iV)^\nu+\epsilon^{(3)}_\mu\Big]}_{=0 \quad \mbox{by}\quad \cL_1\quad\mbox{condition}}\\
 \notag&&+\sqrt{2\al^\prime}\zeta^{*}\cdot k_2\underbrace{\big[\epsilon^{(3)}_{\mu\nu}\eta^{\mu\nu}+\sqrt{2\al^\prime}\epsilon^{(3)}\cdot(
 2  k_3+3i V)\Big]}_{=0 \quad \mbox{by}\quad \cL_2\quad\mbox{condition}}\\
 &=&0.
          \eeqa
  \item Case 5: Type II singlet zero-norm state for $M^{(3)}$\\
  In this case, we replace the polarization tensors as follows: 
  $\displaystyle\epsilon_{\mu\nu}^{(3)} \rightarrow3\al^\prime k_{3\mu}k_{3\nu}+\frac{1}{2}\eta_{\mu\nu}$ and $\displaystyle\epsilon_\mu\rightarrow \sqrt{\frac{\al^\prime}{2}}\Big(5k_{2\mu}-iV_\mu\Big)$
  First note that
              \beqas
              &&2\al^\prime \epsilon^{(3)}_{\rho\si}k^\rho_2k^\si_2-\sqrt{2\al^\prime}\epsilon^{(3)}\cdot k_2\\
              &\Rightarrow&\quad 6\al^{\prime2}(k_2\cdot k_3)^2+\al^\prime k_2^2-5\al^\prime k_2\cdot k_3+i\al^\prime k_2\cdot V=0.
              \eeqas
              \beqa
\notag&&\cA_{PMM}\\
\notag&=&(-6\al^\prime\sqrt{2\al^\prime}\zeta^\ast\cdot k_3 k_2\cdot k_3-\sqrt{2\al^\prime}\zeta^\ast\cdot k_2)\big(2\al^\prime \epsilon^{(2)}_{\mu\nu}k^\mu_3 k^\nu_3-\sqrt{2\al^\prime}\epsilon^{(2)}\cdot k_3\big)\\
\notag&&+\sqrt{2\al^\prime}\zeta^{*}\cdot k_2\left[\begin{array}{rl}&\epsilon^{(2)}_{\mu\nu}\eta^{\mu\nu}+6\al^\prime\epsilon^{(2)}_{\mu\nu}k_3^\mu k_3^\nu+12\al^\prime\sqrt{2\al^\prime}
\epsilon^{(2)}\cdot k_3 k_2\cdot k_3+2\sqrt{2\al^\prime}\epsilon^{(2)}\cdot k_2\\
+&20\al^\prime
\epsilon^{(2)}_{\mu\nu}k_3^\mu k_3^\nu-4\al^\prime i\epsilon^{(2)}_{\mu\nu}k_3^\mu V^\nu -24\al^{\prime2} \epsilon^{(2)}_{\mu\nu}k_3^\mu k_3^\nu k_2\cdot k_3-4\al^\prime \epsilon^{(2)}_{\mu\nu}k_2^\mu k_3^\nu
\\-&15\sqrt{2\al^\prime}\epsilon^{(2)}\cdot k_3+3\sqrt{2\al^\prime}i\epsilon^{(2)}\cdot V\end{array}\right]\\
\notag&&+10\al^\prime\sqrt{2\al^\prime}\epsilon^{(2)}_{\mu\nu}\zeta^{*\mu}k_3^\nu
-2\sqrt{2\al^\prime}i\epsilon^{(2)}_{\mu\nu}\zeta^{*\mu}V^\nu
-12\al^\prime\zeta^{*}\cdot k_3\epsilon^{(2)}\cdot k_3-2\epsilon^{(2)}\cdot \zeta^\ast\\
\notag&&+12\al^\prime\sqrt{2\al^\prime}\zeta^{*}\cdot k_3\epsilon^{(2)}_{\mu\nu}k_3^\mu k^\nu_3
+2\sqrt{2\al^\prime}\epsilon^{(2)}_{\mu\nu}\zeta^{*\mu} k_3^\nu
-12\al^\prime\sqrt{2\al^\prime}\epsilon^{(2)}_{\mu\nu}\zeta^{\ast\mu} k^\nu_3 k_2\cdot k_3
-2\sqrt{2\al^\prime}\epsilon^{(2)}_{\mu\nu}\zeta^{*\mu} k_2^\nu\\
\notag&=&(-\sqrt{2\al^\prime}\zeta^\ast\cdot k_2)\big(2\al^\prime \epsilon^{(2)}_{\mu\nu}k^\mu_3 k^\nu_3-\sqrt{2\al^\prime}\epsilon^{(2)}\cdot k_3\big)-2\zeta^{*\mu}\Big[\sqrt{2\al^\prime}\epsilon^{(2)}_{\mu\nu}(k_2+iV)^\nu+\epsilon^{(2)}_\mu\Big]\\
\notag&&+\sqrt{2\al^\prime}\zeta^{*}\cdot k_2\left[\begin{array}{rl}&\epsilon^{(2)}_{\mu\nu}\eta^{\mu\nu}+26\al^\prime\epsilon^{(2)}_{\mu\nu}k_3^\mu k_3^\nu+12\sqrt{2\al^\prime}
\epsilon^{(2)}\cdot k_3 +2\sqrt{2\al^\prime}\epsilon^{(2)}\cdot k_2\\
&-24\al^{\prime} \epsilon^{(2)}_{\mu\nu}k_3^\mu k_3^\nu \underbrace{-4\al^\prime \epsilon^{(2)}_{\mu\nu}(k_2+iV)^\mu k_3^\nu}_{=2\sqrt{2\al^\prime}\epsilon^{(2)\cdot k_3}}
\\-&15\sqrt{2\al^\prime}\epsilon^{(2)}\cdot k_3+3\sqrt{2\al^\prime}i\epsilon^{(2)}\cdot V\end{array}\right]\\
\notag&=&-2\zeta^{*\mu}\underbrace{\Big[\sqrt{2\al^\prime}\epsilon^{(2)}_{\mu\nu}(k_2+iV)^\nu+\epsilon^{(2)}_\mu\Big]}_{=0 \quad \mbox{by}\quad \cL_1\quad\mbox{condition}}+\sqrt{2\al^\prime}\zeta^{*}\cdot k_2\underbrace{\Big[\epsilon^{(2)}_{\mu\nu}\eta^{\mu\nu}+\sqrt{2\al^\prime}\epsilon^{(2)}\cdot (2 k_2+3iV)\Big]}_{=0 \quad \mbox{by}\quad \cL_2\quad\mbox{condition}}\\
&=&0.
\eeqa
\end{itemize}
 
\subsection{Four-point Functions}
In the calculations of four-point stringy scattering amplitudes, we need to replace the momentum contractions in terms of Mandelstam variables,
\beqa
\label{4PW}\begin{array}{rl}
  2\al^\prime k_1\cdot k_2 &= m_1^2+m_2^2-\al^\prime s \\
  2\al^\prime k_1\cdot k_3 & = -m_2^2-m_4^2+\al^\prime s+\al^\prime t \\
  2\al^\prime k_2\cdot k_3 & = m_2^2+m_3^2-\al^\prime t
\end{array}
\eeqa
\subsubsection{Ward identity for T($k_4$)-T($k_3$)-T($k_2$)-P($\zeta,k_1$)}
  From Eq.\eqref{4PW}, we have
  \beqas
  2\al^\prime k_1\cdot k_2= -\al^\prime s-1 \quad \mbox{and} \quad 2\al^\prime k_1\cdot k_3=\al^\prime t +\al^\prime s+2
  \eeqas
The stringy Ward identities for this process can be checked by substituting the polarization vector for photon, $\zeta\rightarrow\sqrt{2\al^\prime}k_1$,
  \beqa
\notag\hspace{-1cm}&&\cA_{TTTP}\\
\notag&=&\frac{\Ga(-\al^\prime t-1)\Ga(-\al^\prime s-1)}{\Ga(-\al^\prime t-\al^\prime s-1)}\Big[(\al^\prime t+\al^\prime s+2)(2\al^\prime k_1\cdot k_2)+(\al^\prime s+1)(2\al^\prime k_1\cdot k_3)\Big]\\
\notag\hspace{-1cm}&=&\frac{\Ga(-\al^\prime t-1)\Ga(-\al^\prime s-1)}{\Ga(-\al^\prime t-\al^\prime s-1)}\Big[(\al^\prime t+\al^\prime s+2)(-\al^\prime s-1)+(\al^\prime t+\al^\prime s+2)(\al^\prime s+1)\Big]\\
&=&0.
\eeqa
\subsubsection{Ward identities for T($k_4$)-T($k_3$)-P($\zeta_2,k_2$)-P($\zeta_1,k_1$)}
From Eq.\eqref{4PW}, we have
  \beqas
  2\al^\prime k_1\cdot k_2= -\al^\prime s \quad,\quad 2\al^\prime k_1\cdot k_3=\al^\prime t+\al^\prime s+1 \quad \mbox{and} \quad 2\al^\prime k_2\cdot k_3=-\al^\prime t -1
  \eeqas
  First, we substitute the polarization vector for first photon, $\zeta_1\rightarrow\sqrt{2\al^\prime}k_1$,
  \beqa
\notag\hspace{-1cm}&&\cA_{TTPP}\\
\notag&=&\frac{\Ga(-\al^\prime t-1)\Ga(-\al^\prime s-1)}{\Ga(-\al^\prime t-\al^\prime s)}
\left\{\begin{array}{rl}
  (\al^\prime t+1)(\al^\prime t+\al^\prime s+1)\big[\sqrt{2\al^\prime}
  \zeta_2\cdot k_1-&(2\al^\prime)^{\frac{3}{2}} k_1\cdot k_2 \zeta_2\cdot k_1\big]\\
   -(\al^\prime t+1)(\al^\prime s+1)&\big[(2\al^\prime)^{\frac{3}{2}} k_1\cdot k_3 \zeta_2\cdot k_1\big]\\+(\al^\prime t+\al^\prime s+1)(\al^\prime s+1)&\big[(2\al^\prime)^{\frac{3}{2}} k_1\cdot k_2 \zeta_2\cdot k_3\big]\\
   +(\al^\prime s+1)(\al^\prime s)&\big[(2\al^\prime)^{\frac{3}{2}} k_1\cdot k_3 \zeta_2\cdot k_3\big]
\end{array}\right\}\\
\hspace{-1cm}&=&\frac{\Ga(-\al^\prime t-1)\Ga(-\al^\prime s-1)}{\Ga(-\al^\prime t-\al^\prime s)}
\left[\begin{array}{rl}
  \sqrt{2\al^\prime}\zeta_2\cdot k_1&(\al^\prime t+1)(\al^\prime t+\al^\prime s+1)(\al^\prime s+1)\\
   -\sqrt{2\al^\prime}\zeta_2\cdot k_1&(\al^\prime t+1)(\al^\prime s+1)(\al^\prime t+\al^\prime s+1)\\
   +\sqrt{2\al^\prime}\zeta_2\cdot k_3&(\al^\prime t+\al^\prime s+1)(\al^\prime s+1)(-\al^\prime s )\\
   +\sqrt{2\al^\prime}\zeta_2\cdot k_3&(\al^\prime s+1)(\al^\prime s)(\al^\prime t+\al^\prime s+1 )
\end{array}\right]=0.
\eeqa
Then, we check the case with the replacement for the second photon, $\zeta_2\rightarrow\sqrt{2\al^\prime}k_2$,
  \beqa
\notag\hspace{-1cm}&&\cA_{TTPP}\\
\notag&=&\frac{\Ga(-\al^\prime t-1)\Ga(-\al^\prime s-1)}{\Ga(-\al^\prime t-\al^\prime s)}
\left\{\begin{array}{rl}
  (\al^\prime t+1)(\al^\prime t+\al^\prime s+1)\big[\sqrt{2\al^\prime}\zeta_1\cdot k_2-&(2\al^\prime)^{\frac{3}{2}} \zeta_1\cdot k_2 k_1\cdot k_2\big]\\
   -(\al^\prime t+1)(\al^\prime s+1)&\big[(2\al^\prime)^{\frac{3}{2}}\zeta_1\cdot k_3 k_1\cdot k_2\big]\\+(\al^\prime t+\al^\prime s+1)(\al^\prime s+1)&\big[(2\al^\prime)^{\frac{3}{2}}\zeta_1\cdot k_2 k_2\cdot k_3\big]\\
   +(\al^\prime s+1)(\al^\prime s)&\big[(2\al^\prime)^{\frac{3}{2}}\zeta_1\cdot k_3 k_2\cdot k_3\big]
\end{array}\right\}\\
\hspace{-1cm}&=&\frac{\Ga(-\al^\prime t-1)\Ga(-\al^\prime s-1)}{\Ga(-\al^\prime t-\al^\prime s)}
\left[\begin{array}{rl}
  \sqrt{2\al^\prime}\zeta_1\cdot k_2&(\al^\prime t+1)(\al^\prime t+\al^\prime s+1)(\al^\prime s+1)\\
   -\sqrt{2\al^\prime}\zeta_1\cdot k_3&(\al^\prime t+1)(\al^\prime s+1)(-\al^\prime s)\\+\sqrt{2\al^\prime}\zeta_1\cdot k_2&(\al^\prime t+\al^\prime s+1)(\al^\prime s+1)(-\al^\prime t -1)\\
   +\sqrt{2\al^\prime}\zeta_1\cdot k_3&(\al^\prime s+1)(\al^\prime s)(-\al^\prime t -1)
\end{array}\right]=0.
\eeqa
\subsubsection{Ward identities for T($k_4$)-P($\zeta_3,k_3$)-P($\zeta_2,k_2$)-P($\zeta_1,k_1$)}
From Eq.\eqref{4PW}, we have
  \beqas
  2\al^\prime k_1\cdot k_2= -\al^\prime s \quad,\quad 2\al^\prime k_1\cdot k_3=\al^\prime t+\al^\prime s+1 \quad\mbox{and} \quad 2\al^\prime k_2\cdot k_3=-\al^\prime t
  \eeqas
  We replace the polarization vectors of each photon in turn:\\
  
    $\bullet$ Case 1: $\zeta_1\rightarrow\sqrt{2\al^\prime}k_1$\\
    ~\\
    \beqa
 \notag &&\cA_{TPPP}\\
 \notag&=&\frac{\Ga(-\al^\prime t-1)\Ga(-\al^\prime s-1)}{\Ga(-\al^\prime t-\al^\prime s+1)}\times\\
\notag&&\hspace{-1.5cm}\times\left\{\begin{array}{r}
                 (\al^\prime t+\al^\prime s)\left\{
                 \begin{array}{rll}
                   (\al^\prime t+1)(\al^\prime t)&\Big[&(2\al^\prime)^2 k_1\cdot k_2\zeta_2\cdot k_1\zeta_3\cdot k_1-2\al^\prime \zeta_2\cdot k_1\zeta_3\cdot k_1\Big]\\
                   +(\al^\prime t+1)(\al^\prime s+1)&\Big[&-(2\al^\prime)^2 k_1\cdot k_2\zeta_2\cdot k_3\zeta_3\cdot k_1
+(2\al^\prime)^2k_1\cdot k_3\zeta_2\cdot k_1\zeta_3\cdot k_2\Big]\\
                   +(\al^\prime s+1)(\al^\prime s)&\Big[&-(2\al^\prime)^2k_1\cdot k_3\zeta_2\cdot k_3\zeta_3\cdot k_2+2\al^\prime k_1\cdot k_3\zeta_2\cdot\zeta_3\big)\Big]
                 \end{array}\right\}
                 \\
                 +(\al^\prime t+1)(\al^\prime s+1)\left\{\begin{array}{rll}
                 (\al^\prime t)&\Big[&(2\al^\prime)^2 k_1\cdot k_3\zeta_2\cdot k_1\zeta_3\cdot k_1-2\al^\prime\zeta_2\cdot k_1\zeta_3\cdot k_1\Big]\\
                 +(\al^\prime s)&\Big[&-(2\al^\prime)^2k_1\cdot k_3\zeta_2\cdot k_3\zeta_3\cdot k_1+2\al^\prime\zeta_2\cdot k_3\zeta_3\cdot k_1\Big]
                 \end{array}\right\}
                  \\
                  +(\al^\prime t+\al^\prime s+1)(\al^\prime t+\al^\prime s)\left\{
                  \begin{array}{rll}
                  (\al^\prime t+1)&\Big[&(2\al^\prime)^2k_1\cdot k_2\zeta_2\cdot k_1\zeta_3\cdot k_2-2\al^\prime \zeta_2\cdot k_1\zeta_3\cdot k_2\Big]\\
                  +(\al^\prime s+1)&\Big[&-(2\al^\prime)^2k_1\cdot k_2\zeta_2\cdot k_3\zeta_3\cdot k_2+2\al^\prime k_1\cdot k_2\zeta_2\cdot\zeta_3\Big]
                  \end{array}
                  \right\}
               \end{array}\right\}\\
\notag&=&\frac{\Ga(-\al^\prime t-1)\Ga(-\al^\prime s-1)}{\Ga(-\al^\prime t-\al^\prime s+1)}\times\\
&&\hspace{-1.5cm}\times\left\{\begin{array}{l}
                 (\al^\prime t+\al^\prime s)\left[
                 \begin{array}{rl}
                   -(\al^\prime t+1)(\al^\prime t)(\al^\prime s+1)&(2\al^\prime\zeta_2\cdot k_1\zeta_3\cdot k_1)\\
                   +(\al^\prime t+1)(\al^\prime s+1)(\al^\prime s)&(2\al^\prime\zeta_2\cdot k_3\zeta_3\cdot k_1)\\
                   +(\al^\prime t+1)(\al^\prime t+\al^\prime s+1)(\al^\prime s+1)&(2\al^\prime\zeta_2\cdot k_1\zeta_3\cdot k_2)\\
                   -(\al^\prime t+\al^\prime s+1)(\al^\prime s+1)(\al^\prime s)&( 2\al^\prime\zeta_2\cdot k_3\zeta_3\cdot k_2+\zeta_2\cdot\zeta_3)
                 \end{array}\right]
                 \\
                 +(\al^\prime t+1)(\al^\prime s+1)\left[\begin{array}{rl}
                 (\al^\prime t)(\al^\prime t+\al^\prime s)&(2\al^\prime \zeta_2\cdot k_1\zeta_3\cdot k_1)\\
                 -(\al^\prime t+\al^\prime s)(\al^\prime s)&(2\al^\prime\zeta_2\cdot k_3\zeta_3\cdot k_1)
                 \end{array}\right]
                  \\
                  +(\al^\prime t+\al^\prime s+1)(\al^\prime t+\al^\prime s)\left[
                  \begin{array}{rl}
                  -(\al^\prime t+1)(\al^\prime s+1)&(2\al^\prime\zeta_2\cdot k_1\zeta_3\cdot k_2)\\
                  +(\al^\prime s+1)(\al^\prime s)&(2\al^\prime\zeta_2\cdot k_3\zeta_3\cdot k_2+\zeta_2\cdot\zeta_3)
                  \end{array}
                  \right]
               \end{array}\right\}=0.
\eeqa
    $\bullet$ Case 2: $\zeta_2\rightarrow\sqrt{2\al^\prime}k_2$\\
    ~\\
    \beqa
\notag&&\cA_{TPPP}\\
\notag&=&\frac{\Ga(-\al^\prime t-1)\Ga(-\al^\prime s-1)}{\Ga(-\al^\prime t-\al^\prime s+1)}\times\\
\notag&&\hspace{-1.5cm}\times\left\{\begin{array}{r}
                 (\al^\prime t+\al^\prime s)\left\{
                 \begin{array}{rll}
                   (\al^\prime t+1)(\al^\prime t)&\Big[&(2\al^\prime)^2 k_1\cdot k_2\zeta_1\cdot k_2 \zeta_3\cdot k_1-2\al^\prime\zeta_1\cdot k_2\zeta_3\cdot k_1\Big]\\
                   +(\al^\prime t+1)(\al^\prime s+1)&\Big[-& (2\al^\prime)^2k_2\cdot k_3\zeta_1\cdot k_2 \zeta_3\cdot k_1
+(2\al^\prime)^2k_1\cdot k_2\zeta_1\cdot k_3 \zeta_3\cdot k_2\Big]\\
                   +(\al^\prime s+1)(\al^\prime s)&\Big[-& (2\al^\prime)^2k_2\cdot k_3\zeta_1\cdot k_3\zeta_3\cdot k_2+\sqrt{2\al^\prime}\zeta_1\cdot k_3\zeta_3\cdot k_2\big)\Big]
                 \end{array}\right\}
                 \\
                 +(\al^\prime t+1)(\al^\prime s+1)\left\{\begin{array}{rll}
                 (\al^\prime t)&\Big[&(2\al^\prime)^2k_1\cdot k_2\zeta_1\cdot k_3\zeta_3\cdot k_1-2\al^\prime k_1\cdot k_2\zeta_1\cdot\zeta_3\Big]\\
                 +(\al^\prime s)&\Big[-&(2\al^\prime)^2k_2\cdot k_3\zeta_1\cdot k_3\zeta_3\cdot k_1+2\al^\prime k_2\cdot k_3\zeta_1\cdot\zeta_3\Big]
                 \end{array}\right\}
                  \\
                  +(\al^\prime t+\al^\prime s+1)(\al^\prime t+\al^\prime s)\left\{
                  \begin{array}{rll}
                  (\al^\prime t+1)&\Big[&(2\al^\prime)^2k_1\cdot k_2\zeta_1\cdot k_2\zeta_3\cdot k_2-2\al^\prime\zeta_1\cdot k_2\zeta_3\cdot k_2\Big]\\
                  +(\al^\prime s+1)&\Big[-&(2\al^\prime)^2k_2\cdot k_3\zeta_1\cdot k_2\zeta_3\cdot k_2+2\al^\prime\zeta_1\cdot k_2\zeta_3\cdot k_2\Big]
                  \end{array}
                  \right\}
               \end{array}\right\}\\
\notag&=&
               \frac{\Ga(-\al^\prime t-1)\Ga(-\al^\prime s-1)}{\Ga(-\al^\prime t-\al^\prime s+1)}\times\\
&&\hspace{-1.5cm}\times\left\{\begin{array}{r}
                 (\al^\prime t+\al^\prime s)\left[
                 \begin{array}{rl}
                   -(\al^\prime t+1)(\al^\prime t)(\al^\prime s+1)&(2\al^\prime \zeta_1\cdot k_2 \zeta_3\cdot k_1)\\
                   +(\al^\prime t+1)(\al^\prime t)(\al^\prime s+1)&(2\al^\prime \zeta_1\cdot k_2 \zeta_3\cdot k_1)\\
-(\al^\prime t+1)(\al^\prime s+1)(\al^\prime s)&(2\al^\prime\zeta_1\cdot k_3 \zeta_3\cdot k_2)\\
                   +(\al^\prime t+1)(\al^\prime s+1)(\al^\prime s)&( 2\al^\prime\zeta_1\cdot k_3\zeta_3\cdot k_2)
                 \end{array}\right]
                 \\
                 +(\al^\prime t+1)(\al^\prime s+1)\left[\begin{array}{rl}
                 -(\al^\prime t)(\al^\prime s)&(2\al^\prime \zeta_1\cdot k_3\zeta_3\cdot k_1-\zeta_1\cdot\zeta_3)\\
                 +(\al^\prime t)(\al^\prime s)&(2\al^\prime \zeta_1\cdot k_3\zeta_3\cdot k_1-\zeta_1\cdot\zeta_3)
                 \end{array}\right]
                  \\
                  +(\al^\prime t+\al^\prime s+1)(\al^\prime t+\al^\prime s)\left[
                  \begin{array}{rl}
                  -(\al^\prime t+1)(\al^\prime s+1)&(2\al^\prime\zeta_1\cdot k_2\zeta_3\cdot k_2)\\
                  +(\al^\prime t+1)(\al^\prime s+1)&(2\al^\prime \zeta_1\cdot k_2\zeta_3\cdot k_2)
                  \end{array}
                  \right]
               \end{array}\right\}=0.
\eeqa
    $\bullet$ Case 3: $\zeta_3\rightarrow\sqrt{2\al^\prime}k_3$
    \beqa
\notag&&\cA_{TPPP}\\
\notag&=&\frac{\Ga(-\al^\prime t-1)\Ga(-\al^\prime s-1)}{\Ga(-\al^\prime t-\al^\prime s+1)}\times\\
\notag&&\hspace{-1.5cm}\times\left\{\begin{array}{r}
                 (\al^\prime t+\al^\prime s)\left\{
                 \begin{array}{rll}
                   (\al^\prime t+1)(\al^\prime t)&\Big[&(2\al^\prime)^2k_1\cdot k_3\zeta_1\cdot k_2\zeta_2\cdot k_1-2\al^\prime k_1\cdot k_3\zeta_1\cdot\zeta_2\Big]\\
                   +(\al^\prime t+1)(\al^\prime s+1)&\Big[-& (2\al^\prime)^2k_1\cdot k_3\zeta_1\cdot k_2\zeta_2\cdot k_3
+(2\al^\prime)^2 k_2\cdot k_3\zeta_1\cdot k_3\zeta_2\cdot k_1\Big]\\
                   +(\al^\prime s+1)(\al^\prime s)&\Big[-& (2\al^\prime)^2k_2\cdot k_3\zeta_1\cdot k_3\zeta_2\cdot k_3+2\al^\prime \zeta_1\cdot k_3\zeta_2\cdot k_3\big)\Big]
                 \end{array}\right\}
                 \\
                 +(\al^\prime t+1)(\al^\prime s+1)\left\{\begin{array}{rll}
                 (\al^\prime t)&\Big[&(2\al^\prime)^2k_1\cdot k_3\zeta_1\cdot k_3\zeta_2\cdot k_1-2\al^\prime\zeta_1\cdot k_3\zeta_2\cdot k_1\Big]\\
                 +(\al^\prime s)&\Big[-&(2\al^\prime)^2k_1\cdot k_3\zeta_1\cdot k_3\zeta_2\cdot k_3+2\al^\prime\zeta_1\cdot k_3\zeta_2\cdot k_3\Big]
                 \end{array}\right\}
                  \\
                  +(\al^\prime t+\al^\prime s+1)(\al^\prime t+\al^\prime s)\left\{
                  \begin{array}{rll}
                  (\al^\prime t+1)&\Big[&(2\al^\prime)^2k_2\cdot k_3\zeta_1\cdot k_2\zeta_2\cdot k_1-2\al^\prime k_2\cdot k_3\zeta_1\cdot\zeta_2\Big]\\
                  +(\al^\prime s+1)&\Big[-&(2\al^\prime)^2k_2\cdot k_3\zeta_1\cdot k_2\zeta_2\cdot k_3+2\al^\prime\zeta_1\cdot k_2\zeta_2\cdot k_3\Big]
                  \end{array}
                  \right\}
               \end{array}\right\}\\
      \notag         &=&\frac{\Ga(-\al^\prime t-1)\Ga(-\al^\prime s-1)}{\Ga(-\al^\prime t-\al^\prime s+1)}\times\\
&&\hspace{-1.5cm}\times\left\{\begin{array}{r}
                 (\al^\prime t+\al^\prime s)\left[
                 \begin{array}{rl}
                   (\al^\prime t+1)(\al^\prime t)(\al^\prime t+\al^\prime s+1)&(2\al^\prime\zeta_1\cdot k_2\zeta_2\cdot k_1-\zeta_1\cdot\zeta_2)\\
                   -(\al^\prime t+1)(\al^\prime t+\al^\prime s+1)(\al^\prime s+1)&( 2\al^\prime\zeta_1\cdot k_2\zeta_2\cdot k_3)\\
-(\al^\prime t+1)(\al^\prime t)(\al^\prime s+1)&(2\al^\prime\zeta_1\cdot k_3\zeta_2\cdot k_1)\\
                   +(\al^\prime t+1)(\al^\prime s+1)(\al^\prime s)&(2\al^\prime \zeta_1\cdot k_3\zeta_2\cdot k_3)
                 \end{array}\right]
                 \\
                 +(\al^\prime t+1)(\al^\prime s+1)\left[\begin{array}{rl}
                 (\al^\prime t)(\al^\prime t+\al^\prime s)&(2\al^\prime \zeta_1\cdot k_3\zeta_2\cdot k_1)\\
                 -(\al^\prime t+\al^\prime s)(\al^\prime s)&(2\al^\prime\zeta_1\cdot k_3\zeta_2\cdot k_3)
                 \end{array}\right]
                  \\
                  +(\al^\prime t+\al^\prime s+1)(\al^\prime t+\al^\prime s)\left[
                  \begin{array}{rl}
                  -(\al^\prime t+1)(\al^\prime t)&(2\al^\prime\zeta_1\cdot k_2\zeta_2\cdot k_1-\zeta_1\cdot\zeta_2)\\
                  +(\al^\prime t+1)(\al^\prime s+1)&(2\al^\prime\zeta_1\cdot k_2\zeta_2\cdot k_3)
                  \end{array}
                  \right]
               \end{array}\right\}=0.
\eeqa

\subsubsection{Ward identities for T($k_4$)-T($k_3$)-M($\epsilon_{\mu\nu},k_2$)-P($\zeta,k_1$)}
From Eq.\eqref{4PW}, we have
  \beqas
  2\al^\prime k_1\cdot k_2= -\al^\prime s+1 \quad,\quad 2\al^\prime k_1\cdot k_3=\al^\prime t+\al^\prime s \quad\mbox{and} \quad 2\al^\prime k_2\cdot k_3=-\al^\prime t
  \eeqas
  \begin{itemize}
    \item Case 1: $\zeta\rightarrow\sqrt{2\al^\prime}k_1$
      \beqa
\notag&&\cA_{TTMP}\\
\notag&=&\frac{\Ga(-\al^\prime t-1)\Ga(-\al^\prime s-1)}{\Ga(-\al^\prime t-\al^\prime s+1)}\times\\
\notag&&\hspace{-1.5cm}\times\left\{\begin{array}{r}
(\al^\prime t+\al^\prime s)\left\{\begin{array}{rl}(\al^\prime t+1)(\al^\prime t)
&\left[\begin{array}{l}~~2\al^\prime k_1\cdot k_2\big(2\al^\prime\epsilon_{\mu\nu}k^\mu_1k^\nu_1-\sqrt{2\al^\prime}\epsilon\cdot k_1\big)\\
-4\al^\prime\epsilon_{\mu\nu}k_1^\mu k_1^\nu+2\sqrt{2\al^\prime}\epsilon\cdot k_1\end{array}\right]\\
+(\al^\prime t+1)(\al^\prime s+1)&\Big[-2(2\al^\prime)^2 k_1\cdot k_2\epsilon_{\mu\nu}k_1^\mu k_3^\nu+4\al^\prime\epsilon_{\mu\nu}k_1^\mu k_3^\nu\Big]\\
+(\al^\prime s+1)(\al^\prime s)&\Big[~~2\al^\prime k_1\cdot k_2\big(2\al^\prime\epsilon_{\mu\nu}k^\mu_3k^\nu_3-\sqrt{2\al^\prime}\epsilon\cdot k_3\big)\Big]\end{array}\right\}\\
+(\al^\prime s+1)\left\{\begin{array}{rll}(\al^\prime t+1)(\al^\prime t)&\Big[&2\al^\prime k_1\cdot k_3\big(2\al^\prime\epsilon_{\mu\nu}k^\mu_1k^\nu_1-\sqrt{2\al^\prime}\epsilon\cdot k_1\big)\Big]\\
+(\al^\prime t+1)(\al^\prime s)&\Big[-&2(2\al^\prime)^2 k_1\cdot k_3\epsilon_{\mu\nu}k_1^\mu k_3^\nu\Big]\\
+(\al^\prime s)(\al^\prime s-1)&\Big[&2\al^\prime k_1\cdot k_3\big(2\al^\prime\epsilon_{\mu\nu}k^\mu_3k^\nu_3-\sqrt{2\al^\prime}\epsilon\cdot k_3\big)\Big]\end{array}\right\}
\end{array}\right\}\\
\notag&=&\frac{\Ga(-\al^\prime t-1)\Ga(-\al^\prime s-1)}{\Ga(-\al^\prime t-\al^\prime s+1)}\times\\
&&\hspace{-1.5cm}\times\left\{\begin{array}{r}
(\al^\prime t+\al^\prime s)\left[\begin{array}{rl}-(\al^\prime t+1)(\al^\prime t)(\al^\prime s+1)
&(2\al^\prime\epsilon_{\mu\nu}k^\mu_1k^\nu_1-\sqrt{2\al^\prime}\epsilon\cdot k_1)\\
+(\al^\prime t+1)(\al^\prime s+1)(\al^\prime s)&(4\al^\prime \epsilon_{\mu\nu}k_1^\mu k_3^\nu)\\
-(\al^\prime s+1)(\al^\prime s)(\al^\prime s-1)&(2\al^\prime\epsilon_{\mu\nu}k^\mu_3k^\nu_3-\sqrt{2\al^\prime}\epsilon\cdot k_3)\end{array}\right]\\
+(\al^\prime s+1)\left[\begin{array}{rl}(\al^\prime t+1)(\al^\prime t)(\al^\prime t+\al^\prime s)&(2\al^\prime\epsilon_{\mu\nu}k^\mu_1k^\nu_1-\sqrt{2\al^\prime}\epsilon\cdot k_1)\\
-(\al^\prime t+1)(\al^\prime t+\al^\prime s)(\al^\prime s)&(4\al^\prime\epsilon_{\mu\nu}k_1^\mu k_3^\nu)\\
+(\al^\prime t+\al^\prime s)(\al^\prime s)(\al^\prime s-1)&(2\al^\prime\epsilon_{\mu\nu}k^\mu_3k^\nu_3-\sqrt{2\al^\prime}\epsilon\cdot k_3)\end{array}\right]
\end{array}\right\}=0.
\eeqa
    \item Case 2: Type I vector zero norm state\\
    In this case, we replace the polarization tensors as follows: 
    $\displaystyle\epsilon_{\mu\nu} \rightarrow\sqrt{\frac{\al^\prime}{2}}(e_\mu k_{2\nu}+e_\nu k_{2\mu})$ and $\epsilon_\mu\rightarrow e_{\mu}$.\\
    For late convenience, we first simplify the following polarization projections,
     \beqas
     2\al^\prime\epsilon_{\mu\nu}k^\mu_1k^\nu_1-\sqrt{2\al^\prime}\epsilon\cdot k_1\quad
     \Rightarrow& & (2\al^\prime k_1 \cdot k_2)(\sqrt{2\al^\prime}e\cdot k_1)- \sqrt{2\al^\prime}e\cdot k_1\\
     &=& -(\al^\prime s)(\sqrt{2\al^\prime}e\cdot k_1)\\
     -4\al^\prime\epsilon_{\mu\nu}k^\mu_1k^\nu_3\quad
\Rightarrow& & -(2\al^\prime k_1 \cdot k_2)(\sqrt{2\al^\prime}e\cdot k_3)-(2\al^\prime k_2 \cdot k_3)(\sqrt{2\al^\prime}e\cdot k_1)\\
&=&(\al^\prime s-1)(\sqrt{2\al^\prime}e\cdot k_3)+(\al^\prime t)(\sqrt{2\al^\prime}e\cdot k_1)\\
2\al^\prime\epsilon_{\mu\nu}k^\mu_3k^\nu_3-\sqrt{2\al^\prime}\epsilon\cdot k_3\quad
     \Rightarrow& & (2\al^\prime k_2 \cdot k_3)(\sqrt{2\al^\prime}e\cdot k_3)- \sqrt{2\al^\prime}e\cdot k_3\\
     &=& -(\al^\prime t+1)(\sqrt{2\al^\prime}e\cdot k_3)\\
     -2\sqrt{2\al^\prime}\epsilon_{\mu\nu}\zeta^\mu k_1^\nu+2\epsilon\cdot\zeta\quad
     \Rightarrow &&  -2\al^\prime k_1\cdot k_2 \zeta\cdot e-2\al^\prime \zeta \cdot k_2 e\cdot k_1+2\zeta\cdot e\\
     &=&(\al^\prime s+1)\zeta\cdot e-2\al^\prime \zeta \cdot k_2 e\cdot k_1\\
     2\sqrt{2\al^\prime}\epsilon_{\mu\nu}\zeta^\mu k_3^\nu\quad
     \Rightarrow && 2\al^\prime k_2\cdot k_3 \zeta\cdot e+2\al^\prime \zeta \cdot k_2 e\cdot k_3\\
     &=&-(\al^\prime t)\zeta\cdot e+2\al^\prime \zeta \cdot k_2 e\cdot k_3.
     \eeqas
     We have
     \beqa
\notag&&\cA_{TTMP}\\
\notag&=&\frac{\Ga(-\al^\prime t-1)\Ga(-\al^\prime s-1)}{\Ga(-\al^\prime t-\al^\prime s+1)}\times\\
\notag&&\hspace{-1.5cm}\times\left\{\begin{array}{l}
(\al^\prime t+\al^\prime s)
\left\{\begin{array}{rl}
(\al^\prime t+1)(\al^\prime t)&\left[\begin{array}{l}
-(\al^\prime s)(2\al^\prime\zeta\cdot k_2e\cdot k_1)\\
+(\al^\prime s+1)\zeta\cdot e-2\al^\prime \zeta \cdot k_2 e\cdot k_1
     \end{array}\right]\\
+(\al^\prime t+1)(\al^\prime s+1)&\left[\begin{array}{rl}&(\al^\prime s-1)(2\al^\prime\zeta\cdot k_2e\cdot k_3)\\+&(\al^\prime t)(2\al^\prime\zeta\cdot k_2e\cdot k_1)\\
-&(\al^\prime t)\zeta\cdot e+2\al^\prime \zeta \cdot k_2 e\cdot k_3\end{array}\right]\\
+(\al^\prime s+1)(\al^\prime s)&\Big[-
(\al^\prime t+1)(2\al^\prime\zeta\cdot k_2e\cdot k_3)\Big]\end{array}\right\}\\
+(\al^\prime s+1)\left\{\begin{array}{rl}(\al^\prime t+1)(\al^\prime t)&\Big[-(\al^\prime s)(2\al^\prime\zeta\cdot k_3e\cdot k_1)\Big]\\
+(\al^\prime t+1)(\al^\prime s)&\left[\begin{array}{rl}&(\al^\prime s-1)(2\al^\prime\zeta\cdot k_3e\cdot k_3)\\+&(\al^\prime t)(2\al^\prime\zeta\cdot k_3e\cdot k_1)\end{array}\right]\\
+(\al^\prime s)(\al^\prime s-1)&\Big[-(\al^\prime t+1)(2\al^\prime\zeta\cdot k_3e\cdot k_3)\Big]\end{array}\right\}\end{array}\right\}\\
\notag&=&\frac{\Ga(-\al^\prime t-1)\Ga(-\al^\prime s-1)}{\Ga(-\al^\prime t-\al^\prime s+1)}\times\\
&&\hspace{-1.5cm}\times\left\{\begin{array}{l}
(\al^\prime t+\al^\prime s)
\left[\begin{array}{rl}
-(\al^\prime t+1)(\al^\prime t)(\al^\prime s+1)&(2\al^\prime\zeta\cdot k_2e\cdot k_1)\\
+(\al^\prime t+1)(\al^\prime t)(\al^\prime s+1)&(\zeta\cdot e)\\
+(\al^\prime t+1)(\al^\prime s+1)(\al^\prime s)&(2\al^\prime\zeta\cdot k_2e\cdot k_3)\\
+(\al^\prime t+1)(\al^\prime t)(\al^\prime s+1)&(2\al^\prime\zeta\cdot k_2e\cdot k_1)\\
-(\al^\prime t+1)(\al^\prime t)(\al^\prime s+1)&(\zeta\cdot e)\\
-(\al^\prime t+1)(\al^\prime s+1)(\al^\prime s)&(2\al^\prime\zeta\cdot k_2e\cdot k_3)
\end{array}\right]\\
+(\al^\prime s+1)\left[\begin{array}{rl}-(\al^\prime t+1)(\al^\prime t)(\al^\prime s)&(2\al^\prime\zeta\cdot k_3e\cdot k_1)\\
+(\al^\prime t+1)(\al^\prime s)(\al^\prime s-1)&(2\al^\prime\zeta\cdot k_3e\cdot k_3)\\
+(\al^\prime t+1)(\al^\prime t)(\al^\prime s)&(2\al^\prime\zeta\cdot k_3e\cdot k_1)\\
-(\al^\prime t+1)(\al^\prime s)(\al^\prime s-1)&(2\al^\prime\zeta\cdot k_3e\cdot k_3)\end{array}\right]\end{array}\right\}=0.
     \eeqa
    \item Case 3: Type II singlet zero-norm state\\
    In this case, we replace the polarization tensors as follows: 
    $\displaystyle\epsilon_{\mu\nu} \rightarrow3\al^\prime k_{2\mu}k_{2\nu}+\frac{1}{2}\eta_{\mu\nu}$ and $\displaystyle\epsilon_\mu\rightarrow \sqrt{\frac{\al^\prime}{2}}\Big(5k_{2\mu}-iV_\mu\Big)$.\\
    The relevant polarization constraints are given as follows: ($V$ is the dilaton gradient vector, $V^\mu\equiv\partial^\mu\Phi$)
       \beqas
     2\al^\prime\epsilon_{\mu\nu}k^\mu_1k^\nu_1-\sqrt{2\al^\prime}\epsilon\cdot k_1
     \Rightarrow &&6\al^{\prime 2} (k_1\cdot k_2)^2+\al^\prime k_1^2 -5\al^\prime k_1\cdot k_2 +i\al^\prime k_1\cdot V\\
     &=&\frac{3}{2}(\al^\prime s-1)^2+\frac{5}{2}(\al^\prime s-1)=\frac{3}{2}(\al^\prime s)(\al^\prime s-1)+(\al^\prime s-1)\\
     -4\al^\prime\epsilon_{\mu\nu}k^\mu_1k^\nu_3
     \Rightarrow && -12\al^{\prime2}k_1\cdot k_2 k_2\cdot k_3-2\al^\prime k_1\cdot k_3 \\
     &=&-3(\al^\prime t)(\al^\prime s-1)-(\al^\prime t+\al^\prime s)\\
     2\al^\prime\epsilon_{\mu\nu}k^\mu_3k^\nu_3-\sqrt{2\al^\prime}\epsilon\cdot k_3
     \Rightarrow &&6\al^{\prime 2} (k_2\cdot k_3)^2+\al^\prime k_3^2 -5\al^\prime k_2\cdot k_3 +i\al^\prime k_3\cdot V\\
     &=&\frac{3}{2}(\al^\prime t)^2+\frac{5}{2}(\al^\prime t)+1=\frac{3}{2}(\al^\prime t+1)(\al^\prime t)+(\al^\prime t+1)\\
     -2\sqrt{2\al^\prime}\epsilon_{\mu\nu}\zeta^\mu k_1^\nu+2\epsilon\cdot\zeta\quad
     \Rightarrow &&  (-6\al^\prime k_1\cdot k_2) (\sqrt{2\al^\prime} \zeta\cdot k_2)-\sqrt{2\al^\prime}\zeta\cdot k_1+5\sqrt{2\al^\prime}\zeta\cdot k_2-\sqrt{2\al^\prime}i\zeta\cdot V\\
     &=&3(\al^\prime s+1)(\sqrt{2\al^\prime} \zeta\cdot k_2)-\sqrt{2\al^\prime} \zeta\cdot k_2\\
     2\sqrt{2\al^\prime}\epsilon_{\mu\nu}\zeta^\mu k_3^\nu\quad
     \Rightarrow && (6\al^\prime k_2\cdot k_3) (\sqrt{2\al^\prime} \zeta\cdot k_2)+\sqrt{2\al^\prime}\zeta\cdot k_3\\
     &=&-3(\al^\prime t) (\sqrt{2\al^\prime} \zeta\cdot k_2)+\sqrt{2\al^\prime}\zeta\cdot k_3
     \eeqas
     Notice that the dependence on $V$ in all these invariant combination are absorbed into Mandelstam variables.
    \beqa
\notag&&\cA_{TTMP}\\
\notag&=&\frac{\Ga(-\al^\prime t-1)\Ga(-\al^\prime s-1)}{\Ga(-\al^\prime t-\al^\prime s+1)}\times\\
\notag&&\hspace{-1.5cm}\times\left\{\begin{array}{r}
(\al^\prime t+\al^\prime s)\left\{\begin{array}{rl}(\al^\prime t+1)(\al^\prime t)
&\left[\begin{array}{rl}&\displaystyle\frac{3}{2}(\al^\prime s)(\al^\prime s-1)(\sqrt{2\al^\prime}\zeta\cdot k_2)\\+&(\al^\prime s-1)(\sqrt{2\al^\prime}\zeta\cdot k_2)\\
+&3(\al^\prime s+1)(\sqrt{2\al^\prime} \zeta\cdot k_2)-\sqrt{2\al^\prime} \zeta\cdot k_2\end{array}\right]\\
+(\al^\prime t+1)(\al^\prime s+1)&
\left[\begin{array}{rl}-& 3(\al^\prime t)(\al^\prime s-1)(\sqrt{2\al^\prime}\zeta\cdot k_2)\\
-&(\al^\prime t+\al^\prime s)(\sqrt{2\al^\prime}\zeta\cdot k_2)\\
-&3(\al^\prime t) (\sqrt{2\al^\prime} \zeta\cdot k_2)+\sqrt{2\al^\prime}\zeta\cdot k_3\end{array}\right]\\
+(\al^\prime s+1)(\al^\prime s)&\left[\begin{array}{rl}&\displaystyle\frac{3}{2}(\al^\prime t)(\al^\prime t+1)(\sqrt{2\al^\prime}\zeta\cdot k_2)\\
+&(\al^\prime t+1)(\sqrt{2\al^\prime}\zeta\cdot k_2)\end{array}\right]\end{array}\right\}\\
+(\al^\prime s+1)
\left\{\begin{array}{rl}
(\al^\prime t+1)(\al^\prime t)&
\left[\begin{array}{rl}
&(\al^\prime s)(\al^\prime s-1)(\displaystyle\frac{3}{2}\sqrt{2\al^\prime}\zeta\cdot k_3)\\
+&(\al^\prime s-1)(\sqrt{2\al^\prime}\zeta\cdot k_3)\end{array}\right]\\
+(\al^\prime t+1)(\al^\prime s)&
\left[\begin{array}{rl}
-&(\al^\prime t)(\al^\prime s-1)(3\sqrt{2\al^\prime}\zeta\cdot k_3)\\
-&(\al^\prime t+\al^\prime s)(\sqrt{2\al^\prime}\zeta\cdot k_3)
\end{array}\right]\\
+(\al^\prime s)(\al^\prime s-1)&\left[\begin{array}{rl}&(\al^\prime t+1)(\al^\prime t)(\displaystyle\frac{3}{2}\sqrt{2\al^\prime}\zeta\cdot k_3)\\
+&(\al^\prime t+1)(\sqrt{2\al^\prime}\zeta\cdot k_3)\end{array}\right]
\end{array}\right\}
\end{array}\right\}
\eeqa
\beqa
\notag&=&\frac{\Ga(-\al^\prime t-1)\Ga(-\al^\prime s-1)}{\Ga(-\al^\prime t-\al^\prime s+1)}\times\\
\notag&&\hspace{-1.5cm}\times\left\{\begin{array}{r}
(\al^\prime t+\al^\prime s)\left[\begin{array}{rl}
(\al^\prime t+1)(\al^\prime t)(\al^\prime s)(\al^\prime s+1)&(\displaystyle\frac{3}{2}\sqrt{2\al^\prime}\zeta\cdot k_2)\\
+(\al^\prime t+1)(\al^\prime t)(\al^\prime s+1)&(\sqrt{2\al^\prime}\zeta\cdot k_2)\\
-(\al^\prime t+1)(\al^\prime t)(\al^\prime s+1)(\al^\prime s)&(3\sqrt{2\al^\prime}\zeta\cdot k_2)\\
-(\al^\prime t+1)(\al^\prime t)(\al^\prime s+1)&(\sqrt{2\al^\prime}\zeta\cdot k_2)\\
-(\al^\prime t+1)(\al^\prime s+1)(\al^\prime s)&(\sqrt{2\al^\prime}\zeta\cdot k_2)\\
+(\al^\prime t+1)(\al^\prime s+1)&(\sqrt{2\al^\prime}\zeta\cdot k_3)\\
+(\al^\prime t+1)(\al^\prime t)(\al^\prime s+1)(\al^\prime s)&(\displaystyle\frac{3}{2}\sqrt{2\al^\prime}\zeta\cdot k_2)\\
(\al^\prime t+1)(\al^\prime s+1)(\al^\prime s)&(\sqrt{2\al^\prime}\zeta\cdot k_2)\end{array}\right]\\
+(\al^\prime s+1)
\left[\begin{array}{rl}
(\al^\prime t+1)(\al^\prime t)(\al^\prime s)(\al^\prime s-1)&
(\displaystyle\frac{3}{2}\sqrt{2\al^\prime}\zeta\cdot k_3)\\
+(\al^\prime t+1)(\al^\prime t)(\al^\prime s-1)&(\sqrt{2\al^\prime}\zeta\cdot k_3)\\
-(\al^\prime t+1)(\al^\prime t)(\al^\prime s)(\al^\prime s-1)&(3\sqrt{2\al^\prime}\zeta\cdot k_3)\\
-(\al^\prime t+1)(\al^\prime t+\al^\prime s)(\al^\prime s)&(\sqrt{2\al^\prime}\zeta\cdot k_3)\\
+(\al^\prime t+1)(\al^\prime t)(\al^\prime s)(\al^\prime s-1)&(\displaystyle\frac{3}{2}\sqrt{2\al^\prime}\zeta\cdot k_3)\\
+(\al^\prime t+1)(\al^\prime s)(\al^\prime s-1)&(\sqrt{2\al^\prime}\zeta\cdot k_3)
\end{array}\right]
\end{array}\right\}\\
\notag&=&\frac{\Ga(-\al^\prime t-1)\Ga(-\al^\prime s-1)}{\Ga(-\al^\prime t-\al^\prime s+1)}\times\\
&&\hspace{-1.5cm}\times\left\{\begin{array}{r}
(\al^\prime t+\al^\prime s)\Big[
(\al^\prime t+1)(\al^\prime s+1)(\sqrt{2\al^\prime}\zeta\cdot k_3)\Big]\\
+(\al^\prime s+1)
\left[\begin{array}{rl}
+(\al^\prime t+1)(\al^\prime t)(\al^\prime s-1)&(\sqrt{2\al^\prime}\zeta\cdot k_3)\\
-(\al^\prime t+1)(\al^\prime t+\al^\prime s)(\al^\prime s)&(\sqrt{2\al^\prime}\zeta\cdot k_3)\\
+(\al^\prime t+1)(\al^\prime s)(\al^\prime s-1)&(\sqrt{2\al^\prime}\zeta\cdot k_3)
\end{array}\right]
\end{array}\right\}=0.
\eeqa
  \end{itemize}
\subsubsection{Ward identities for T($k_4$)-P($\zeta_3,k_3$)-M($\epsilon_{\mu\nu},k_2$)-P($\zeta_1,k_1$)}
From Eq.\eqref{4PW}, we have
  \beqas
  2\al^\prime k_1\cdot k_2= -\al^\prime s+1 \quad,\quad 2\al^\prime k_1\cdot k_3=\al^\prime t+\al^\prime s \quad\mbox{and} \quad 2\al^\prime k_2\cdot k_3=-\al^\prime t+1
  \eeqas
  
  \begin{itemize}
    \item Case 1: $\zeta_1\rightarrow\sqrt{2\al^\prime}k_1$
    \beqas
\notag&&\cA_{TPMP}\\
\notag&=&\frac{\Ga(-\al^\prime t-1)\Ga(-\al^\prime s-1)}{\Ga(-\al^\prime t-\al^\prime s+2)}\times\\
\notag&&\hspace{-3.5cm}\times\left\{\begin{array}{l}
(\al^\prime t+1)(\al^\prime s+1)
\left\{\begin{array}{rl}(\al^\prime t)(\al^\prime t-1)&\left[\begin{array}{rl}&\sqrt{2\al^\prime}\zeta_3\cdot k_1\\-&(2\al^\prime)^{\frac{3}{2}}k_1\cdot k_3 \zeta_3\cdot k_1\end{array}\right]\big(2\al^\prime\epsilon_{\mu\nu}k^\mu_1k^\nu_1-\sqrt{2\al^\prime}\epsilon\cdot k_1\big)\\
+(\al^\prime t)(\al^\prime s)&\left[\begin{array}{rl}&\sqrt{2\al^\prime}\zeta_3\cdot k_1\\-&(2\al^\prime)^{\frac{3}{2}}k_1\cdot k_3 \zeta_3\cdot k_1\end{array}\right]\big(-4\al^\prime \epsilon_{\mu\nu}k_1^\mu k_3^\nu\big)\\
+(\al^\prime s)(\al^\prime s-1)&\left[\begin{array}{rl}&\sqrt{2\al^\prime}\zeta_3\cdot k_1\\-&(2\al^\prime)^{\frac{3}{2}}k_1\cdot k_3 \zeta_3\cdot k_1\end{array}\right]\big(2\al^\prime\epsilon_{\mu\nu}k^\mu_3k^\nu_3-\sqrt{2\al^\prime}\epsilon\cdot k_3\big)
\end{array}\right\}\\
+(\al^\prime t+\al^\prime s-1)\left\{\begin{array}{rl}
(\al^\prime t+1)(\al^\prime t)(\al^\prime t-1)&\left[\begin{array}{rl}
-&(2\al^\prime)^{\frac{3}{2}}k_1\cdot k_2\zeta_3\cdot k_1\big(2\al^\prime\epsilon_{\mu\nu}k^\mu_1k^\nu_1-\sqrt{2\al^\prime}\epsilon\cdot k_1\big)\\
+&2\sqrt{2\al^\prime} \zeta_3\cdot k_1\big(2\al^\prime\epsilon_{\mu\nu}k^\mu_1k_1^\nu-\sqrt{2\al^\prime}\epsilon\cdot k_1\big)\end{array}\right]\\
+(\al^\prime t+1)(\al^\prime t)(\al^\prime s+1)&\left[\begin{array}{rl}-&(2\al^\prime)^{\frac{3}{2}}k_1\cdot k_3\zeta_3\cdot k_2\big(2\al^\prime\epsilon_{\mu\nu}k^\mu_1k^\nu_1-\sqrt{2\al^\prime}\epsilon\cdot k_1\big)\\
+&2(2\al^\prime)^{\frac{5}{2}} k_1\cdot k_2 \zeta_3\cdot k_1\epsilon_{\mu\nu}k^\mu_1k^\nu_3\\
-&2(2\al^\prime)^{\frac{3}{2}}\zeta_3\cdot k_1\epsilon_{\mu\nu}k^\mu_1k^\nu_3\end{array}\right]\\
+(\al^\prime t+1)(\al^\prime s+1)(\al^\prime s)&\left[\begin{array}{rl}-&(2\al^\prime)^{\frac{3}{2}}k_1\cdot k_2\zeta_3\cdot k_1\big(2\al^\prime\epsilon_{\mu\nu}k^\mu_3k^\nu_3-\sqrt{2\al^\prime}\epsilon\cdot k_3\big)
\\+&2(2\al^\prime)^{\frac{5}{2}}k_1\cdot k_3 \zeta_3\cdot k_2\epsilon_{\mu\nu}k^\mu_1k^\nu_3\\
-&2(2\al^\prime)^{\frac{3}{2}}k_1\cdot k_3\epsilon_{\mu\nu}k^\mu_1\zeta^\nu_3\end{array}\right]\\
+(\al^\prime s+1)(\al^\prime s)(\al^\prime s-1)&\left[\begin{array}{rl}-&(2\al^\prime)^{\frac{3}{2}}k_1\cdot k_3\zeta_3\cdot k_2\big(2\al^\prime\epsilon_{\mu\nu}k^\mu_3k^\nu_3-\sqrt{2\al^\prime}\epsilon\cdot k_3\big)
\\+&4\al^\prime k_1\cdot k_3\big(\sqrt{2\al^\prime}\epsilon_{\mu\nu}k^\mu_3\zeta_3^\nu-\epsilon\cdot\zeta_3\big)\end{array}\right]
\end{array}\right\}\\
+(\al^\prime t+\al^\prime s)(\al^\prime t+\al^\prime s-1)\left\{\begin{array}{rl}
&(\al^\prime t+1)(\al^\prime t)\left[\begin{array}{rl}-&(2\al^\prime)^{\frac{3}{2}}k_1\cdot k_2\zeta_3\cdot k_2\big(2\al^\prime\epsilon_{\mu\nu}k^\mu_1k^\nu_1-\sqrt{2\al^\prime}\epsilon\cdot k_1\big)
\\+&2\sqrt{2\al^\prime} \zeta_3\cdot k_2\big(2\al^\prime\epsilon_{\mu\nu}k^\mu_1 k_1^\nu-\sqrt{2\al^\prime}\epsilon\cdot k_1\big)\end{array}\right]\\
+&(\al^\prime t+1)(\al^\prime s+1)\left[\begin{array}{rl}&2(2\al^\prime)^{\frac{5}{2}}k_1\cdot k_2\zeta_3\cdot k_2 \epsilon_{\mu\nu}k^\mu_1k^\nu_3\\
-&2(2\al^\prime)^{\frac{3}{2}} k_1\cdot k_2 \epsilon_{\mu\nu}k^\mu_1\zeta^\nu_3\\
+&2\sqrt{2\al^\prime}\epsilon_{\mu\nu}k_1^\mu\zeta_3^\nu-2(2\al^\prime)^{\frac{3}{2}} \zeta_3\cdot k_2 \epsilon_{\mu\nu}k^\mu_1 k^\nu_3\end{array}\right]\\
+&(\al^\prime s+1)(\al^\prime s)\left[\begin{array}{rl}-&(2\al^\prime)^{\frac{3}{2}}k_1\cdot k_2\zeta_3\cdot k_2\big(2\al^\prime\epsilon_{\mu\nu}k^\mu_3k^\nu_3-\sqrt{2\al^\prime}\epsilon\cdot k_3\big)
\\+&4\al^\prime k_1\cdot k_2\big(\sqrt{2\al^\prime}\epsilon_{\mu\nu}k^\mu_3\zeta_3^\nu-\epsilon\cdot\zeta_3\big)\end{array}\right]
\end{array}\right\}\end{array}
\right\}\\
\eeqas
\beqa
\notag&=&\frac{\Ga(-\al^\prime t-1)\Ga(-\al^\prime s-1)}{\Ga(-\al^\prime t-\al^\prime s+2)}\times\\
\notag&&\hspace{-3cm}\times\left\{\begin{array}{l}
(\al^\prime t+1)(\al^\prime s+1)
\left[\begin{array}{rl}-(\al^\prime t)(\al^\prime t-1)(\al^\prime t+\al^\prime s-1)&(\sqrt{2\al^\prime}\zeta_3\cdot k_1)(2\al^\prime\epsilon_{\mu\nu}k^\mu_1k^\nu_1-\sqrt{2\al^\prime}\epsilon\cdot k_1)\\
+(\al^\prime t)(\al^\prime t+\al^\prime s-1)(\al^\prime s)&(\sqrt{2\al^\prime}\zeta_3\cdot k_1)(4\al^\prime \epsilon_{\mu\nu}k_1^\mu k_3^\nu)\\
-(\al^\prime t+\al^\prime s-1)(\al^\prime s)(\al^\prime s-1)&(\sqrt{2\al^\prime}\zeta_3\cdot k_1)(2\al^\prime\epsilon_{\mu\nu}k^\mu_3k^\nu_3-\sqrt{2\al^\prime}\epsilon\cdot k_3)
\end{array}\right]\\
+(\al^\prime t+\al^\prime s-1)\left[\begin{array}{rl}
(\al^\prime t+1)(\al^\prime t)(\al^\prime t-1)(\al^\prime s+1)&(\sqrt{2\al^\prime}\zeta_3\cdot k_1)(2\al^\prime\epsilon_{\mu\nu}k^\mu_1k^\nu_1-\sqrt{2\al^\prime}\epsilon\cdot k_1)\\
-(\al^\prime t+1)(\al^\prime t)(\al^\prime t+\al^\prime s)(\al^\prime s+1)&(\sqrt{2\al^\prime}\zeta_3\cdot k_2)(2\al^\prime\epsilon_{\mu\nu}k^\mu_1k^\nu_1-\sqrt{2\al^\prime}\epsilon\cdot k_1)\\
-(\al^\prime t+1)(\al^\prime t)(\al^\prime s+1)(\al^\prime s)&(\sqrt{2\al^\prime}\zeta_3\cdot k_1)(4\al^\prime\epsilon_{\mu\nu}k^\mu_1k^\nu_3)\\
+(\al^\prime t+1)(\al^\prime s+1)(\al^\prime s)(\al^\prime s-1)&(\sqrt{2\al^\prime}\zeta_3\cdot k_1)(2\al^\prime\epsilon_{\mu\nu}k^\mu_3k^\nu_3-\sqrt{2\al^\prime}\epsilon\cdot k_3)\\
+(\al^\prime t+1)(\al^\prime t+\al^\prime s)(\al^\prime s+1)(\al^\prime s)&(\sqrt{2\al^\prime}\zeta_3\cdot k_2)(4\al^\prime\epsilon_{\mu\nu}k^\mu_1k^\nu_3)\\
-(\al^\prime t+1)(\al^\prime t+\al^\prime s)(\al^\prime s+1)(\al^\prime s)&(\sqrt{2\al^\prime}\epsilon_{\mu\nu}k^\mu_1\zeta^\nu_3)\\
-(\al^\prime t+\al^\prime s)(\al^\prime s+1)(\al^\prime s)(\al^\prime s-1)&(\sqrt{2\al^\prime}\zeta_3\cdot k_2)(2\al^\prime\epsilon_{\mu\nu}k^\mu_3k^\nu_3-\sqrt{2\al^\prime}\epsilon\cdot k_3)\\
+(\al^\prime t+\al^\prime s)(\al^\prime s+1)(\al^\prime s)(\al^\prime s-1)&(2\sqrt{2\al^\prime}\epsilon_{\mu\nu}k^\mu_3\zeta_3^\nu-2\epsilon\cdot\zeta_3)\end{array}\right]\\
+(\al^\prime t+\al^\prime s)(\al^\prime t+\al^\prime s-1)\left[\begin{array}{rl}
(\al^\prime t+1)(\al^\prime t)(\al^\prime s+1)&(\sqrt{2\al^\prime}\zeta_3\cdot k_2)(2\al^\prime\epsilon_{\mu\nu}k^\mu_1k^\nu_1-\sqrt{2\al^\prime}\epsilon\cdot k_1)\\
-(\al^\prime t+1)(\al^\prime s+1)(\al^\prime s)&(\sqrt{2\al^\prime}\zeta_3\cdot k_2) (4\al^\prime\epsilon_{\mu\nu}k^\mu_1k^\nu_3)\\
+(\al^\prime t+1)(\al^\prime s+1)(\al^\prime s)&(2\sqrt{2\al^\prime} \epsilon_{\mu\nu}k^\mu_1\zeta^\nu_3)\\
+(\al^\prime s+1)(\al^\prime s)(\al^\prime s-1)&(\sqrt{2\al^\prime}\zeta_3\cdot k_2)(2\al^\prime\epsilon_{\mu\nu}k^\mu_3k^\nu_3-\sqrt{2\al^\prime}\epsilon\cdot k_3)\\
-(\al^\prime s+1)(\al^\prime s)(\al^\prime s-1)&(2\sqrt{2\al^\prime}\epsilon_{\mu\nu}k^\mu_3\zeta_3^\nu-2\epsilon\cdot\zeta_3)
\end{array}\right]\end{array}\right\}\\
&&\hspace{-3cm}=0.
\eeqa
    \item Case 2: $\zeta_3\rightarrow\sqrt{2\al^\prime}k_3$
    \beqas
\notag&&\cA_{TPMP}\\
\notag&=&\frac{\Ga(-\al^\prime t-1)\Ga(-\al^\prime s-1)}{\Ga(-\al^\prime t-\al^\prime s+2)}\times\\
\notag&&\hspace{-3.5cm}\times\left\{\begin{array}{l}
(\al^\prime t+1)(\al^\prime s+1)
\left\{\begin{array}{rl}(\al^\prime t)(\al^\prime t-1)&\left[\begin{array}{rl}&\sqrt{2\al^\prime}\zeta_1\cdot k_3\\-&(2\al^\prime)^{\frac{3}{2}}k_1\cdot k_3\zeta_1\cdot k_3 \end{array}\right]\big(2\al^\prime\epsilon_{\mu\nu}k^\mu_1k^\nu_1-\sqrt{2\al^\prime}\epsilon\cdot k_1\big)\\
+(\al^\prime t)(\al^\prime s)&\left[\begin{array}{rl}&\sqrt{2\al^\prime}\zeta_1\cdot k_3\\-&(2\al^\prime)^{\frac{3}{2}}k_1\cdot k_3\zeta_1\cdot k_3 \end{array}\right]\big(-4\al^\prime \epsilon_{\mu\nu}k_1^\mu k_3^\nu\big)\\
+(\al^\prime s)(\al^\prime s-1)&\left[\begin{array}{rl}&\sqrt{2\al^\prime}\zeta_1\cdot k_3\\-&(2\al^\prime)^{\frac{3}{2}}k_1\cdot k_3\zeta_1\cdot k_3 \end{array}\right]\big(2\al^\prime\epsilon_{\mu\nu}k^\mu_3k^\nu_3-\sqrt{2\al^\prime}\epsilon\cdot k_3\big)
\end{array}\right\}\\
+(\al^\prime t+\al^\prime s-1)\left\{\begin{array}{rl}
(\al^\prime t+1)(\al^\prime t)(\al^\prime t-1)&\left[\begin{array}{rl}
-&(2\al^\prime)^{\frac{3}{2}}k_1\cdot k_3\zeta_1\cdot k_2 \big(2\al^\prime\epsilon_{\mu\nu}k^\mu_1k^\nu_1-\sqrt{2\al^\prime}\epsilon\cdot k_1\big)\\
+&4\al^\prime k_1\cdot k_3\big(\sqrt{2\al^\prime}\epsilon_{\mu\nu}\zeta^\mu_1k_1^\nu-\epsilon\cdot \zeta_1\big)\end{array}\right]\\
+(\al^\prime t+1)(\al^\prime t)(\al^\prime s+1)&\left[\begin{array}{rl}-&(2\al^\prime)^{\frac{3}{2}}k_2\cdot k_3\zeta_1\cdot k_3 \big(2\al^\prime\epsilon_{\mu\nu}k^\mu_1k^\nu_1-\sqrt{2\al^\prime}\epsilon\cdot k_1\big)\\
+&2(2\al^\prime)^{\frac{5}{2}}k_1\cdot k_3 \zeta_1\cdot k_2 \epsilon_{\mu\nu}k^\mu_1k^\nu_3\\
-&2(2\al^\prime)^{\frac{3}{2}}k_1\cdot k_3\epsilon_{\mu\nu}\zeta^\mu_1k^\nu_3\end{array}\right]\\
+(\al^\prime t+1)(\al^\prime s+1)(\al^\prime s)&\left[\begin{array}{rl}-&(2\al^\prime)^{\frac{3}{2}}k_1\cdot k_3\zeta_1\cdot k_2\big(2\al^\prime\epsilon_{\mu\nu}k^\mu_3k^\nu_3-\sqrt{2\al^\prime}\epsilon\cdot k_3\big)
\\+&2(2\al^\prime)^{\frac{5}{2}}k_2\cdot k_3 \zeta_1\cdot k_3\epsilon_{\mu\nu}k^\mu_1k^\nu_3\\
-&2(2\al^\prime)^{\frac{3}{2}}\zeta_1\cdot k_3\epsilon_{\mu\nu}k^\mu_1 k^\nu_3\end{array}\right]\\
+(\al^\prime s+1)(\al^\prime s)(\al^\prime s-1)&\left[\begin{array}{rl}-&(2\al^\prime)^{\frac{3}{2}}k_2\cdot k_3\zeta_1\cdot k_3\big(2\al^\prime\epsilon_{\mu\nu}k^\mu_3k^\nu_3-\sqrt{2\al^\prime}\epsilon\cdot k_3\big)
\\+&2\sqrt{2\al^\prime} \zeta_1\cdot k_3\big(2\al^\prime\epsilon_{\mu\nu}k^\mu_3 k_3^\nu-\sqrt{2\al^\prime}\epsilon\cdot k_3\big)\end{array}\right]
\end{array}\right\}\\
+(\al^\prime t+\al^\prime s)(\al^\prime t+\al^\prime s-1)\left\{\begin{array}{rl}
&(\al^\prime t+1)(\al^\prime t)\left[\begin{array}{rl}-&(2\al^\prime)^{\frac{3}{2}}k_2\cdot k_3\zeta_1\cdot k_2\big(2\al^\prime\epsilon_{\mu\nu}k^\mu_1k^\nu_1-\sqrt{2\al^\prime}\epsilon\cdot k_1\big)
\\+&2\sqrt{2\al^\prime} k_2\cdot k_3\big(\sqrt{2\al^\prime}\epsilon_{\mu\nu}\zeta^\mu_1 k_1^\nu-\epsilon\cdot \zeta_1\big)\end{array}\right]\\
+&(\al^\prime t+1)(\al^\prime s+1)\left[\begin{array}{rl}&2(2\al^\prime)^{\frac{5}{2}}k_2\cdot k_3\zeta_1\cdot k_2 \epsilon_{\mu\nu}k^\mu_1k^\nu_3\\
-&2(2\al^\prime)^{\frac{3}{2}} \zeta_1\cdot k_2 \epsilon_{\mu\nu}k^\mu_1 k^\nu_3\\
+&2\sqrt{2\al^\prime}\epsilon_{\mu\nu}\zeta_1^\mu k_3^\nu-2(2\al^\prime)^{\frac{3}{2}} k_2\cdot k_3 \epsilon_{\mu\nu}\zeta^\mu_1 k^\nu_3\end{array}\right]\\
+&(\al^\prime s+1)(\al^\prime s)\left[\begin{array}{rl}-&(2\al^\prime)^{\frac{3}{2}}k_2\cdot k_3\zeta_1\cdot k_2\big(2\al^\prime\epsilon_{\mu\nu}k^\mu_3k^\nu_3-\sqrt{2\al^\prime}\epsilon\cdot k_3\big)
\\+&2\sqrt{2\al^\prime} \zeta_1\cdot k_2\big(2\al^\prime\epsilon_{\mu\nu}k^\mu_3 k_3^\nu-\sqrt{2\al^\prime}\epsilon\cdot k_3\big)\end{array}\right]
\end{array}\right\}\end{array}
\right\}
\eeqas
\beqa
\notag&=&\frac{\Ga(-\al^\prime t-1)\Ga(-\al^\prime s-1)}{\Ga(-\al^\prime t-\al^\prime s+2)}\times\\
\notag&&\hspace{-3cm}\times\left\{\begin{array}{l}
(\al^\prime t+1)(\al^\prime s+1)
\left[\begin{array}{rl}-(\al^\prime t)(\al^\prime t-1)(\al^\prime t+\al^\prime s-1)&(\sqrt{2\al^\prime}\zeta_1\cdot k_3)(2\al^\prime\epsilon_{\mu\nu}k^\mu_1k^\nu_1-\sqrt{2\al^\prime}\epsilon\cdot k_1)\\
+(\al^\prime t)(\al^\prime t+\al^\prime s-1)(\al^\prime s)&(\sqrt{2\al^\prime}\zeta_1\cdot k_3)(4\al^\prime \epsilon_{\mu\nu}k_1^\mu k_3^\nu)\\
-(\al^\prime t+\al^\prime s-1)(\al^\prime s)(\al^\prime s-1)&(\sqrt{2\al^\prime}\zeta_1\cdot k_3)(2\al^\prime\epsilon_{\mu\nu}k^\mu_3k^\nu_3-\sqrt{2\al^\prime}\epsilon\cdot k_3)
\end{array}\right]\\
+(\al^\prime t+\al^\prime s-1)\left[\begin{array}{rl}
-(\al^\prime t+1)(\al^\prime t)(\al^\prime t-1)(\al^\prime t+\al^\prime s)&(\sqrt{2\al^\prime}\zeta_1\cdot k_2)(2\al^\prime\epsilon_{\mu\nu}k^\mu_1k^\nu_1-\sqrt{2\al^\prime}\epsilon\cdot k_1)\\
+(\al^\prime t+1)(\al^\prime t)(\al^\prime t-1)(\al^\prime t+\al^\prime s)&(2\sqrt{2\al^\prime}\epsilon_{\mu\nu}\zeta^\mu_1k_1^\nu-2\epsilon\cdot \zeta_1)\\
+(\al^\prime t+1)(\al^\prime t)(\al^\prime t-1)(\al^\prime s+1)&(\sqrt{2\al^\prime}\zeta_1\cdot k_3) (2\al^\prime\epsilon_{\mu\nu}k^\mu_1k^\nu_1-\sqrt{2\al^\prime}\epsilon\cdot k_1)\\
+(\al^\prime t+1)(\al^\prime t)(\al^\prime t+\al^\prime s)(\al^\prime s+1)&(\sqrt{2\al^\prime}\zeta_1\cdot k_2) (4\al^\prime\epsilon_{\mu\nu}k^\mu_1k^\nu_3)\\
-(\al^\prime t+1)(\al^\prime t)(\al^\prime t+\al^\prime s)(\al^\prime s+1)&(2\sqrt{2\al^\prime}\epsilon_{\mu\nu}\zeta^\mu_1k^\nu_3)\\
-(\al^\prime t+1)(\al^\prime t+\al^\prime s)(\al^\prime s+1)(\al^\prime s)&(\sqrt{2\al^\prime}\zeta_1\cdot k_2)(2\al^\prime\epsilon_{\mu\nu}k^\mu_3k^\nu_3-\sqrt{2\al^\prime}\epsilon\cdot k_3)\\
-(\al^\prime t+1)(\al^\prime t)(\al^\prime s+1)(\al^\prime s)&(\sqrt{2\al^\prime}\zeta_1\cdot k_3)(4\al^\prime\epsilon_{\mu\nu}k^\mu_1 k^\nu_3)\\
+(\al^\prime t+1)(\al^\prime s+1)(\al^\prime s)(\al^\prime s-1)&(\sqrt{2\al^\prime}\zeta_1\cdot k_3)(2\al^\prime\epsilon_{\mu\nu}k^\mu_3k^\nu_3-\sqrt{2\al^\prime}\epsilon\cdot k_3)\\
\end{array}\right]\\
+(\al^\prime t+\al^\prime s)(\al^\prime t+\al^\prime s-1)\left[\begin{array}{rl}
+(\al^\prime t+1)(\al^\prime t)(\al^\prime t-1)&(\sqrt{2\al^\prime}\zeta_1\cdot k_2)(2\al^\prime\epsilon_{\mu\nu}k^\mu_1k^\nu_1-\sqrt{2\al^\prime}\epsilon\cdot k_1)\\
-(\al^\prime t+1)(\al^\prime t)(\al^\prime t-1)&(2\sqrt{2\al^\prime}\epsilon_{\mu\nu}\zeta^\mu_1 k_1^\nu-2\epsilon\cdot \zeta_1\big)\\
-(\al^\prime t+1)(\al^\prime t)(\al^\prime s+1)&(\sqrt{2\al^\prime}\zeta_1\cdot k_2) (4\al^\prime\epsilon_{\mu\nu}k^\mu_1k^\nu_3)\\
+(\al^\prime t+1)(\al^\prime t)(\al^\prime s+1)&(2\sqrt{2\al^\prime}\epsilon_{\mu\nu}\zeta^\mu_1 k^\nu_3)\\
+(\al^\prime t)(\al^\prime s+1)(\al^\prime s)&(\sqrt{2\al^\prime}\zeta_1\cdot k_2)(2\al^\prime\epsilon_{\mu\nu}k^\mu_3k^\nu_3-\sqrt{2\al^\prime}\epsilon\cdot k_3)
\end{array}\right]
\end{array}\right\}\\
&&\hspace{-3cm}=0.
\eeqa
    \item Case 3: Type I vector zero norm state\\
    In this case, we replace the polarization tensors as follows: 
    $\displaystyle\epsilon_{\mu\nu} \rightarrow\sqrt{\frac{\al^\prime}{2}}(e_\mu k_{2\nu}+e_\nu k_{2\mu})$ and $\epsilon_\mu\rightarrow e_{\mu}$.\\
    First note that
     \beqas
     2\al^\prime\epsilon_{\mu\nu}k^\mu_1k^\nu_1-\sqrt{2\al^\prime}\epsilon\cdot k_1\quad
     \Rightarrow& & (2\al^\prime k_1 \cdot k_2)(\sqrt{2\al^\prime}e\cdot k_1)- \sqrt{2\al^\prime}e\cdot k_1\\
     &=& -(\al^\prime s)(\sqrt{2\al^\prime}e\cdot k_1)\\
     -4\al^\prime\epsilon_{\mu\nu}k^\mu_1k^\nu_3\quad
\Rightarrow& & -(2\al^\prime k_1 \cdot k_2)(\sqrt{2\al^\prime}e\cdot k_3)-(2\al^\prime k_2 \cdot k_3)(\sqrt{2\al^\prime}e\cdot k_1)\\
&=&(\al^\prime s-1)(\sqrt{2\al^\prime}e\cdot k_3)+(\al^\prime t-1)(\sqrt{2\al^\prime}e\cdot k_1)\\
2\al^\prime\epsilon_{\mu\nu}k^\mu_3k^\nu_3-\sqrt{2\al^\prime}\epsilon\cdot k_3\quad
     \Rightarrow& & (2\al^\prime k_2 \cdot k_3)(\sqrt{2\al^\prime}e\cdot k_3)- \sqrt{2\al^\prime}e\cdot k_3\\
     &=& -(\al^\prime t)(\sqrt{2\al^\prime}e\cdot k_3)\\
     -2\sqrt{2\al^\prime}\epsilon_{\mu\nu}\zeta_1^\mu k_1^\mu+2\epsilon\cdot\zeta_1\quad
     \Rightarrow &&  -2\al^\prime k_1\cdot k_2 \zeta_1\cdot e-2\al^\prime \zeta_1 \cdot k_2 e\cdot k_1+2\zeta_1\cdot e\\
     &=&(\al^\prime s+1)\zeta_1\cdot e-2\al^\prime \zeta_1 \cdot k_2 e\cdot k_1\\
     2\sqrt{2\al^\prime}\epsilon_{\mu\nu}\zeta_1^\mu k_3^\nu\quad
     \Rightarrow && 2\al^\prime k_2\cdot k_3 \zeta_1\cdot e+2\al^\prime \zeta_1 \cdot k_2 e\cdot k_3\\
     &=&-(\al^\prime t+1)\zeta_1\cdot e+2\al^\prime \zeta_1 \cdot k_2 e\cdot k_3\\
     -2\sqrt{2\al^\prime}\epsilon_{\mu\nu}\zeta_3^\mu k_3^\mu+2\epsilon\cdot\zeta_3\quad
     \Rightarrow &&  -2\al^\prime k_2\cdot k_3 \zeta_3\cdot e-2\al^\prime \zeta_3 \cdot k_2 e\cdot k_3+2\zeta_3\cdot e\\
     &=&(\al^\prime t-1)\zeta_3\cdot e-2\al^\prime \zeta_3 \cdot k_2 e\cdot k_3\\
     2\sqrt{2\al^\prime}\epsilon_{\mu\nu}\zeta_3^\mu k_1^\nu\quad
     \Rightarrow && 2\al^\prime k_1\cdot k_2 \zeta_3\cdot e+2\al^\prime \zeta_3 \cdot k_2 e\cdot k_1\\
     &=&-(\al^\prime s-1)\zeta_3\cdot e+2\al^\prime \zeta_3 \cdot k_2 e\cdot k_1\\
     -2\epsilon_{\mu\nu}\zeta_1^\mu\zeta_3^\nu \quad \Rightarrow &&-\sqrt{2\al^\prime}\zeta_1 \cdot k_2 \zeta_3 \cdot e-\sqrt{2\al^\prime}\zeta_1\cdot e\zeta_3\cdot k_2.
     \eeqas
     \beqas
\notag&&\cA_{TPMP}=\frac{\Ga(-\al^\prime t-1)\Ga(-\al^\prime s-1)}{\Ga(-\al^\prime t-\al^\prime s+2)}\times\\
\notag&&\hspace{-3cm}\times\left\{\begin{array}{l}
(\al^\prime t+1)(\al^\prime s+1)
\left\{\begin{array}{rl}-(\al^\prime t)(\al^\prime t-1)&(\zeta_1\cdot\zeta_3-2\al^\prime\zeta_1\cdot k_3 \zeta_3\cdot k_1)(\al^\prime s)(\sqrt{2\al^\prime}e\cdot k_1)\\
+(\al^\prime t)(\al^\prime s)&(\zeta_1\cdot\zeta_3-2\al^\prime\zeta_1\cdot k_3 \zeta_3\cdot k_1)\left[\begin{array}{rl}&(\al^\prime s-1)(\sqrt{2\al^\prime}e\cdot k_3)\\+&(\al^\prime t-1)(\sqrt{2\al^\prime}e\cdot k_1)\end{array}\right]\\
-(\al^\prime s)(\al^\prime s-1)&(\zeta_1\cdot\zeta_3-2\al^\prime\zeta_1\cdot k_3 \zeta_3\cdot k_1)(\al^\prime t)(\sqrt{2\al^\prime}e\cdot k_3)
\end{array}\right\}\\
+(\al^\prime t+\al^\prime s-1)\left\{\begin{array}{rl}
(\al^\prime t+1)(\al^\prime t)(\al^\prime t-1)&\left[\begin{array}{rl}
&(\al^\prime s)(2\al^\prime\zeta_1\cdot k_2\zeta_3\cdot k_1) (\sqrt{2\al^\prime}e\cdot k_1)\\
-&(\al^\prime s+1)(\sqrt{2\al^\prime}\zeta_3\cdot k_1\zeta_1\cdot e)\\
+&(2\al^\prime \zeta_1 \cdot k_2 \zeta_3\cdot k_1)(\sqrt{2\al^\prime}e\cdot k_1)\end{array}\right]\\
+(\al^\prime t+1)(\al^\prime t)(\al^\prime s+1)&
\left[\begin{array}{rl}
 &(\al^\prime s)(2\al^\prime\zeta_1\cdot k_3\zeta_3\cdot k_2)(\sqrt{2\al^\prime}e\cdot k_1)\\
-&(\al^\prime s-1)(2\al^\prime\zeta_1\cdot k_2 \zeta_3\cdot k_1)(\sqrt{2\al^\prime}e\cdot k_3)\\
-&(\al^\prime t-1)(2\al^\prime\zeta_1\cdot k_2 \zeta_3\cdot k_1)(\sqrt{2\al^\prime}e\cdot k_1)\\
+&(\al^\prime t-1)(2\sqrt{\al^\prime}\zeta_3\cdot k_1\zeta_1\cdot e)\\
-&(2\al^\prime \zeta_1 \cdot k_2 \zeta_3\cdot k_1)(2\sqrt{\al^\prime}e\cdot k_3)\end{array}\right]\\
+(\al^\prime t+1)(\al^\prime s+1)(\al^\prime s)&\left[\begin{array}{rl}&(\al^\prime t)(2\al^\prime\zeta_1\cdot k_2\zeta_3\cdot k_1)(\sqrt{2\al^\prime}e\cdot k_3)\\
-&(\al^\prime s-1)(2\al^\prime\zeta_1\cdot k_3 \zeta_3\cdot k_2)(\sqrt{2\al^\prime}e\cdot k_3)\\
-&(\al^\prime t-1)(2\al^\prime\zeta_1\cdot k_3 \zeta_3\cdot k_2)(\sqrt{2\al^\prime}e\cdot k_1)\\
+&(\al^\prime s-1)(\sqrt{2\al^\prime}\zeta_1\cdot k_3\zeta_3\cdot e)\\
-&(2\al^\prime \zeta_1\cdot k_3\zeta_3 \cdot k_2)( \sqrt{2\al^\prime} e\cdot k_1)
\end{array}\right]\\
+(\al^\prime s+1)(\al^\prime s)(\al^\prime s-1)&\left[\begin{array}{rl}&(\al^\prime t)(2\al^\prime\zeta_1\cdot k_3\zeta_3\cdot k_2)(\sqrt{2\al^\prime}e\cdot k_3)
\\-&(\al^\prime t+1)(\sqrt{2\al^\prime} \zeta_1\cdot k_3\zeta_3\cdot e)\\
+&(2\al^\prime \zeta_1\cdot k_3\zeta_3\cdot k_2)( \sqrt{2\al^\prime} e\cdot k_3)
\end{array}\right]
\end{array}\right\}\\
+(\al^\prime t+\al^\prime s)(\al^\prime t+\al^\prime s-1)\left\{\begin{array}{rl}
(\al^\prime t+1)(\al^\prime t)&\left[\begin{array}{rl}&(\al^\prime s)(2\al^\prime\zeta_1\cdot k_2\zeta_3\cdot k_2)(\sqrt{2\al^\prime}e\cdot k_1)
\\-&(\al^\prime s+1)(\sqrt{2\al^\prime} \zeta_3\cdot k_2\zeta_1\cdot e)\\
+&(2\al^\prime \zeta_1 \cdot k_2\zeta_3\cdot k_2)(\sqrt{2\al^\prime} e\cdot k_1)
\end{array}\right]\\
+(\al^\prime t+1)(\al^\prime s+1)&\left[\begin{array}{rl}
-&(\al^\prime s-1)(2\al^\prime\zeta_1\cdot k_2\zeta_3\cdot k_2)(\sqrt{2\al^\prime}e\cdot k_3)\\
-&(\al^\prime t-1)(2\al^\prime\zeta_1\cdot k_2\zeta_3\cdot k_2)(\sqrt{2\al^\prime}e\cdot k_1) \\
+&(\al^\prime s-1)(\sqrt{2\al^\prime} \zeta_1\cdot k_2 \zeta_3\cdot e)\\
-&(2\al^\prime \zeta_1\cdot k_2\zeta_3 \cdot k_2)( \sqrt{2\al^\prime}e\cdot k_1)\\
+&\sqrt{2\al^\prime}\zeta_1 \cdot k_2 \zeta_3 \cdot e\\
+&\sqrt{2\al^\prime}\zeta_3\cdot k_2\zeta_1\cdot e\\
+& (\al^\prime t-1)(\sqrt{2\al^\prime}\zeta_3\cdot k_2 \zeta_1\cdot e)\\
-&(2\al^\prime \zeta_1 \cdot k_2\zeta_3\cdot k_2)(\sqrt{2\al^\prime} e\cdot k_3)\end{array}\right]\\
+(\al^\prime s+1)(\al^\prime s)&\left[\begin{array}{rl}&(\al^\prime t)(2\al^\prime\zeta_1\cdot k_2\zeta_3\cdot k_2)(\sqrt{2\al^\prime}e\cdot k_3)
\\-&(\al^\prime t+1)(\sqrt{2\al^\prime} \zeta_1\cdot k_2\zeta_3\cdot e)\\
+&(2\al^\prime \zeta_1\cdot k_2\zeta_3\cdot k_2) (\sqrt{2\al^\prime} e\cdot k_3)
\end{array}\right]
\end{array}\right\}\end{array}
\right\}
\eeqas
\beqa
\notag&=&\frac{\Ga(-\al^\prime t-1)\Ga(-\al^\prime s-1)}{\Ga(-\al^\prime t-\al^\prime s+2)}\times\\
\notag&&\hspace{-3cm}\times\left\{\begin{array}{l}
+(\al^\prime t+\al^\prime s-1)\left\{\begin{array}{rl}
(\al^\prime t+1)(\al^\prime t)(\al^\prime t-1)(\al^\prime s+1)&\left[\begin{array}{rl}
&(2\al^\prime\zeta_1\cdot k_2\zeta_3\cdot k_1) (\sqrt{2\al^\prime}e\cdot k_1)\\
-&(\sqrt{2\al^\prime}\zeta_3\cdot k_1\zeta_1\cdot e)\end{array}\right]\\
+(\al^\prime t+1)(\al^\prime t)(\al^\prime s+1)&
\left[\begin{array}{rl}
 &(\al^\prime s)(2\al^\prime\zeta_1\cdot k_3\zeta_3\cdot k_2)(\sqrt{2\al^\prime}e\cdot k_1)\\
-&(\al^\prime s)(2\al^\prime\zeta_1\cdot k_2 \zeta_3\cdot k_1)(\sqrt{2\al^\prime}e\cdot k_3)\\
-&(\al^\prime t-1)(2\al^\prime\zeta_1\cdot k_2 \zeta_3\cdot k_1)(\sqrt{2\al^\prime}e\cdot k_1)\\
+&(\al^\prime t-1)(2\sqrt{\al^\prime}\zeta_3\cdot k_1\zeta_1\cdot e)\end{array}\right]\\
+(\al^\prime t+1)(\al^\prime s+1)(\al^\prime s)&\left[\begin{array}{rl}&(\al^\prime t)(2\al^\prime\zeta_1\cdot k_2\zeta_3\cdot k_1)(\sqrt{2\al^\prime}e\cdot k_3)\\
-&(\al^\prime t)(2\al^\prime\zeta_1\cdot k_3 \zeta_3\cdot k_2)(\sqrt{2\al^\prime}e\cdot k_1)\\
-&(\al^\prime s-1)(2\al^\prime\zeta_1\cdot k_3 \zeta_3\cdot k_2)(\sqrt{2\al^\prime}e\cdot k_3)\\
+&(\al^\prime s-1)(\sqrt{2\al^\prime}\zeta_1\cdot k_3\zeta_3\cdot e)
\end{array}\right]\\
+(\al^\prime s+1)(\al^\prime s)(\al^\prime s-1)&\left[\begin{array}{rl}&(\al^\prime t)(2\al^\prime\zeta_1\cdot k_3\zeta_3\cdot k_2)(\sqrt{2\al^\prime}e\cdot k_3)
\\-&(\al^\prime t)(\sqrt{2\al^\prime} \zeta_1\cdot k_3\zeta_3\cdot e)\\
+&(2\al^\prime \zeta_1\cdot k_3\zeta_3\cdot k_2)( \sqrt{2\al^\prime} e\cdot k_3)
\end{array}\right]
\end{array}\right\}\\
+(\al^\prime t+\al^\prime s)(\al^\prime t+\al^\prime s-1)\left\{\begin{array}{rl}
(\al^\prime t+1)(\al^\prime t)(\al^\prime s+1)&\left[\begin{array}{rl}&(2\al^\prime\zeta_1\cdot k_2\zeta_3\cdot k_2)(\sqrt{2\al^\prime}e\cdot k_1)
\\-&(\sqrt{2\al^\prime} \zeta_3\cdot k_2\zeta_1\cdot e)
\end{array}\right]\\
+(\al^\prime t+1)(\al^\prime s+1)&\left[\begin{array}{rl}
-&(\al^\prime t)(2\al^\prime\zeta_1\cdot k_2\zeta_3\cdot k_2)(\sqrt{2\al^\prime}e\cdot k_1) \\
+& (\al^\prime t)(\sqrt{2\al^\prime}\zeta_3\cdot k_2 \zeta_1\cdot e)\\
-&(\al^\prime s)(2\al^\prime\zeta_1\cdot k_2\zeta_3\cdot k_2)(\sqrt{2\al^\prime}e\cdot k_3)\\
+&(\al^\prime s)(\sqrt{2\al^\prime} \zeta_1\cdot k_2 \zeta_3\cdot e)
\end{array}\right]\\
+(\al^\prime s+1)(\al^\prime s)&\left[\begin{array}{rl}&(\al^\prime t+1)(2\al^\prime\zeta_1\cdot k_2\zeta_3\cdot k_2)(\sqrt{2\al^\prime}e\cdot k_3)
\\-&(\al^\prime t+1)(\sqrt{2\al^\prime} \zeta_1\cdot k_2\zeta_3\cdot e)
\end{array}\right]
\end{array}\right\}\end{array}\right\}\\
&&\hspace{-3cm}=0.
\eeqa
    \item Case 4: Type II singlet zero-norm state
    In this case, we replace the polarization tensors as follows: 
    $\displaystyle\epsilon_{\mu\nu} \rightarrow3\al^\prime k_{2\mu}k_{2\nu}+\frac{1}{2}\eta_{\mu\nu}$ and $\displaystyle\epsilon_\mu\rightarrow \sqrt{\frac{\al^\prime}{2}}\Big(5k_{2\mu}-iV_\mu\Big)$
    \beqas
     2\al^\prime\epsilon_{\mu\nu}k^\mu_1k^\nu_1-\sqrt{2\al^\prime}\epsilon\cdot k_1
     \Rightarrow &&6\al^{\prime 2} (k_1\cdot k_2)^2+\al^\prime k_1^2 -5\al^\prime k_1\cdot k_2 +i\al^\prime k_1\cdot V\\
     &=&\frac{3}{2}(\al^\prime s-1)^2+\frac{5}{2}(\al^\prime s-1)=\frac{3}{2}(\al^\prime s)(\al^\prime s-1)+(\al^\prime s-1)\\
     -4\al^\prime\epsilon_{\mu\nu}k^\mu_1k^\nu_3
     \Rightarrow && -12\al^{\prime2}k_1\cdot k_2 k_2\cdot k_3-2\al^\prime k_1\cdot k_3 \\
     &=&-3(\al^\prime t-1)(\al^\prime s-1)-(\al^\prime t+\al^\prime s)\\
     2\al^\prime\epsilon_{\mu\nu}k^\mu_3k^\nu_3-\sqrt{2\al^\prime}\epsilon\cdot k_3
     \Rightarrow &&6\al^{\prime 2} (k_2\cdot k_3)^2+\al^\prime k_3^2 -5\al^\prime k_2\cdot k_3 +i\al^\prime k_3\cdot V\\
     &=&\frac{3}{2}(\al^\prime t-1)^2+\frac{5}{2}(\al^\prime t-1)=\frac{3}{2}(\al^\prime t)(\al^\prime t-1)+(\al^\prime t-1)\\
     -2\sqrt{2\al^\prime}\epsilon_{\mu\nu}\zeta_1^\mu k_1^\nu+2\epsilon\cdot\zeta_1\quad
     \Rightarrow &&  (-6\al^\prime k_1\cdot k_2) (\sqrt{2\al^\prime} \zeta_1\cdot k_2)-\sqrt{2\al^\prime}\zeta_1\cdot k_1\\
     &&+5\sqrt{2\al^\prime}\zeta_1\cdot k_2-\sqrt{2\al^\prime}i\zeta_1\cdot V\\
     &=&3(\al^\prime s+1)(\sqrt{2\al^\prime} \zeta_1\cdot k_2)-\sqrt{2\al^\prime} \zeta_1\cdot k_2\\
     2\sqrt{2\al^\prime}\epsilon_{\mu\nu}\zeta_1^\mu k_3^\nu\quad
     \Rightarrow && (6\al^\prime k_2\cdot k_3) (\sqrt{2\al^\prime} \zeta_1\cdot k_2)+\sqrt{2\al^\prime}\zeta_1\cdot k_3\\
     &=&-3(\al^\prime t-1) (\sqrt{2\al^\prime} \zeta_1\cdot k_2)+\sqrt{2\al^\prime}\zeta_1\cdot k_3\\
     -2\sqrt{2\al^\prime}\epsilon_{\mu\nu}\zeta_3^\mu k_3^\nu+2\epsilon\cdot\zeta_3\quad
     \Rightarrow &&  (-6\al^\prime k_2\cdot k_3) (\sqrt{2\al^\prime} \zeta_3\cdot k_2)-\sqrt{2\al^\prime}\zeta_3\cdot k_3\\
     &&+5\sqrt{2\al^\prime}\zeta_3\cdot k_2-\sqrt{2\al^\prime}i\zeta_3\cdot V\\
     &=&3(\al^\prime t+1)(\sqrt{2\al^\prime} \zeta_3\cdot k_2)-\sqrt{2\al^\prime} \zeta_3\cdot k_2\\
     2\sqrt{2\al^\prime}\epsilon_{\mu\nu}\zeta_3^\mu k_1^\nu\quad
     \Rightarrow && (6\al^\prime k_1\cdot k_2) (\sqrt{2\al^\prime} \zeta_3\cdot k_2)+\sqrt{2\al^\prime}\zeta_3\cdot k_1\\
     &=&-3(\al^\prime s-1) (\sqrt{2\al^\prime} \zeta_3\cdot k_2)+\sqrt{2\al^\prime}\zeta_3\cdot k_1\\
     -2\epsilon_{\mu\nu}\zeta_1^\mu\zeta_3^\nu \quad \Rightarrow && -6\al^\prime \zeta_1\cdot k_2\ \zeta_3\cdot k_2-\zeta_1\cdot \zeta_3.
     \eeqas
\beqas
\notag&&\cA_{TPMP}=\frac{\Ga(-\al^\prime t-1)\Ga(-\al^\prime s-1)}{\Ga(-\al^\prime t-\al^\prime s+2)}\times\\
&&\hspace{-3cm}\times\left\{\begin{array}{l}
(\al^\prime t+1)(\al^\prime s+1)
\left\{\begin{array}{rl}
&(\al^\prime t)(\al^\prime t-1)(\zeta_1\cdot\zeta_3-2\al^\prime\zeta_1\cdot k_3 \zeta_3\cdot k_1)\Big[\frac{3}{2}(\al^\prime s)(\al^\prime s-1)+(\al^\prime s-1)\Big]\\
+&(\al^\prime t)(\al^\prime s)(\zeta_1\cdot\zeta_3-2\al^\prime\zeta_1\cdot k_3 \zeta_3\cdot k_1)\left[\begin{array}{rl}-&3(\al^\prime t-1)(\al^\prime s-1)\\-&(\al^\prime t+\al^\prime s)\end{array}\right]\\
+&(\al^\prime s)(\al^\prime s-1)(\zeta_1\cdot\zeta_3-2\al^\prime\zeta_1\cdot k_3 \zeta_3\cdot k_1)\Big[\frac{3}{2}(\al^\prime t)(\al^\prime t-1)+(\al^\prime t-1)\Big]
\end{array}\right\}\\
+(\al^\prime t+\al^\prime s-1)\left\{\begin{array}{rl}
\left[\begin{array}{c}(\al^\prime t+1)(\al^\prime t)\\(\al^\prime t-1)\end{array}\right]&\left\{\begin{array}{r}
-2\al^\prime\zeta_1\cdot k_2\zeta_3\cdot k_1\Big[\frac{3}{2}(\al^\prime s)(\al^\prime s-1)+(\al^\prime s-1)\Big]\\
+\sqrt{2\al^\prime} \zeta_3\cdot k_1\left[\begin{array}{rl}-&3(\al^\prime s+1)(\sqrt{2\al^\prime} \zeta_1\cdot k_2)\\+&\sqrt{2\al^\prime} \zeta_1\cdot k_2\end{array}\right]\end{array}\right\}\\
+\left[\begin{array}{c}(\al^\prime t+1)(\al^\prime t)\\(\al^\prime s+1)\end{array}\right]&\left\{\begin{array}{rl}-&2\al^\prime\zeta_1\cdot k_3\zeta_3\cdot k_2\Big[\frac{3}{2}(\al^\prime s)(\al^\prime s-1)+(\al^\prime s-1)\Big]\\
+&2\al^\prime\zeta_1\cdot k_2 \zeta_3\cdot k_1\left[\begin{array}{rl}&3(\al^\prime t-1)(\al^\prime s-1)\\+&(\al^\prime t+\al^\prime s)\end{array}\right]\\
-&\sqrt{2\al^\prime}\zeta_3\cdot k_1\left[\begin{array}{rl}
-&3(\al^\prime t-1) (\sqrt{2\al^\prime} \zeta_1\cdot k_2)\\+&\sqrt{2\al^\prime}\zeta_1\cdot k_3
\end{array}\right]\end{array}\right\}\\
+\left[\begin{array}{c}(\al^\prime t+1)\\(\al^\prime s+1)(\al^\prime s)\end{array}\right]&\left\{\begin{array}{rl}-&2\al^\prime\zeta_1\cdot k_2\zeta_3\cdot k_1\Big[\frac{3}{2}(\al^\prime t)(\al^\prime t-1)+(\al^\prime t-1)\Big]
\\+&2\al^\prime\zeta_1\cdot k_3 \zeta_3\cdot k_2\left[\begin{array}{rl}&3(\al^\prime t-1)(\al^\prime s-1)\\+&(\al^\prime t+\al^\prime s)\end{array}\right]\\-&\sqrt{2\al^\prime}\zeta_1\cdot k_3\left[\begin{array}{rl}
-&3(\al^\prime s-1) (\sqrt{2\al^\prime} \zeta_3\cdot k_2)\\+&\sqrt{2\al^\prime}\zeta_3\cdot k_1
\end{array}\right]\end{array}\right\}\\
+\left[\begin{array}{c}(\al^\prime s+1)(\al^\prime s)\\(\al^\prime s-1)\end{array}\right]&\left\{\begin{array}{r}-2\al^\prime\zeta_1\cdot k_3\zeta_3\cdot k_2\Big[\frac{3}{2}(\al^\prime t)(\al^\prime t-1)+(\al^\prime t-1)\Big]
\\+\sqrt{2\al^\prime} \zeta_1\cdot k_3\left[\begin{array}{rl}-&3(\al^\prime t+1)(\sqrt{2\al^\prime} \zeta_3\cdot k_2)\\+&\sqrt{2\al^\prime} \zeta_3\cdot k_2\end{array}\right]\end{array}\right\}
\end{array}\right\}\\
+\left[\begin{array}{c}(\al^\prime t+\al^\prime s)\\(\al^\prime t+\al^\prime s-1)\end{array}\right]\left\{\begin{array}{rl}
(\al^\prime t+1)(\al^\prime t)&\left\{\begin{array}{r}-2\al^\prime\zeta_1\cdot k_2\zeta_3\cdot k_2\Big[\frac{3}{2}(\al^\prime s)(\al^\prime s-1)+(\al^\prime s-1)\Big]
\\+\sqrt{2\al^\prime} \zeta_3\cdot k_2\left[\begin{array}{rl}-&3(\al^\prime s+1)(\sqrt{2\al^\prime} \zeta_1\cdot k_2)\\+&\sqrt{2\al^\prime} \zeta_1\cdot k_2\end{array}\right]\end{array}\right\}\\
+\left[\begin{array}{c}(\al^\prime t+1)\\(\al^\prime s+1)\end{array}\right]&\left\{\begin{array}{rl}&2\al^\prime\zeta_1\cdot k_2\zeta_3\cdot k_2 \Big[3(\al^\prime t-1)(\al^\prime s-1)+(\al^\prime t+\al^\prime s)\Big]\\
-&\sqrt{2\al^\prime} \zeta_1\cdot k_2
\left[\begin{array}{rl}
-&3(\al^\prime s-1) (\sqrt{2\al^\prime} \zeta_3\cdot k_2)\\+&\sqrt{2\al^\prime}\zeta_3\cdot k_1
\end{array}\right]\\
+&6\al^\prime \zeta_1\cdot k_2 \zeta_3\cdot k_2+\zeta_1\cdot \zeta_3-4\al^\prime \zeta_3\cdot k_2\\
-&\sqrt{2\al^\prime}\zeta_3\cdot k_2\left[\begin{array}{rl}
-&3(\al^\prime t-1) (\sqrt{2\al^\prime} \zeta_1\cdot k_2)\\+&\sqrt{2\al^\prime}\zeta_1\cdot k_3
\end{array}\right]\end{array}\right\}\\
+(\al^\prime s+1)(\al^\prime s)&\left\{\begin{array}{r}-2\al^\prime\zeta_1\cdot k_2\zeta_3\cdot k_2\Big[\frac{3}{2}(\al^\prime t)(\al^\prime t-1)+(\al^\prime t-1)\Big]
\\+\sqrt{2\al^\prime} \zeta_1\cdot k_2\left[\begin{array}{rl}-&3(\al^\prime t+1)(\sqrt{2\al^\prime} \zeta_3\cdot k_2)\\+&\sqrt{2\al^\prime} \zeta_3\cdot k_2\end{array}\right]\end{array}\right\}
\end{array}\right\}\end{array}
\right\}
\eeqas

\beqa
\notag&=&\frac{\Ga(-\al^\prime t-1)\Ga(-\al^\prime s-1)}{\Ga(-\al^\prime t-\al^\prime s+2)}\times\\
\notag&&\hspace{-3cm}\times\left\{\begin{array}{l}
(\al^\prime t+1)(\al^\prime s+1)
\left[\begin{array}{rl}
&(\al^\prime t)(\al^\prime t-1)(\al^\prime s-1)(\zeta_1\cdot\zeta_3-2\al^\prime\zeta_1\cdot k_3 \zeta_3\cdot k_1)\\
-&(\al^\prime t)(\al^\prime s)(\al^\prime t+\al^\prime s)(\zeta_1\cdot\zeta_3-2\al^\prime\zeta_1\cdot k_3 \zeta_3\cdot k_1)\\
+&(\al^\prime s)(\al^\prime s-1)(\al^\prime t-1)(\zeta_1\cdot\zeta_3-2\al^\prime\zeta_1\cdot k_3 \zeta_3\cdot k_1)
\end{array}\right]\\
+(\al^\prime t+\al^\prime s-1)
\left[\begin{array}{rl}
-(\al^\prime t+1)(\al^\prime t)(\al^\prime t-1)(\al^\prime s+1)&
(2\al^\prime\zeta_1\cdot k_2\zeta_3\cdot k_1)\\
-(\al^\prime t+1)(\al^\prime t)(\al^\prime s+1)(\al^\prime s-1)&(2\al^\prime\zeta_1\cdot k_3\zeta_3\cdot k_2)\\
+(\al^\prime t+1)(\al^\prime t)(\al^\prime s+1)(\al^\prime t+\al^\prime s)&(2\al^\prime\zeta_1\cdot k_2 \zeta_3\cdot k_1)\\
-(\al^\prime t+1)(\al^\prime t)(\al^\prime s+1)&(2\al^\prime\zeta_1\cdot k_3\zeta_3\cdot k_1)\\
-(\al^\prime t+1)(\al^\prime s+1)(\al^\prime s)(\al^\prime t-1)&(2\al^\prime\zeta_1\cdot k_2\zeta_3\cdot k_1)\\
+(\al^\prime t+1)(\al^\prime s+1)(\al^\prime s)(\al^\prime t+\al^\prime s)&(2\al^\prime\zeta_1\cdot k_3 \zeta_3\cdot k_2)\\
-(\al^\prime t+1)(\al^\prime s+1)(\al^\prime s)&(2\al^\prime\zeta_1\cdot k_3\zeta_3\cdot k_1)\\

-(\al^\prime s+1)(\al^\prime s)(\al^\prime s-1)(\al^\prime t+1)&(2\al^\prime\zeta_1\cdot k_3\zeta_3\cdot k_2)
\end{array}\right]\\
+(\al^\prime t+\al^\prime s)(\al^\prime t+\al^\prime s-1)\left[\begin{array}{rl}
-(\al^\prime t+1)(\al^\prime t)(\al^\prime s+1)&(2\al^\prime\zeta_1\cdot k_2\zeta_3\cdot k_2)\\
+(\al^\prime t+1)(\al^\prime s+1)(\al^\prime t+\al^\prime s)&(2\al^\prime\zeta_1\cdot k_2\zeta_3\cdot k_2)\\
-(\al^\prime t+1)(\al^\prime s+1)&(2\al^\prime\zeta_1\cdot k_2 \zeta_3\cdot k_1)\\
-(\al^\prime t+1)(\al^\prime s+1)&(2\al^\prime\zeta_1\cdot k_3 \zeta_3\cdot k_2)\\
+(\al^\prime t+1)(\al^\prime s+1)&\zeta_1\cdot\zeta_3\\
-(\al^\prime s+1)(\al^\prime s)(\al^\prime t+1)&(2\al^\prime\zeta_1\cdot k_2\zeta_3\cdot k_2)
\end{array}\right]
\end{array}\right\}\\
&&\hspace{-3cm}=0.
\eeqa
  \end{itemize}
  
\section{Summary and Discussion}
\par In this paper, we study the bosonic open string theory in the linear dilaton background. Based on the operator formulation, we calculate the physical state spectrum of the stringy excitations up to the first massive level and the tree-level scattering amplitudes among these physical states. To be specific, we obtain the following new results:
\begin{enumerate}
  \item[(1)] We obtain the physical state conditions in the old covariant first quantization spectrum up to the first massive level.
  \item[(2)] We obtain the normal-ordered vertex operators corresponding to the physical states and show that they satisfy the conformal algebra.
  \item[(3)] Based on oscillator representation and coherent state method, we calculate three-point and four point stringy scattering amplitudes among tachyon, photon, and spin-two tensors.
  \item[(4)] We verify the decoupling of zero-norm states in the on-shell stringy scattering amplitudes for the three and four-point functions under consideration. These stringy Ward identities can be viewed as consistent checks for our calculations of stringy amplitudes. It also supports the idea of modified energy-momentum conservation rule in the definition of inner product for the center of mass degrees of freedom through inner product Eq.\eqref{in}
  \end{enumerate}
\par   The physical motivation for studying this particular time-dependent background is that this is the simplest example one can study the issue of high-energy symmetry and the background independence of string theory in a quantitative way. For detailed explanations and the physical implications, please refer to \cite{Chan:2009kb}. In order to realize the universal property of the high-energy stringy symmetry, one can connect two fixed-points in the moduli space of string theory by a continuous path, and we should be able to map out the spectral flow of stringy spectrum and observe the deformation of the high-energy stringy symmetry as we move from one fixed point to another.
\par
Indeed, our results in this paper already provide some hints of this idea. Specifically, from the result (1), we can see that there is a clear deformation of physical state condition as we change the gradient of dilaton field $V^\mu$. One can also see that if we turn off the linear dilaton field, all the vertex operators in result (2) are identical to the standard flat space-time vertex operators. Finally, with the general definitions of (deformed) kinematic variables, including Mandelstam variables, Eq.\eqref{MadV}, we find that there exists no apparent dependence of the dilaton gradient in the stringy scattering amplitudes. Consequently, the zero-dilaton limits of the stringy scattering amplitudes in the result (3) naturally coincide wit those of flat space-time.
\par
Our explicit checks of the stringy Ward identities (result (4)) provides a concrete example of deformation of stringy gauge symmetry (including massive excitations) as we migrate in the moduli space. To see the connection with high-energy stringy symmetry, we need to specify the kinematic set-up and perform a systematic expansion of all kinematic variables to obtain the high-energy limits. These calculations are reported in \cite{Chan:2009kb}.
\appendix
\section{Useful Commutation Relations}\label{ApxB}
Based on fundamental commutation relations
\beqas
[\al^\mu_n.\al^\nu_m]=n\de_{n+m}\eta^{\mu\nu} \quad\mbox{and}\quad [x^\mu,p^\nu]=i\eta^{\mu\nu}.
\eeqas
One can derive the following equalities
\beqa
\Big[\al_0^\mu,\exp\big(ik\cdot X_{0}\big)\Big]&=&\sqrt{2\al^\prime}k^\mu\exp\big(ik\cdot X_{0}\big)\\
\Big[\al_n^\mu,\exp\big(ik\cdot X_{-}\big)\Big]&=&\sqrt{2\al^\prime}k^\mu e^{in\tau}\exp\big(ik\cdot X_{-}\big)\qquad(n>0)\\
\label{AB3}\Big[\al_n^\mu,\exp\big(ik\cdot X_{+}\big)\Big]&=&\sqrt{2\al^\prime}k^\mu e^{in\tau}\exp\big(ik\cdot X_{+}\big)\qquad(n<0).
\eeqa
From this it is easy to see
\beqa
\Big[\al^\mu_n,V_T\Big]=\sqrt{2\al^\prime}k^\mu e^{in\tau}V_T\qquad(\forall n\in Z).
\eeqa
For normal-ordering calculation,
\beqa
\notag\Big[\dot X^\mu_+, \exp\big(ik\cdot X_{-}\big)\Big]&=&\Big[\sqrt{2\al^\prime}\sum_{n=1}^\infty\al_{n}^\mu e^{-in\tau},\exp\Big(\sqrt{2\al^\prime}k\cdot\sum_{m=1}^{\infty}\frac{\al_{-m}}{m}e^{im\tau}\Big)\Big]\\
&=&2\al^\prime k^\mu\exp\big(ik\cdot X_{-}\big)\sum_{n=1}^{\infty}\mathbf{1}=-\al^\prime k^\mu\exp\big(ik\cdot X_{-}\big)\\
\notag\Big[\dot X^\mu_+, \dot X^\nu_-\Big]&=&
\Big[\sqrt{2\al^\prime}\sum_{n=1}^{\infty}\al_n^\mu e^{-in\tau},\sqrt{2\al^\prime}\sum_{n=1}^{\infty}\al_{-m}^\nu e^{im\tau}\Big]
 \\&=&2\al^\prime\eta^{\mu\nu}\Big(\sum_{n,m=1}^{\infty}n\Big)=-\frac{\al^\prime}{6}\eta^{\mu\nu}\\
\notag \Big[\ddot X^\mu_+, \exp\big(ik\cdot X_{-}\big)\Big]&=&\Big[ -i\sqrt{2\al^\prime}\sum_{n=1}^\infty n\al_{n}^\mu e^{-in\tau},\exp\big(\sqrt{2\al^\prime}k\cdot\sum_{m=1}^{\infty}
 \frac{\al_{-m}}{m}e^{im\tau}\big)\Big]\\
 &=&-2i\al^\prime k^\mu\exp\big(ik\cdot X_{-}\big)\Big(\sum_{n=1}^{\infty}n\Big)=\frac{ik^\mu\al^\prime}{6}\exp\big(ik\cdot X_{-}\big).
\eeqa
In the calculation of the 4-point scattering amplitudes, the following formulae are useful:
\beqa
\notag\Big[\exp\big[ik\cdot X_+(1)\big],\exp\big[ik^\prime\cdot X_-(y)\big]\Big]&=&\exp\Big(-2\al^\prime k\cdot k^\prime\sum_{n=1}^{\infty}\frac{y^n}{n}\Big)\exp\big[ik^\prime\cdot X_-(y)\big]\\
&=&(1-y)^{2\al^\prime k\cdot k^\prime}\exp\big[ik^\prime\cdot X_-(y)\big]\\
\notag\Big[\exp\big[ik\cdot X_+(1)\big],\dot X_-^\mu(y) \Big]&=&-2\al^\prime k^\mu\sum_{n=1}^\infty y^n \exp[ik\cdot X_+(1)]\\
&=&\frac{-2\al^\prime k^\mu y}{1-y}\exp\big[ik\cdot X_+(1)\big]\\
\notag\Big[\exp\big[ik\cdot X_+(1)\big],\ddot X_-^\mu(y) \Big]&=&-2i\al^\prime k^\mu\sum_{n=1}^\infty ny^n \exp\big[ik\cdot X_+(1)\big]\\
&=&\frac{-2i\al^\prime k^\mu y}{(1-y)^2}\exp\big[ik\cdot X_+(1)\big]\\
\notag\Big[\dot X_0^\mu(1),\exp\big[ik\cdot X_0(y)\big] \Big]&=&[2\al^\prime p^\mu, \exp\big(ik\cdot x\big)]y^{2\al^\prime k\cdot p}y^{\al^\prime k^2}\\
&=&2\al^\prime k^\mu \exp\big[ik\cdot X_0(y)\big]\\
\notag\Big[\zeta\cdot\dot X_+(1),\dot X^\mu_-(y)\Big]&=&\Big[\sqrt{2\al^\prime}\sum_{n=1}^\infty \zeta\cdot \al_n,\sqrt{2\al^\prime}\sum_{m=1}^\infty \al_{-m}^\mu y^m\Big]\\
&=&2\al^\prime\zeta^\mu\sum_{n=1}^\infty ny^n=\frac{2\al^\prime \zeta^\mu y}{(1-y)^2}\\
\notag\Big[\zeta\cdot\dot X_+(1),\ddot X^\mu_-(y)\Big]
\notag&=&\Big[\sqrt{2\al^\prime}\sum_{n=1}^\infty \al_n,i\sqrt{2\al^\prime}\sum_{m=1}^\infty m\al_{-m}y^m\Big]\\
&=&2\al^\prime i\zeta^\mu\sum_{n=1}^\infty n^2y^n=\frac{2\al^\prime i\zeta^\mu (1+y)y}{(1-y)^3}\\
\Big[\zeta\cdot\dot X_+(1).\exp\big[ik\cdot X_-(y)\big]\Big]&=&\frac{2\al^\prime\zeta\cdot k y}{1-y}
\eeqa

\section{Useful Four-point Functions}\label{ApxC}
\beqas
A_4&=&ZW, \\
Z&:&\mbox{zero mode part of $A_4$}\\
W&:&\mbox{nonzero mode part of $A_4$}
\eeqas
\beqas
Z
&=&\bra{0;k_4}\exp(ik_3\cdot x)\exp(ik_2\cdot x)\exp(-\al^\prime k_2\cdot V\tau)\exp(2i\al^\prime k_2\cdot p\tau)\exp(i\al^\prime k_2^2\tau)\ket{0;k_4}\\
&=&\bra{0;k_1}\exp(ik_3\cdot x)\exp(ik_2\cdot x)\ket{0;k_4}\exp(2\al^\prime k_1\cdot k_2\tau)\exp\big[i\al^\prime k_2\cdot(k_2+iV)\tau\big]\\
&=&\exp[i(2\al^\prime k_1\cdot k_2-m_2^2)\tau]\de(k_4^\ast-k_3-k_2-k_1)\\
&=&y^{2\al^\prime k_1\cdot k_2-m_2^2}\de(k_4^\ast-k_3-k_2-k_1)\\
&=&y^{-\al^\prime s+m_1^2}
\eeqas
\beqas
W&=&\bra{0}\exp(\sum_{n=1}^{\infty}\sqrt{2\al^\prime}k_2\cdot\frac{\al_{-n}}{n})
\exp(-\sum_{n=1}^{\infty}\sqrt{2\al^\prime}k_2\cdot\frac{\al_n}{n})\times\\
&&\times\exp(\sum_{m=1}^{\infty}\sqrt{2\al^\prime}k_3\cdot\frac{\al_{-m}}{m} y^m)
\exp(-\sum_{m=1}^{\infty}\sqrt{2\al^\prime}k_3\cdot\frac{\al_m}{m} y^{-m})
\ket{0}\\
&=&\bra{0}
\exp(-\sum_{n=1}^{\infty}\sqrt{2\al^\prime}k_2\cdot\frac{\al_n}{n})\exp(\sum_{m=1}^{\infty}\sqrt{2\al^\prime}k_3\cdot\frac{\al_{-m}}{m} y^m)
\ket{0}\\
&=&\exp(-2\al^\prime\sum_{n=1}^\infty k_2\cdot k_3\frac{y^n}{n})\\
&=&\exp[2\al^\prime k_2\cdot k_3\ln(1-y)]\\
&=&(1-y)^{2\al^\prime k_2\cdot k_3}\\
&=&(1-y)^{-\al^\prime t+m_2^2+m_3^2}
\eeqas
\beqa
A_4=y^{-\al^\prime s+m_1^2}(1-y)^{-\al^\prime t+m_2^2+m_3^2}.
\eeqa
\beqa
\notag\mathbb{P}&\equiv&\bra{0,k_4}\cV_T(k_3,1)\cV_T(k_2,y)\zeta\cdot\al_{-1}\ket{0,k_1}\\
\notag&=&\bra{0,k_4}\cV_T(k_3,1)\Big[\cV_T(k_2,y),\zeta\cdot\al_{-1}\Big]\ket{0,k_1}+
\bra{0,k_4}\Big[\cV_T(k_3,1),\zeta\cdot\al_{-1}\Big]\cV_T(k_2,y)\ket{0,k_1}\\
\label{AC2}&=&-\sqrt{2\al^\prime}(\zeta\cdot k_2 y^{-1}+\zeta\cdot k_3)A_4.
\eeqa
In the second line, we use the Eq.\eqref{AB3} in Appendix \ref{ApxB}. 
\begin{acknowledgments}
The authors wish to acknowledge Pravina Borhade, Pei-Ming Ho, Jen-Chi Lee, Shunsuke Teraguchi, Yi Yang, and Chi-Hsien Yeh for early collaborations in shaping the idea of this work. This work is supported by the
 National Science Council of Taiwan under the contract 96-2112-M-029-002-MY3 and the string focus group under the  National Center for Theoretical Sciences.
\end{acknowledgments}

{}
\end{document}